\RequirePackage{rotating}
\documentclass[reprint,
showkeys,
nofootinbib,
 amsmath,amssymb,
 aps,
floatfix,natbib,16pt, twocolumn
]{revtex4-1}
\setcitestyle{authoryear,round}
\usepackage{graphicx}
\usepackage{natbib}
\usepackage{bm}%
\usepackage{rotating}
\usepackage{xcolor,colortbl}
\usepackage{booktabs}
\usepackage{wrapfig}
\usepackage{setspace}
\usepackage{cellspace}
\usepackage{latexsym,color}
\usepackage{amssymb}
\usepackage[T1]{fontenc}
\usepackage{amsmath,amssymb,mathrsfs,amsfonts,dsfont}
\usepackage{color}
\usepackage{xcolor}
\definecolor{LinkBlue}{rgb}{0.00,0.00,1.00}

\usepackage[colorlinks=true,linkcolor=LinkBlue,urlcolor=Grree,	 citecolor=LinkBlue,citecolor=LinkBlue,hyperfootnotes=true]{hyperref}

\definecolor{lightgray}{rgb}{0.75,0.75,0.75}

\def\be{\begin{equation}}
\def\ee{\end{equation}}
\def\bea{\begin{eqnarray}}
\newcommand{\cc}{\mathrm{C}}
\newcommand{\apss}{Ap\&SS}
\newcommand{\apjs}{APJS}
\newcommand{\mnras}{MNRAS}
\newcommand{\jj}{\mathrm{J}}
\newcommand{\aap}{A\&A}
\def\eea{\end{eqnarray}}

\newcommand{\pp}{\textbf{()}}
\newcommand{\non}[1]{\text{{{$\not$}}}{#1}}
\newcommand{\mso}{\mathrm{mso}}
\newcommand{\mbo}{\mathrm{mbo}}
\newcommand{\Qa}{\mathcal{Q}}
\newcommand{\il}{~}

    \newcommand{\oo}{\mathrm{O}}

\count\footins = 1000

\begin{document}

\title{
Relating   Kerr \textbf{SMBHs} in Active Galactic Nuclei to \textbf{RAD} configurations}

\author{D. Pugliese, Z. Stuchl\'{\i}k}

\affiliation{
Institute of Physics and Research Centre of Theoretical Physics and Astrophysics, Faculty of Philosophy\&Science, Silesian University in Opava,
 Bezru\v{c}ovo n\'{a}m\v{e}st\'{i} 13, CZ-74601 Opava, Czech Republic\\
 \email{d.pugliese.physics@gmail.com;zdenek.stuchlik@physics.cz} }

 \date{Received XXXX xx,XXXX; accepted YYYY yy, YYYY}


  \begin{abstract}
{{There is strong  observational evidence that many   active galactic nuclei ({\textbf{AGNs}})   harbour    super-massive  black holes ({\textbf{SMBHs}}), demonstrating  multi-accretion episodes during their life-time.}
In such  \textbf{AGNs},  corotating and counterrotating tori, or strongly misaligned disks, as related  to the central Kerr {\textbf{SMBH}} spin, can report traces of the \textbf{AGNs} evolution.}
{Here we concentrate on aggregates of accretion disks structures, \textbf{r}inged \textbf{a}ccretion \textbf{d}isks  ({\textbf{RADs}}) orbiting a central Kerr \textbf{SMBH}, assuming that  each torus of the \textbf{RADs} is centered in the equatorial  plane of the attractor, {tori  are \emph{coplanar} and axi-symmetric}. Many of the  \textbf{RAD} aspects  are governed mostly by the spin  of the Kerr geometry.}
{We classify Kerr black holes  (\textbf{BHs}) due to their dimensionless spin,  according to possible combinations of corotating and counterrotating  equilibrium or unstable (accreting) tori composing the \textbf{RADs}.
The number of accreting tori in \textbf{RADs} cannot exceed $n=2$. We present list  of $14$ characteristic values  of the Kerr \textbf{BH} dimensionless spin $a$ governing the classification in  whole    the black hole range $0\leq a\leq M$, uniquely constrained by  the {\textbf{RAD}} properties.
}
{The spin values  are remarkably close providing an accurate characterization  of the  Kerr attractors   based on the \textbf{RAD} properties.   \textbf{RAD} dynamics is richer in the spacetimes of  high spin values.
One of the critical predictions states that a
\textbf{RAD} tori couple formed by  an  outer accreting corotating   and  an inner  accreting counterrotating torus is  expected to be    observed only around slowly   spinning ({$a<0.46M$}) \textbf{BHs}.
The analysis  strongly binds the fluid and \textbf{BH} characteristics providing  indications  on the situations where to search for \textbf{RADs} observational evidences.
  Obscuring and  screening tori, possibly evident  as traces   in X-ray spectrum emission, are strongly constrained, eventually ruling out many  assumptions used in the current investigations of the screening effects. %
  }
{ We expect relevance  of our classification of Kerr spacetimes  in relation to astrophysical phenomena arising in different  stages  of \textbf{AGNs} life  that could be observed by   the planned X-ray satellite  observatory \textbf{ATHENA} (\textbf{A}dvanced \textbf{T}elescope for \textbf{H}igh \textbf{EN}ergy \textbf{A}strophysics).}
\end{abstract}

\date{\today}

\maketitle

\section{Introduction}

Black hole  (\textbf{BH})  physics    and the \textbf{BH} accretion disk investigation   is developing in the last years.  The launch of new satellite  observatories in the near future allows  an unprecedented close look at situations and contexts which were  inconceivable only a few years ago.
Observable features  on the accreting  disks, focusing on  morphology of accretion processes and associated  jet  emissions provide increasingly more detailed and focused pictures of these objects.  The sensational opening of a new observational era  represented by gravitational waves (\textbf{GWs}) detection  allows us to focus on questions of broader  interest involving  more deeply  the \textbf{BH} physics.
On the other hand, the theoretical modeling seems to resort deeply to these improvements. Concerning the \textbf{BH} accretion disks processes, there is  great  expectation   towards  the X-ray  emission sector, with several missions as \textbf{XMM-Newton} (X-ray Multi-Mirror Mission)\footnote{{\textbf{http://sci.esa.int/science-e/www/area/index.cfm?fareaid=23}}},  \textbf{RXTE} (Rossi X-ray Timing
Explorer)\footnote{{\textbf{http://heasarc.gsfc.nasa.gov/docs/xte/xtegof.html}}} or \textbf{ATHENA}\footnote{\textbf{http://the-athena-x-ray-observatory.eu/}}. Recent studies  point out  also   an interesting  possible  connection between accretion  processes and \textbf{GWs} \citep{PRL}.
In this scenario, however,  many questions remain still  unresolved, leaving them as  intriguing  problems of observational astrophysics,  appearing  to demand even a greater effort from the point of view of the model development,  as the   still missing solution of \textbf{g}amma \textbf{r}ay \textbf{b}ursts (\textbf{GRBs}) origin, the jet launch, the \textbf{q}uasi-\textbf{p}eriodic \textbf{o}scillations (QPOs), the formation  of  \textbf{SMBHs}  in \textbf{AGNs}.
In general, the   theoretical  investigation is increasingly oriented towards the attempt to find a correlation between different phenomena and a broader embedding  environment, {creating a  general framework of analysis envisaging the \textbf{\textbf{BH}}-disk
system as a whole.} In this sense we may talk about an  {Environmental}  {Astrophysics}.
 Evidences of this fact are the
 debates on the jet-accretion correlation,
the \textbf{BH} accretion rate-disk luminosity issue, the \textbf{BH} growth--accretion
disk and \textbf{BH}--spin shift--accretion disk correlation and   \textbf{BH} populations and galaxy age correlation--see for example \citet{Hamb1,Sadowski:2015jaa,Madau(1988),Ricci:2017wmr,Narayan:2013gca,Mewes:2015gma,Morningstar:2014hea,
Yu:2015lqj,Volonteri:2002vz,Yoshida,Regan:2017vre,Yang:2017slb,Xie:2017jbz}.

In this regard, a crucial aspect  to establish    correlation  between \textbf{SMBHs} and their environment is first the  recognition of the  \textbf{BH} attractor.
Although  the unambiguous \textbf{SMBH}  identification     reduces to assign the \textbf{BH}  spin $a$ and mass $M$ parameters or even, for many purposes, { only the \textbf{BH}}  spin-mass ratio $a/M$, this task is still  controversial and debated issue of the  \textbf{BH} astrophysics.
We shall see    in fact  that many aspects of the \textbf{BH}  accretion disk physics depend only  on the ratio $a/M$.
The issue to  identify a  rotating \textbf{BH} by determining its  intrinsic rotation, or spin-mass ratios, is a rather  difficult issue, a complex task which is challenged by  different observational and theoretical approaches. All these    methods  are continuously debated and confronted--see for example \citet{Capellupo:2017qpt,McClintock:2006xd,Daly:2008zk}.
It should be noted then that the evaluation of the \textbf{SMBHs} spin is strictly  correlated with the ``mass-problem'': the  assessment of the precise value of the spin parameter of the \textbf{BH}  is connected with  the evaluation of  the main features  of the \textbf{BH} accretion disk system,  as the \textbf{BH} accretion rate or the location of the inner edge of the accretion disk.
The \textbf{GWs}  detection from coalescence of  \textbf{BHs} in a binary system  may serve, in   future, as a further possible  method to fix a \textbf{BH} spin parameter \citep{Farr:2017uvj,vanPutten:2016wpa,vanPutten:2012vr,vanPutten:2015eda}.
However, nowadays  this task is often  approached  in \textbf{BH}-accretion disk framework,  by considering the  evaluation of the  mass accretion rate (connected with the disk luminosity) or, for example, the location of  the   inner edge of an  accreting disk. Nevertheless all these aspects are certainly not settled    in one univocal picture; for example, even the definition of the inner edge of an  accreting disk  is  very controversial -see for example \citet{Krolik:2002ae,BMP98,2010A&A...521A..15A,Agol:1999dn,Paczynski:2000tz,open,long}.

{Therefore, GW observations
 have the ability to provide a ``disk''-independent way to trace back the (dimensionless) \textbf{BH} spin  and more generally an evaluation of the mass and spin parameters of the  black holes --  for   GW   methods  for  the evaluation  of the \textbf{BH} spin see for example \cite{Farr:2017uvj,vanPutten:2015eda,vanPutten:2016wpa,2016PhRvD..93h4042P, 2016ApJ...826...91A}.
}

In this paper, we face the problem of the black hole identification, proposing an  approach which we believe can be promising especially for \textbf{SMBHs} in  \textbf{AGNs}.
Our investigation is essentially centered  on the exploitation of a  special connection between \textbf{SMBHs} and their  accretion tori; we show  that the dimensionless spin of a central \textbf{BH} ($a/M$) and the morphological  and equilibrium  properties of   its accreting  disks, are strongly related.
 Recently, various analyses have shown, in the  general  relativistic regime, that  completely axis-symmetric and coplanar configurations are strongly restricted with regards to their formation,  kinematic characteristics (as range of variation of angular momentum)  or the emergence  of instability. Their existence is  constrained  according to  different evolutionary phases of the individual configurations and, more importantly, by the  properties of the central Kerr attractor  \citep{pugtot,ringed,open,dsystem,long}.

More generally, there is a strong relation between the galaxy dynamics and its
super-massive guest, specially in the accretion processes. It is  expected that such  \textbf{SMBHs} in \textbf{AGNs}
are characterized by a series of multi-accreting episodes
during their life-time as a consequence of interaction with the
galactic environment, made up by stars and dust, being influenced by the galaxy dynamics.  Further example of the complex and rich black hole-active  galaxy  interaction is  known as \emph{feedback \textbf{AGN}}-- \citep{Ricci:2017abh,Ricci,Ricci:2017wmr,Yoshida,Komossa:2015qya}. These activities may leave
traces in the form of matter remnants orbiting the central attractors.
Thus, chaotical, discontinuous   accretion episodes  can  produce sequences of orbiting toroidal structures  with strongly   differing features as,  for example, different rotation orientations with respect to the central Kerr \textbf{BH} where corotating and counterrotating accretion stages can be mixed \citep{Dyda:2014pia,Aligetal(2013),Carmona-Loaiza:2015fqa,Lovelace:1996kx,Romanova}. Strongly misaligned disks may appear  with respect to the central \textbf{SMBH} spin \citep{Nixon:2013qfa,2015MNRAS.449.1251D,Bonnerot:2015ara,Aly:2015vqa}.
However, in this work classes of rotating  \textbf{BHs}, identified by their  dimensionless spin, are associated  to particular features of the \textbf{BH} orbiting accreting tori, namely    aggregates of  toroidal axis-symmetric  accretion disks, also known as Ringed Accretion Disks  (\textbf{RADs}), centered on the equatorial plane of a Kerr \textbf{SMBH}. The \textbf{RAD} aggregate  is composed by both corotating and counterrotating tori, the limiting  case of  single accretion torus orbiting  the \textbf{SMBH} is also addressed as a special case of the \textbf{RAD}. Each \textbf{RAD}  toroidal  component is modeled by a  perfect fluid  with barotropic equation  of state and constant specific angular momentum  distribution \citep{Pu:mnrasPD,pugtot,abrafra,Pugliese:2013hp}.
\textbf{RAD} models follow the possibility that several accretion tori can  be formed around very compact objects as \textbf{SMBHs}
($10^6-10^9 M_{\odot}$, $M_{\odot}$
being solar masses) in \textbf{AGNs}. \textbf{RAD} may be also originated after different accretion phases in some binary systems or \textbf{BH} kick-out, or by local clouds accretion. {We mention  also \cite{Bonnell}  for an analysis of the  massive cloud spiraling into the  \textbf{SMBH} Galaxy.}

{More generally, multiple  structures orbiting around \textbf{SMBHs} can  be created   from   several different processes involving the interaction between the \textbf{BH} attractors and their  environment.
  An original Keplerian disk can split into two (or more) components  (toroids) for some destructive effects where for example the \emph{self-gravity } of the disk becomes relevant. The  impact of the  disk self-gravity  in different aspects of the \textbf{BH} accretion  is  discussed in Sec.\il\ref{Sec:gothc}. Formation of more tori  is also one of the possible endings of a misaligned disk, for example in a binary system, where the torque and warping is   relevant to induce  a disk fragmentation.  In all these cases, the rotational law of newly  formed tori can be also very different. In fact, multiple systems   can  originate   in  several periods  of the accretion life from different material embeddings.}
{In an early phases of their evolution, accretion disks  can   be misaligned with respect to the equatorial plane of the Kerr attractor and in many cases such disks   are expected to be  ``warped'' and ``twisted'' accretion disks.  Although the misaligned or warped case is not covered here, we also discuss the occurrence of this possibility within the \textbf{RAD}  frame  in Sec.\il\ref{Sec:gothc}, where we  also address  possible   instability processes when  \textbf{RAD} is extended to consider aggregates with the contribution of the magnetic field.   However, misaligned disks, for  specific  values of the characteristic parameters, will  eventually end in a steady state with an inner aligned disk. It has been shown that counterrotating tori can  derive also from  highly misaligned disks after  galaxy mergers, with galactic planes  strongly inclined.  }

{Furthermore, in a warped disk scenario the analysis of inner region of accretion disk connected with the  jet emission is  also  considered for the assessment of the central \textbf{BH} spin, assuming that  the  jet directions are indicative of the direction of the \textbf{BH} spin. We consider this briefly in Sec.\il\ref{Sec:gothc}. In the case of misaligned  disks, these studies focus on   radio--jet direction in \textbf{AGN} along orbital plane direction.
More generally,  \textbf{BH} and  jet emission connection (for example  radio and  X-ray emission) are considered in galactic embedding to test correlation between the  galactic host and jets;  the presence of a warped disk can also explain the jet emission  orientation   with respect to the galactic plane showing also a strong  misalignment. Without going into details of this aspect of the accreting process, going  beyond the scope of the present work, we mention in particular the case of galaxy \textbf{NGC4258}  
\citep{Moran:2008dv,
Humphreys:2007ir,
Doeleman:2008qh,
Qin:2008uu,
Rodriguez:2008bi,
Kondratko:2008gr}.}

{Jet emission, in fact,  constitutes  the   third ingredient  in the     unified   \textbf{BH}--accretion disk framework. Almost any \textbf{BH} is associated to a jet emission.
How exactly the jet emission and morphology (collimation along the axis, rapidity of emission launch, chemical composition) can be precisely linked or  induced by the mechanisms of accretion remains to be clarified. Jets are supposed to be  connected with  the dynamics of the  inner region of the disk in accretion. The role of jet in the \textbf{BH}-accretion disk systems and accretion physics is multiple, altering the energetic  of  the accretion processes  with the extraction of the rotational   energy of the central \textbf{BH}  and the rotational energy of the  disk, and  changing the accretion disk inner edge.
The inner--edge-jet correlation   has to be then framed in the case of the  misaligned disks, where
the  location   of the  edge  is  related to the  strength of
the jet-- \cite{NL2009,Fender:2004aw,Fender:2009re,Soleri:2010fz,Tetarenko18,AGNnoBH,FangMurase2018,1998NewAR..42..593F,1998MNRAS.300..573F,1999MNRAS.304..865F,Fender(2001)}.
For an analysis of the energetic X-ray transient with associated relativistic jets,
  updated investigations  on jet emission detection see
 \cite{NatureMa,Maraschi:2002pp,Chen:2015cga,Yu:2015lqj,Zhang:2015eka,Sbarrato:2014uxa,MSBNNW2009,
 Ghisellini:2014pwa,BMP98}.
 For  jet-accretion disk  correlation in  \textbf{AGN}  see for example
\cite{liska,Caproni:2017nsh,Inoue:2017bgt,Gandhi:2017dix,Duran:2016wdi,Vedantham:2017kyb,Bogdan:2017pgi,Banados:2017unc,DAmmando:2017ufp}
and  also
\cite{NL2009,Fender:2004aw,Fender:2009re,Soleri:2010fz,Tetarenko18,AGNnoBH,FangMurase2018}.
}

{Jet production, as an important additional aspect of the physics of accretion  disk, fits into the  \textbf{RAD}  context  in many ways. Firstly, more points of accretion  can be present in the \textbf{RAD} \emph{inside} the ringed structure.
It has been shown in \cite{ringed} that a \textbf{RAD}, as an aggregated body of orbiting  tori, can be considered as  a single disk orbiting around a central Kerr attractor on its equatorial plane, with axi-symmetric    but knobby surface  and  diversified  specific  angular momentum distribution, due to the different contributions of each torus of the \textbf{RAD} agglomerate that can be  either corotating or counterrotating with respect to the central attractor.  Only for the \textbf{RAD} systems satisfying
 certain constraints on the \textbf{BH} spin and on the specific angular momentum of the tori, a double accretion phase can occur. Such a situation implies the concomitant presence of two coplanar accreting tori of the \textbf{RAD} in the same period of the \textbf{BH} life. In the frame of  accretion disk--jet correlation, this would imply the presence of a  shell of double jets, one  from an outer counterrotating torus  and one associated to the inner corotating accreting torus. Moreover, the  geometrically thick tori  considered as  \textbf{RAD} components  are known to be associated to   several species of  open surfaces (proto-jets) related  to emission of matter funnels in jets--\cite{abrafra}. These configurations have been discussed in literature in several contexts. The \textbf{RAD} model inherits  this  characteristic of the accretion  torus. The proto-jets,  as related to the  critical points of the hydrostatic pressure in the force balance of the \textbf{RAD} tori, can give  raise to complicated sets of jets funnels either from corotating or counterrotating fluids in more points of the tori agglomeration. We do not consider directly in this investigation the open surfaces, but  they will be considered for the classification of the attractors. A more focused analysis on  proto-jets  in the \textbf{RAD} framework can be found in \cite{open,long,app}. }

 The plan of this article is as follows. Section \il\ref{Sec:Taa-DISK} introduces the model of \textbf{RAD}  orbiting   a central  Kerr attractor: in Sec.\il\ref{Sec:prop} we examine  main properties of the Kerr \textbf{BH} exact solution.
  \textbf{RAD} model is then developed in Sec.\il\ref{Eq:tori} where the  toroidal configurations are discussed, and  main properties and  characteristics of the tori and the macro-structure are presented.
  The \textbf{RAD}  is  a relatively new model,  this analysis has therefore required the introduction of new concepts adapted to the model. For this purpose it is  convenient to introduce a reference  Section\il\ref{Sec:notation-sec}  where we give relevant model details   grouping  together  the main definitions used over the course of this work.
  We made use also of Table\il\ref{Table:pol-cy-multi} and Table\il\ref{Table:def-flour}   where
     the  symbols and relevant notation  used throughout this article are listed.
 Section\il\ref{Sec:RADs-sis} encloses the main results of this work.  Tables\il\ref{Table:nature-Att} shows the major classes of attractors considered in this section and, in a compact form,  main properties the associated \textbf{RADs}, as investigated in  Sec.\il\ref{Sec:notation-sec}.
 {In Sec.\il(\ref{Sec:nw})  we concentrate on the double accretion occurring in \textbf{RAD} couples}.
 In Sec.\il\ref{Sec:gothc}
general discussion on the  \textbf{RAD} instabilities and  generalizations of the \textbf{RAD} model is presented. We also discuss the possible significance of the tori self-gravity, tori misalignment, viscosity and magnetic field in the \textbf{RAD}.
 Concluding remarks are in Sec.\il\ref{Sec:Open-Concl}, where
 we  report a brief summary of the results of our analysis, followed by considerations on the impact of the   \textbf{RAD} hypothesis  in the \textbf{AGN} environments and on  some aspects of the phenomenology connected with the \textbf{RAD}.
 Finally, Appendix  follows, where  further details on \textbf{BH} spin properties are provided.
 \section{Ringed accretion disks orbiting    Kerr attractors}\label{Sec:Taa-DISK}
\subsection{Geometry and test-particle motion}\label{Sec:prop}
%
The Kerr  metric line element can be
written   in the Boyer-Lindquist (BL)  coordinates
\( \{t,r,\theta ,\phi \}\) as follows
%
\bea \label{alai}&& ds^2=-dt^2+\frac{\Sigma}{\Delta}dr^2+\Sigma
d\theta^2+(r^2+a^2)\sin^2\theta
d\phi^2
\\
&&\nonumber+\frac{2M}{\Sigma}r(dt-a\sin^2\theta d\phi)^2\, \\
\nonumber
&&
 \mbox{where}\quad\Sigma\equiv r^2+a^2\cos\theta^2,\Delta\equiv r^2-2 M r+a^2,
\eea
and    $0<a=J/M\leq M$  is the specific angular momentum, $J$ is the
total angular momentum of the gravitational source and $M$ is the  gravitational mass parameter.
The non-rotating  limiting case $a=0$ corresponds to  the   Schwarzschild metric, while the extreme Kerr Black hole  has dimensionless spin $a/M=1$.
The horizons, $r_-<r_+$, and the outer static limit $r_{\epsilon}^+$, are respectively given by\footnote{We adopt the
geometrical  units $c=1=G$ and  the $(-,+,+,+)$ signature, Greek indices run in $\{0,1,2,3\}$.  The   four-velocity  satisfies $u^{\alpha} u_{\alpha}=-1$. The radius $r$ has unit of
mass $[M]$, and the angular momentum  has unit of $[M]^2$, the velocities  $[u^t]=[u^r]=1$
and $[u^{\varphi}]=[u^{\theta}]=[M]^{-1}$ with $[u^{\varphi}/u^{t}]=[M]^{-1}$ and
$[u_{\varphi}/u_{t}]=[M]$. For the seek of convenience, we always consider the
dimensionless  energy and effective potential $[V_{eff}]=1$ and an angular momentum per
unit of mass $[L]/[M]=[M]$.}:
\bea
r_{\pm}\equiv M\pm\sqrt{M^2-a^2};\quad r_{\epsilon}^{+}\equiv M+\sqrt{M^2- a^2 \cos\theta^2};
\eea
where $r_+<r_{\epsilon}^+$ on   $\theta\neq0$  and  $r_{\epsilon}^+=2M$  in the equatorial plane $(\theta=\pi/2)$. The region $\Sigma_{\epsilon}^+\equiv[r_{\epsilon}^+, r_+[$ is known as ergoregion, and the static limit $r_{\epsilon}^+$ is also known as outer ergosurface.
 Because of the symmetries of the  Kerr  geometry, the quantities
\be\label{Eq:after}
{E} \equiv -g_{\alpha \beta}\xi_{t}^{\alpha} p^{\beta}=-p_t,\quad L \equiv
g_{\alpha \beta}\xi_{\phi}^{\alpha}p^{\beta}=p_{\phi}\, \ee
are  constants of motion,   $p^{\alpha}$  is the particle four--momentum  and  $\xi_{\phi}=\partial_{\phi} $  and  $\xi_{t}=\partial_{t} $ are the rotational  Killing vector field and the
time Killing vector  field representing the stationarity of the  spacetime \cite{1973CMaPh..31..161B}.
The     test particle dynamics is  invariant under the mutual transformation of the parameters
$(a,L)\rightarrow(-a,-L)$. As a consequence of this,
 we    restrict  the  analysis of the test particle circular motion to the case of  positive values of $a$
for corotating  $(L>0)$ and counterrotating   $(L<0)$ orbits.
\subsection{Toroidal configurations}\label{Eq:tori}
In this work we consider   perfect fluid  toroidal  configurations    orbiting a    Kerr \textbf{BH}.
We take the   energy momentum tensor  for   one-species particle perfect  fluid system
\be\label{E:Tm}
T_{\alpha \beta}=(\varrho +p) u_{\alpha} u_{\beta}+\  p g_{\alpha \beta},
\ee
where $u^{\alpha}$  is
a timelike flow vector field and  $\varrho$ and $p$ are  the total energy density and
pressure  respectively, as measured by an observer comoving with the fluid with velocity $u^{\alpha}$.
We set up the
problem symmetries  assuming there is $\partial_t \mathbf{Q}=0$ and
$\partial_{\varphi} \mathbf{Q}=0$,  with $\mathbf{Q}$ being a generic spacetime tensor.
Accordingly,  the  continuity equation
is  identically satisfied and the  fluid dynamics  is  governed by the \emph{Euler equation} only:
\bea\label{E:1a0}
(p+\varrho)u^\alpha \nabla_\alpha u^\gamma+ \ h^{\beta\gamma}\nabla_\beta p=0\,, \eea
where  $\nabla_\alpha g_{\beta\gamma}=0$,  $h_{\alpha \beta}=g_{\alpha \beta}+ u_\alpha u_\beta$ is  the projection tensor   \citep{Pugliese:2011aa,pugtot}.
We then assume  a barotropic equation of state $p=p(\varrho)$, the  orbital motion with  $u^{\theta}=0$ and
$u^r=0$. Within  these conditions, (\ref{E:1a0}) can be written as an equation for the barotropic  pressure $p$ as follows:
\bea\label{Eq:scond-d}
&&\frac{\partial_{\mu}p}{\varrho+p}=-{\partial_{\mu }W}+\frac{\Omega \partial_{\mu}\ell}{1-\Omega \ell},\quad \ell\equiv \frac{L}{E},
 \\\nonumber
 &&W\equiv\ln V_{eff}(\ell),\quad V_{eff}(\ell)=u_t= \pm\sqrt{\frac{g_{\phi t}^2-g_{tt} g_{\phi \phi}}{g_{\phi \phi}+2 \ell g_{\phi t} +\ell^2g_{tt}}}.
\eea
In (\ref{Eq:scond-d}),  $\Omega=u^{\phi}/u^{t}$ is the fluid relativistic angular frequency related to distant observers, while  $W(r;\ell,a)$  is
the Paczy{\'n}ski-Wiita  (P-W) potential, expressed  in terms of the  fluid  \emph{effective potential}   $V_{eff}(r;\ell,a)$.

The effective potential $V_{eff}(r;\ell,a)$ reflects the background  Kerr geometry through  the parameter $a$, and the centrifugal effects through the fluid specific angular momenta $\ell$,  here assumed  constant  and conserved-- \citep{Lei:2008ui,Abramowicz:2008bk}.
The fluid equilibrium is therefore  regulated by  the balance of the    gravitational and  pressure  terms versus    centrifugal  factors arising  due to the fluid  rotation and  the curvature  effects  of the  Kerr background, encoded in the effective potential function $V_{eff}$.
Analogously to the   test particle dynamics, as the fluid  effective potential    function   is invariant under the mutual transformation of  the parameters
$(a,\ell)\rightarrow(-a,-\ell)$,    we  can assume $a>0$ and consider $\ell>0$
for \emph{corotating}   and $\ell<0$  for  \emph{counterrotating} fluids, within  the notation $(\mp)$   respectively.

\begin{figure*}[ht!]
\centering
\resizebox{\hsize}{!}
   {
\begin{tabular}{cccc}
\includegraphics[width=0.23\columnwidth]{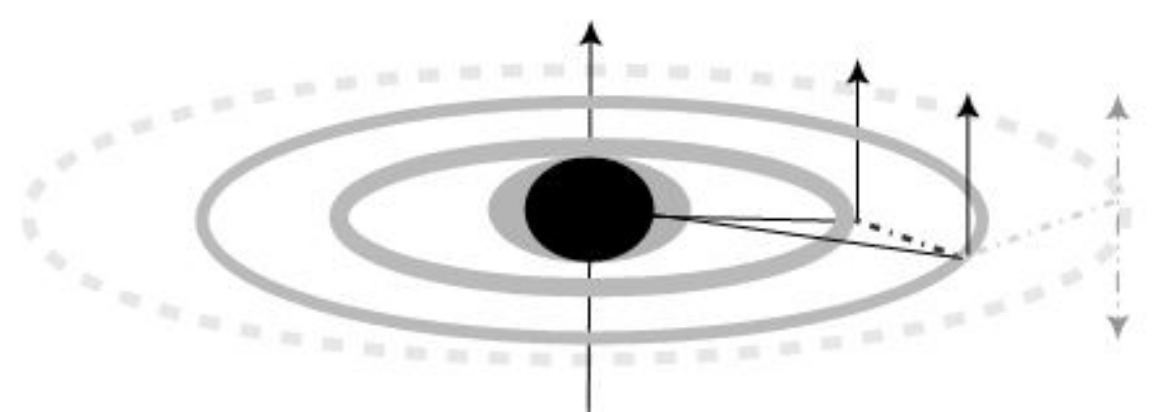}&
\includegraphics[width=0.23\columnwidth]{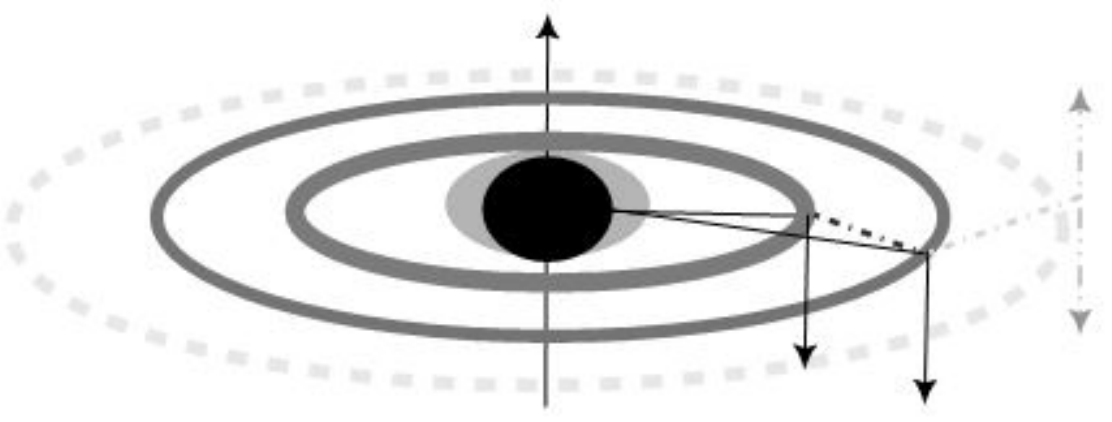}&
\includegraphics[width=0.23\columnwidth]{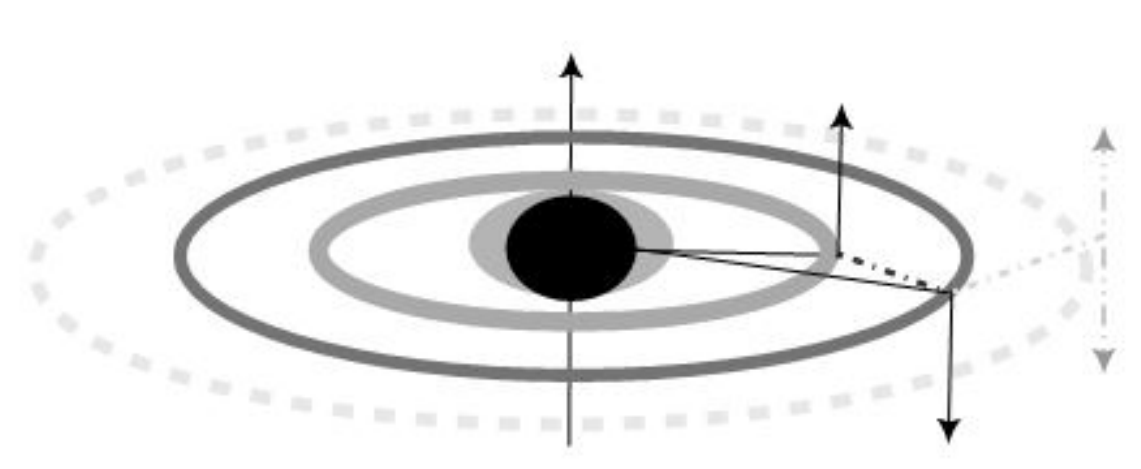}&
\includegraphics[width=0.23\columnwidth]{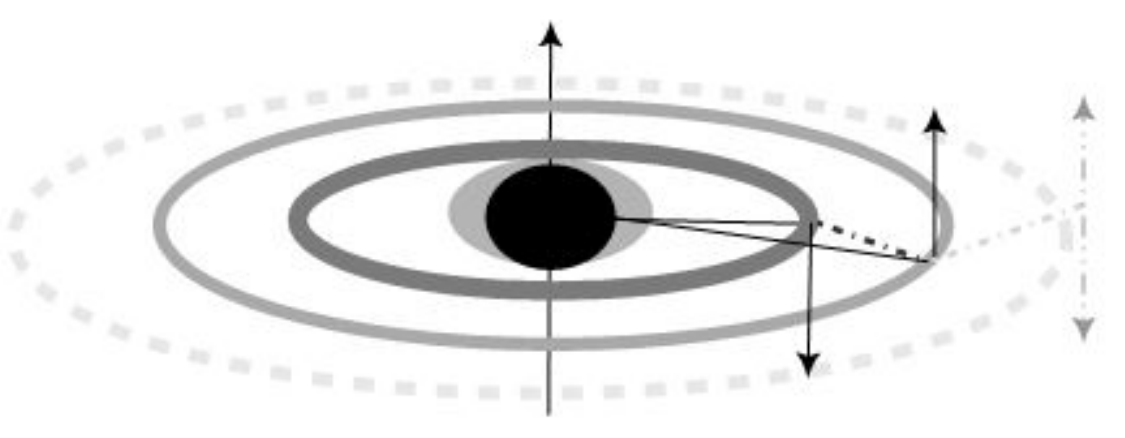}\\
{\textbf{(1)}}&{\textbf{(2)}}&{{\textbf{(3)}}}&{{\textbf{(4)}}}
\end{tabular}}
\caption{Pictorial schemes of a  system of {coplanar toroids  (rings)} orbiting a Kerr black hole  attractor. Black region is the black hole, gray region is the ergosphere.  The distances between the tori and attractor are not in scale. Rings  are schematically represented  as two-dimensional objects corresponding to the equilibrium topology.
The arrows represent the rotation: the dimensionless spin of the attractor $a/M\geq0$ is considered always positive, \textbf{\sffamily$(\mathbf{+;\cdot,\cdot,\cdot})$}, ``spin-up'' in the  picture,  or vanishing for the limiting case of the  static Schwarzschild solution. The fluid specific angular momentum of  an accretion disk $\ell$ can be positive, $\ell a>0$, for corotating $(-)$ (``spin-up'') or negative (light gray line), $\ell a<0$, for counterrotating $(+)$ (``spin-down'')  (gray line) with respect to the central black hole. The outer third ring is represented  by a dashed  and double arrowed line, as may be corotating or couterrotating.
  $\ell$corotating  rings $\ell_i\ell_o>0$ of corotating  tori, {\sffamily$(\mathbf{+;+,+,\pm})$}, is presented in scheme \textbf{(1)}-this  represents the inner  configurations of Fig.\il\ref{Fig:but-s-sou}-\textbf{ (b)}.  $\ell$corotating  rings $\ell_i\ell_o>0$ of counterrotating   tori, \textbf{$(\mathbf{+;-,-,\pm})$}, is presented in scheme \textbf{(2)} and this represents the inner configurations of Fig.\il\ref{Fig:but-s-sou}-\textbf{(d)}.   $\ell$counterrotating if $\ell_i\ell_o<0$ are   scheme \textbf{(3)}, $(\mathbf{+;+,-,\pm})$, and  scheme \textbf{(4)}, $(\mathbf{+;-,+,\pm})$-see also Fig.\il\ref{Fig:but-s-sou}-\textbf{(a)}  and \textbf{(c)} respectively. }\label{Figs:GROUND-scheme}
\end{figure*}

\begin{figure}[h!]
\centering
\includegraphics[width=.7\columnwidth,angle=90]{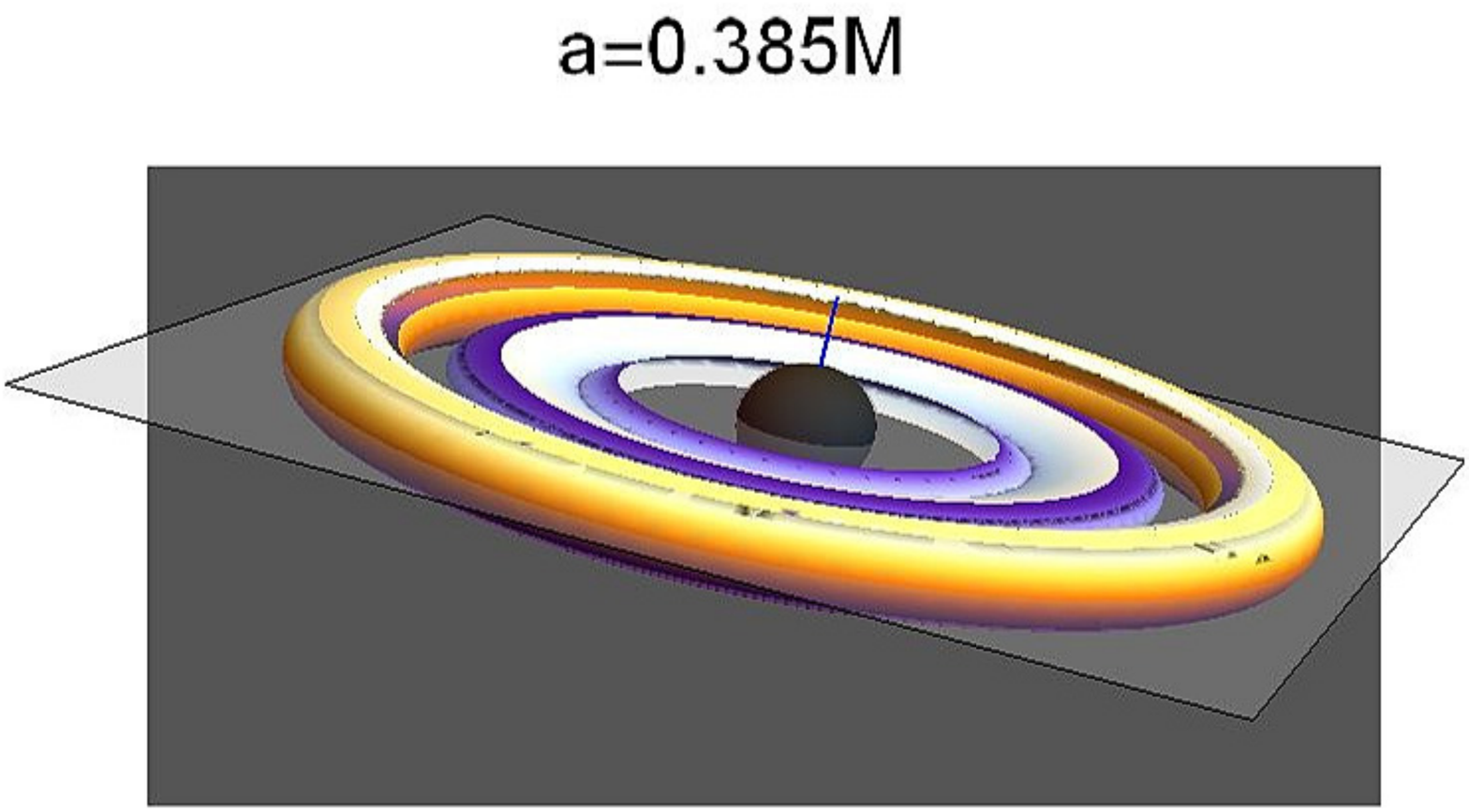}\\
\includegraphics[width=.57\columnwidth,angle=90]{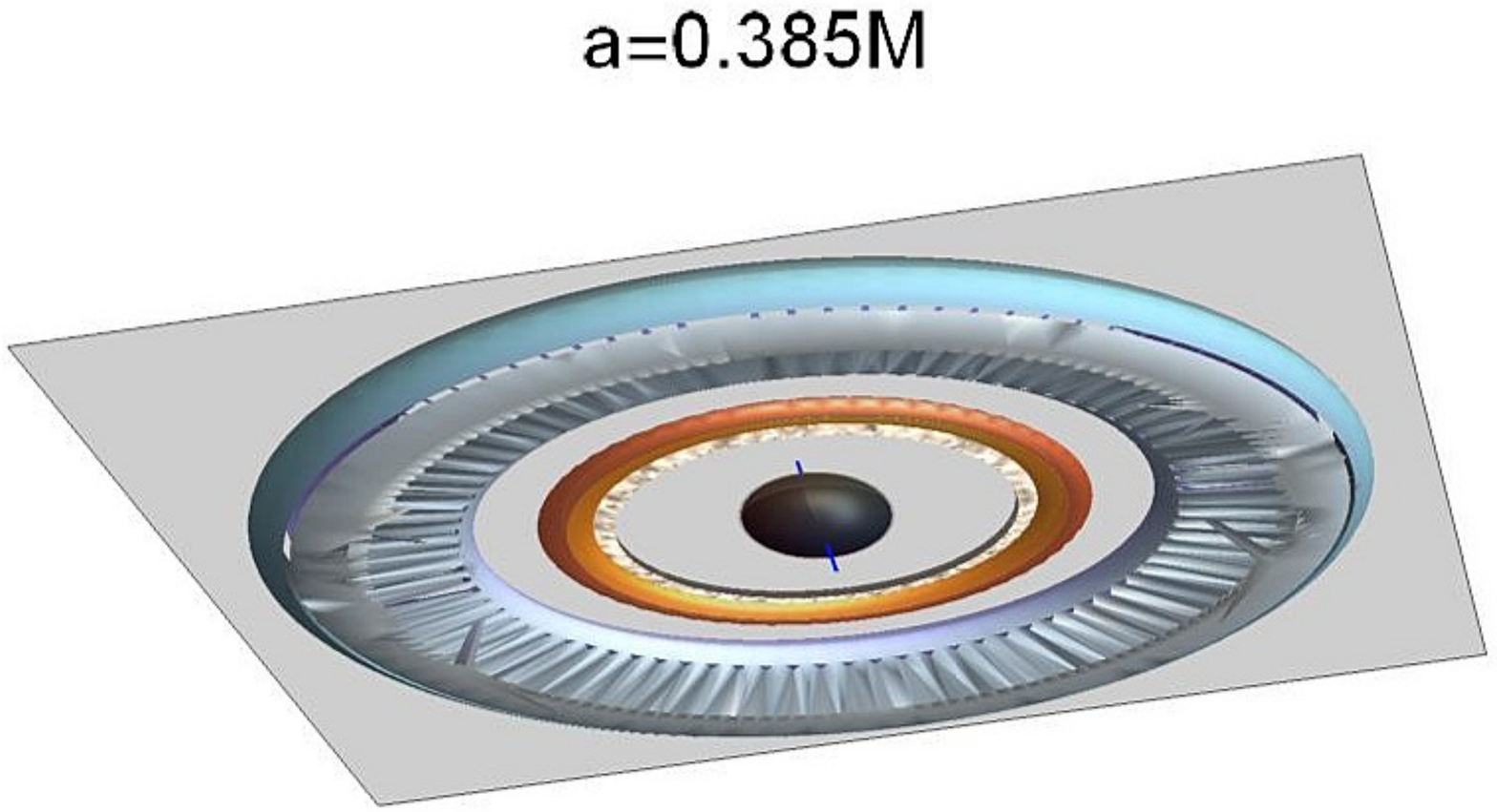}\\
\includegraphics[width=.5\columnwidth,angle=90]{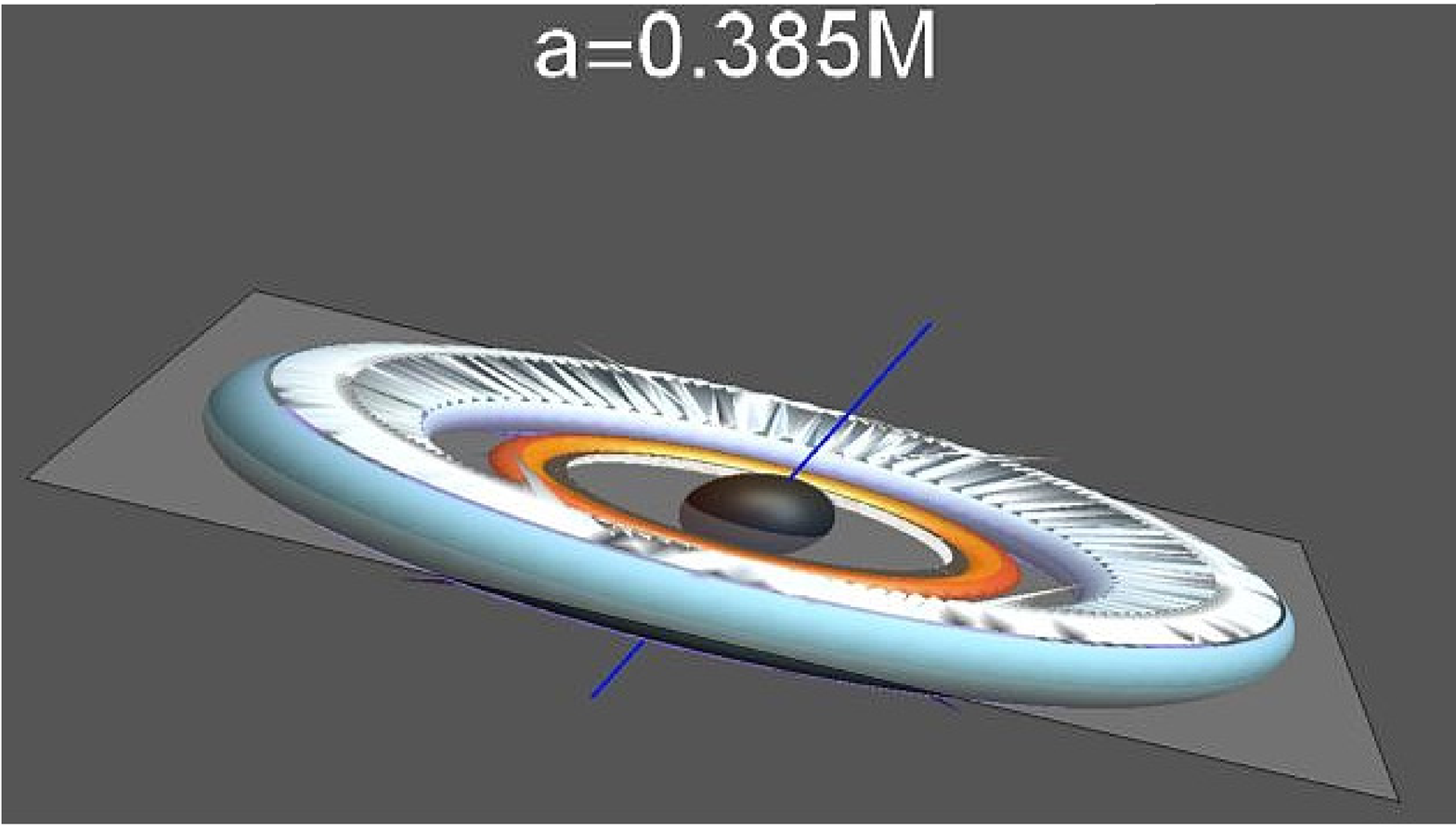}
\caption{\textbf{{GRHD}}-Numerical 3D integration of   tori density surfaces. {Tori are coplanar and  orbiting around a central  the  \textbf{BH}, on the equatorial plane.} Colors are chosen according to improved visual effect, integration is stopped at the emerging of disks collision. Black region is $r < r_+$, $r_+$ is the outer horizon
of the black hole of spin $a = 0.385M$, gray region is the outer
ergosurface.  Top panel  pictures two {coplanar} $\ell$counterrotating tori $\cc^+_\times<\cc^-$,  the inner counterrotating  torus is accreting onto the central \textbf{BH}, accretion flux is stopped during \textbf{RAD}  integration. Center and bottom panels  show  a \textbf{RAD} sequence $\cc^+<\cc_o^-\leq\cc_i^-$, made by three  $\ell$counterrotating tori: an  outer counterrotating torus  and an inner couple of colliding corotating toroids. Bottom panel  focuses on a different \textbf{RAD} view,  featuring the collision of the inner couple, and the separation with respect to the outer counterrotating  torus. {The entire system \textbf{BH}-\textbf{RAD} is tilted with respect to the observer. The \textbf{RAD}-\textbf{BH} equatorial plane and the spin axis  are also shown.}}\label{Fig:Fig3D}
\end{figure}

We can   summarize the  properties of the   ringed accretion disk  (\textbf{RAD}) model  introduced in   \citet{ringed} as a  general relativistic model  of  toroidal  configurations, $\mathbf{C}^n=\bigcup^n\cc_i$, consisting of a collection of  $n$ sub-configurations  (configuration \emph{order}  $n$) of  complex   corotating and counterrotating toroids orbiting in   the equatorial plane of    a  Kerr attractor. \textbf{RAD} features an axially symmetry, ``knobby'' accretion disk centered on a Kerr \textbf{BH}.
 \textbf{RAD} tori  can be corotating or counterrotating with respect to the central Kerr \textbf{BH} therefore, assuming     a $(\cc_a, \cc_b)$ couple    with specific angular momentum $(\ell_a, \ell_b)$ respectively,   we need to introduce   the concept  of
 \emph{$\ell$corotating} toroids,  {or ``\textbf{ell}''corotating}, defined by  the condition $\ell_{a}\ell_{b}>0$, and { \emph{$\ell$counterrotating},  {or ``\textbf{ell}''counterrotating }} toroids defined  by the relation   $\ell_{a}\ell_{b}<0$.  Two $\ell$corotating tori  can be  corotating, $\ell a>0$, or counterrotating,  $\ell a<0$, with respect to the central attractor--see  schemes in Figs\il\ref{Figs:GROUND-scheme} and   tori in Figs\il\ref{Fig:Fig3D} and  Figs\il\ref{Fig:but-s-sou}.
The construction of the ringed configurations  is actually in many features quite independent of the model  adopted for the single \textbf{RAD} torus (sub-configuration or ring).
   In fact, in situations where the curvature  effects of the Kerr geometry are significant,  results are largely independent of the specific characteristics of the model for the  single toroidal structure,  being primarily  based on the characteristics of the   geodesic structure  the Kerr spacetime  related to  the matter distribution.
 To simplify discussion,  we consider here each toroid of the ringed  disk  to be governed by the   general relativistic
   hydrodynamic (\textbf{GRHD}) Boyer  condition  for  equilibrium configurations  of rotating perfect fluids. {However, these tori are effectively used as initial  configurations  also for integration in more complex \textbf{GRMHD} models --see for example \cite{Lei:2008ui,abrafra,Luci}}.
In this approach,
the toroidal surfaces  are the equipotential surfaces, $V_{eff}=K=$constant, of the effective potential $V_{eff}(\ell,r)$,  \citep{Boy:1965:PCPS:,KJA78} corresponding  also to the surfaces of constant density, specific angular momentum $\ell$, and constant  relativistic angular frequency  $\Omega$, where $\Omega=\Omega(\ell)$  as a consequence of the von Zeipel theorem  \citep{M.A.Abramowicz,Zanotti:2014haa,KJA78}.
Therefore,  each torus is  uniquely identified by the couple of $(\ell,K)$ parameters.
Assuming a constant specific angular momentum and considering the parameter $K$ related to fluid density, we focus on the solution of the Euler equations associated to    the critical points of the  effective potential. These solutions, when they exist, represent orbiting  configurations which may be closed, quiescent or non accreting $\cc$, and  cusped  $\cc_{\times}$ (accreting)  toroids, or open, $\oo_{\times}$,  critical configurations which are associated to some proto-jet matter  \citep{open}. In general, we use the notation $\pp$ and $\pp_{\times}$ to indicate any equilibrium or critical configuration  without any further specification of its topology.

The minimum point, $r_{\min}$, of the effective potential (the maximum point for the hydrostatic  pressure)  corresponds to the center $r_{cent}$ of each toroid. For the cases where a  maximum  point, $r_{\max}$,  exists for the effective potential (the minimum point for the  pressure) it  corresponds to the critical points,  $r_{\times}$, of an accreting torus or, $r_{J}$, for a proto-jet. {The open equipotential surfaces have  been  variously related   to  ``proto jet-shell'' structures \citep{KJA78,Sadowski:2015jaa,Lasota:2015bii,Lyutikov(2009),Blaschke:2016uyo,Madau(1988),Sikora(1981),Okuda:2015dua}. An analysis of these configurations in the \textbf{RAD} framework, has been directly addressed in  \citet{open,long}.}
The inner edge  of the Boyer surface  is at $r_{in}\in[r_{\max},r_{\min}[$  on the equatorial plane,
while the outer edge  is at $r_{out}>r_{\min}$  on the equatorial plane\footnote{For the  geometrically thick configurations it is generally assumed that
 the time-scale of the dynamical processes $\tau_{dyn}$  (regulated by the gravitational and inertial forces, the timescale for  pressure to balance the  gravitational and centrifugal force) is much lower than the time-scale of the thermal ones $\tau_{the}$  (i.e. heating and cooling processes, timescale of  radiation entropy redistribution) that is lower than the time-scale of the viscous processes $\tau_{vis}$, and the effects of strong gravitational fields are dominant with respect to the  dissipative ones and predominant to determine  the unstable phases of the systems \citep{F-D-02,Igumenshchev,Pac-Wii}.
Moreover, we should note that the Paczy\'nski accretion mechanics from  a  Roche lobe overflow  induces   the mass loss  from tori being an important
 local stabilizing mechanism  against thermal and viscous instabilities, and globally   against the Papaloizou-Pringle instability (for a review we refer to  \citet{abrafra}).
 The effects of strong gravitational fields dominate  on  the  dissipative ones and the functional form of the angular
momentum and entropy distribution depends on the initial conditions of the system and on
the details of the dissipative processes only, during the evolution of dynamical processes,  \citep{Abramowicz:2008bk}.}  as in
 Figs\il\ref{Fig:but-s-sou}.

The location of the inner and outer edges of the disks is  strongly constrained\footnote{It is worth to note that these constraints may be equally applied for almost any model of accretion  disk around a Kerr attractor. For a  discussion on the definition and location of the inner edge of the accreting torus see \citet{Krolik:2002ae,BMP98,2010A&A...521A..15A,Agol:1999dn,Paczynski:2000tz}.  For possible restriction of the outer edge of the toroid in de Sitter  spacetimes see \citet{there-s000s,there-s005s,HAL-s,Wa-SH9,Stuchlik:2008dv,Stuchlik:2008ea,Stuchlik:2009jv}.} by the geodesic structure of the Kerr spacetime, consisting of the   union of the   orbital regions with boundaries  at the notable radii  $R^{\pm}\equiv \{r_{\gamma}^{\pm}, r_{\mbo}^{\pm},r_{\mso}^{\pm}\}$.

\begin{figure}[h!]
\centering
\begin{tabular}{c}
\includegraphics[width=1\columnwidth]{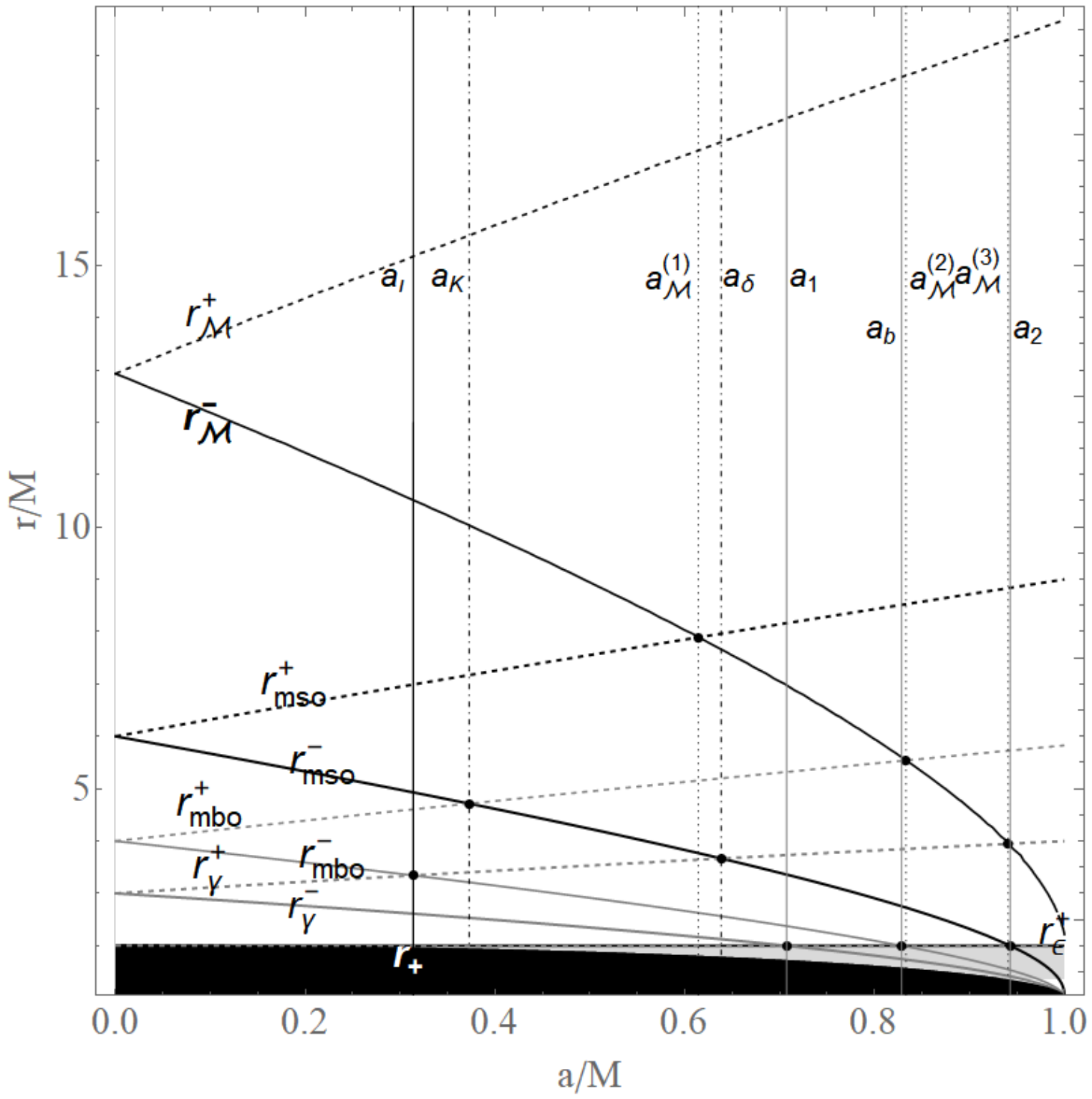}\\
\includegraphics[width=1\columnwidth]{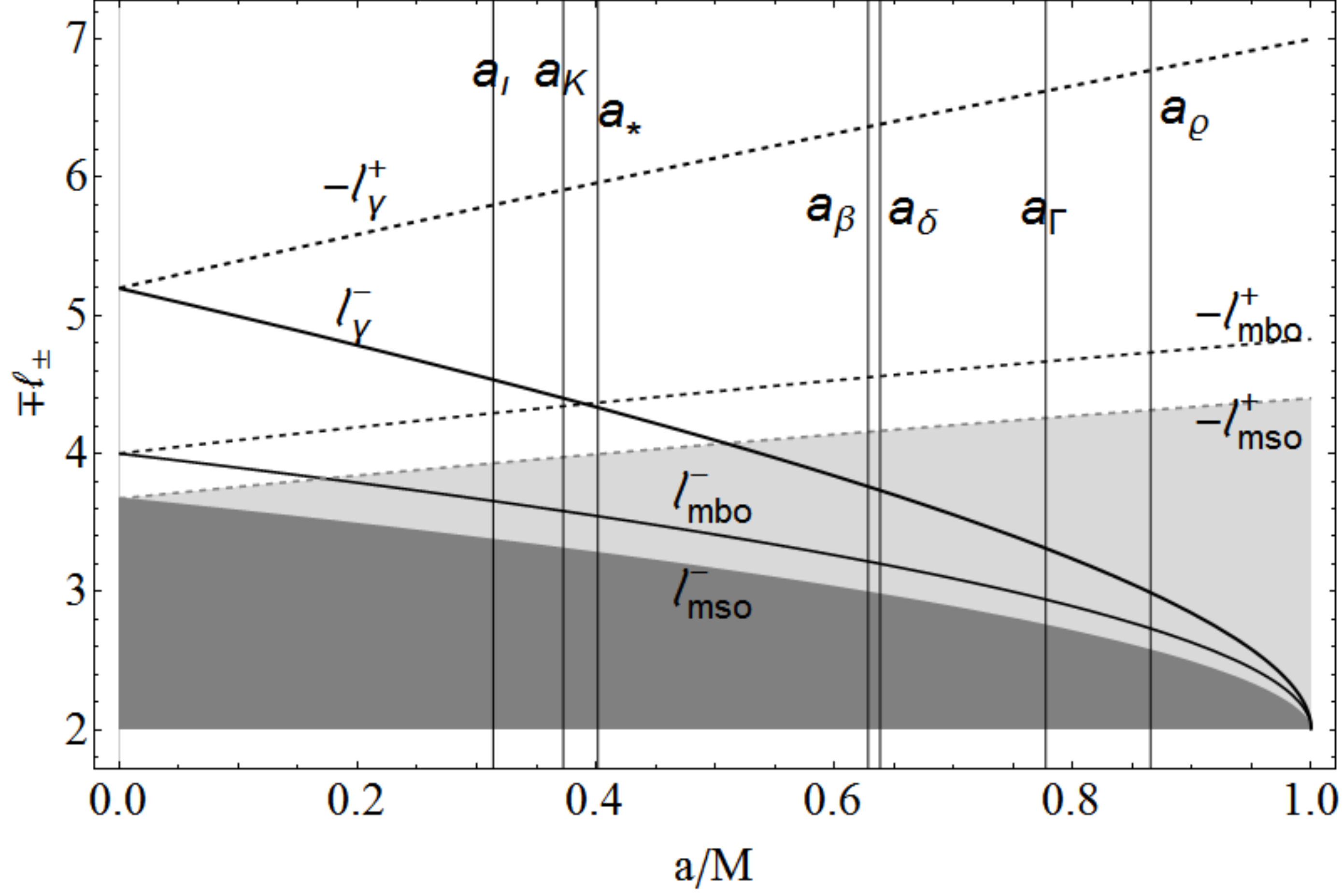}
\end{tabular}
\caption{Geodesic structure of the Kerr geometry: notable radii  $R\equiv \{r_{\gamma}^{\pm}, r_{\mbo}^{\pm},r_{\mso}^{\pm}\}$ (upper panel), and  the respective  fluid specific angular momenta $\ell^{\pm}_i=\ell^{\pm}(r^{\pm}_i)$   where $r^{\pm}_i\in R^{\pm}$,  $r_{\mathcal{M}}^{\pm}$ is the maximum point    of
derivative $\partial_r(\mp \ell^{\pm})$ for $a/M$ respectively. Some notable spacetime spin-mass ratios are also plotted,  a list can  be found in Table\il\ref{Table:nature-Att}. Black region is $r<r_+$, $r_+$ being the outer horizon of the Kerr geometry, gray region is $r<r_{\epsilon}^+$, $r_{\epsilon}^+$ is the outer ergosurface.}\label{Figs:GROUND-StA0}
\end{figure}

\begin{figure*}[ht!]
\begin{tabular}{cc}
\includegraphics[width=1\columnwidth]{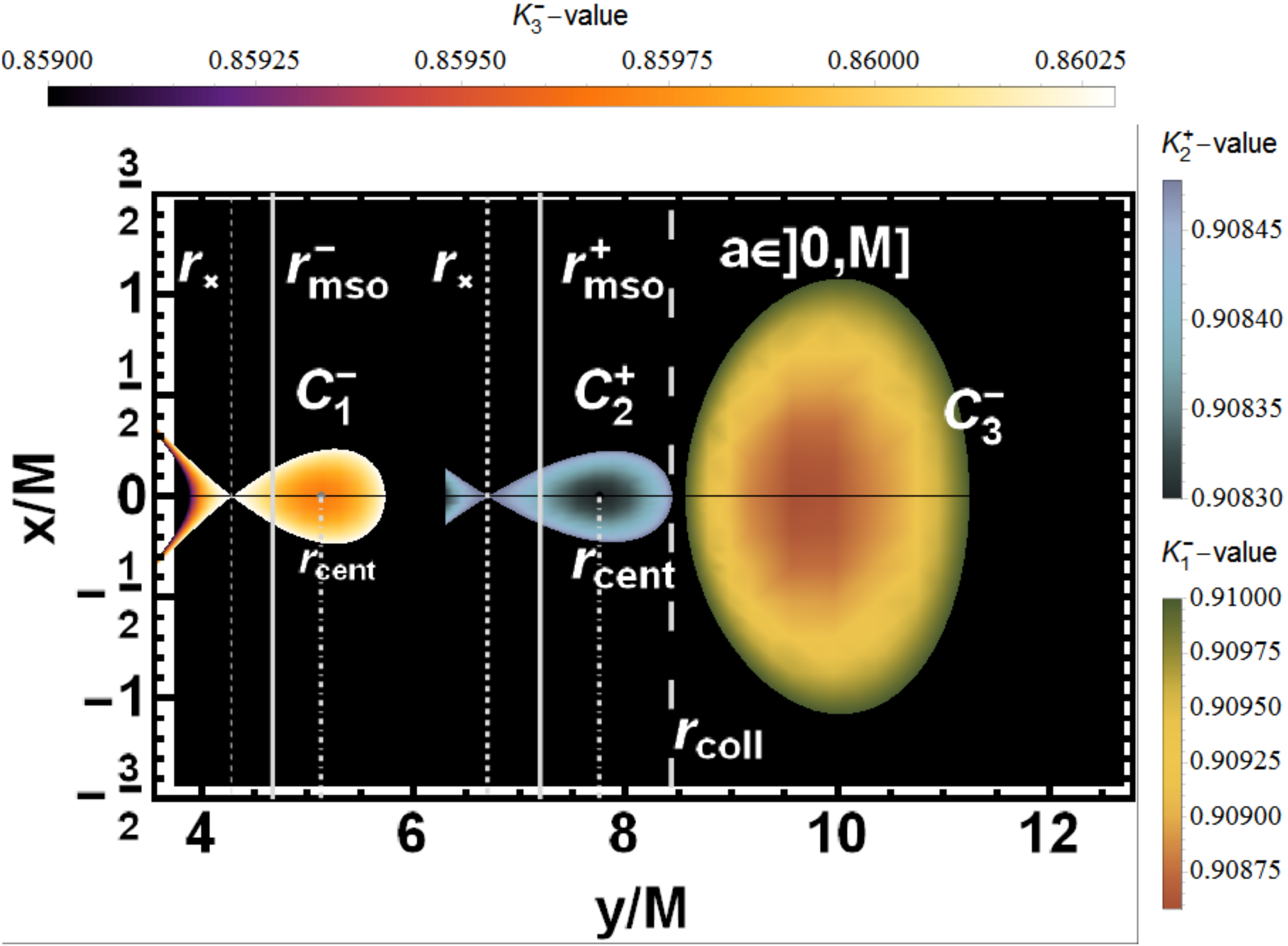}&
\includegraphics[width=1\columnwidth]{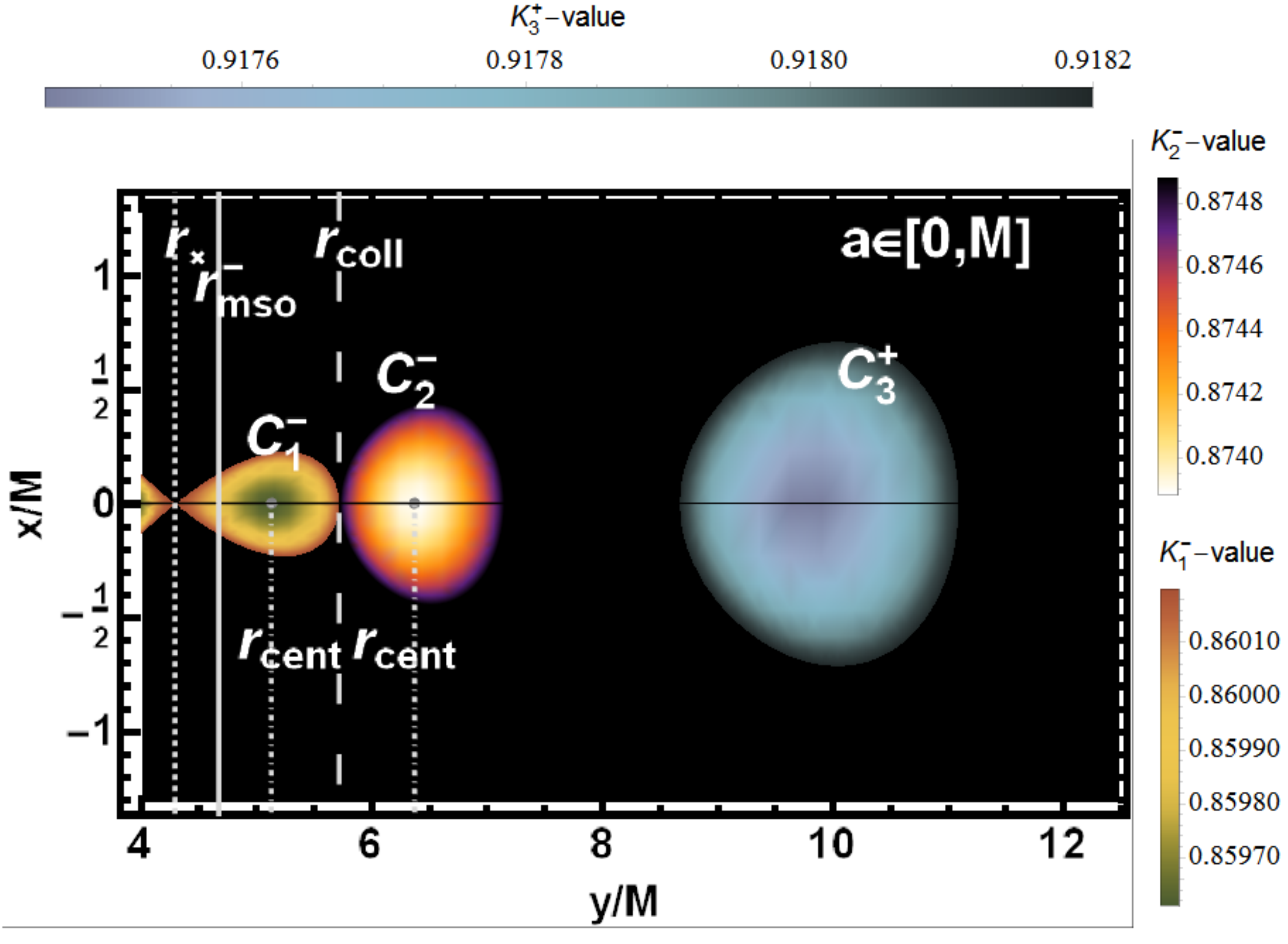}\\
{\LARGE\textbf{(a)}}&{\LARGE\textbf{(b)}}\\
\includegraphics[width=1\columnwidth]{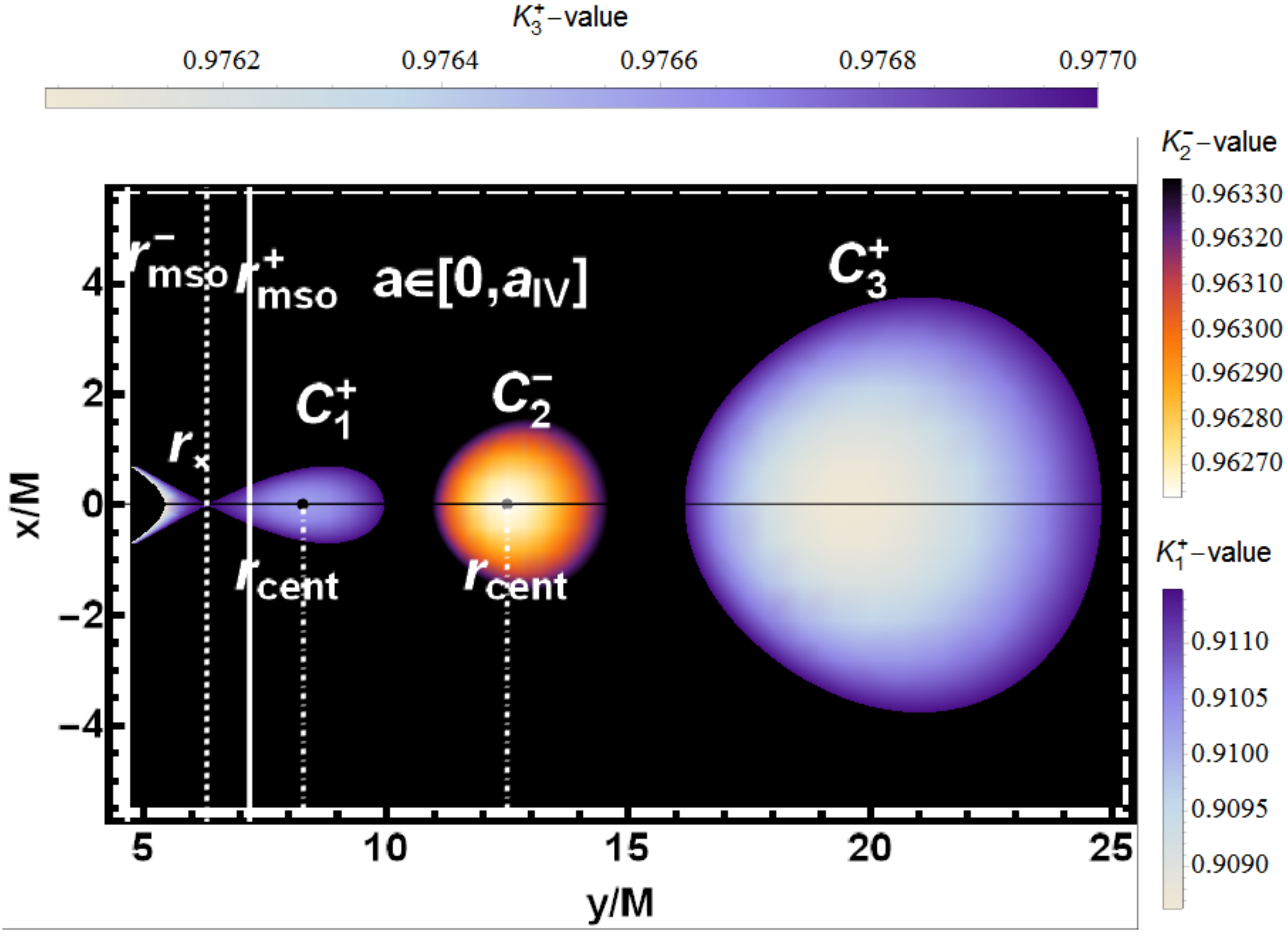}&
\includegraphics[width=1\columnwidth]{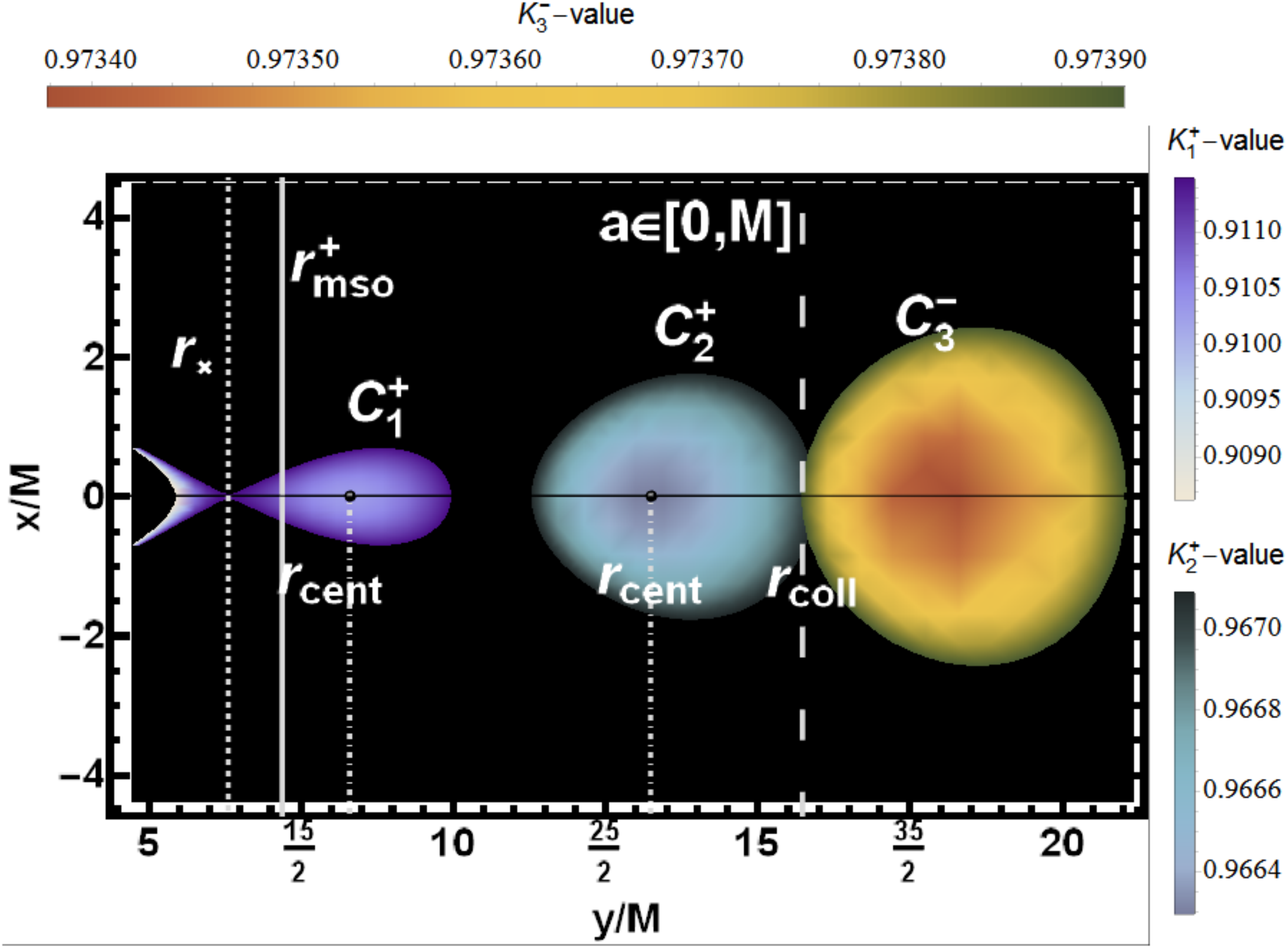}
\\
{\LARGE\textbf{(c)}}&{\LARGE\textbf{(d)}}
\end{tabular}
\caption{
\textbf{GRHD} numerical 2D  integration  of  triple toroidal configurations (\textbf{RADs} of the order $n=3$). {Tori are coplanar and  orbiting on the equatorial plane of  a central  the  \textbf{BH}.} \emph{\textbf{(a)}-Panel}:
sequence $\cc_{\times}^-<\cc_{\times}^+
\leq\cc^-$. Couple  $\cc_{\times}^-<\cc_{\times}^+$
may form only around  attractors with spin $a>0$,
no tori may  form between the attractor and the inner accreting corotating torus, while  corotating equilibrium tori between the two accreting configurations are possible, see (\ref{Eq:spar-bay-cbc}).
 \emph{\textbf{(b)}-Panel}: sequence
$\cc_{\times}^-\leq\cc^-<\cc^+$. \emph{\textbf{(c) }-Panel:} sequence
$\cc_{\times}^+<\cc^-<\cc^+$.
Only   around the Kerr \textbf{BHs}  with spin $a\in[0,a_{IV}[$ the couple $\cc_{\times}^+<\cc^-$ can form, no other tori are possible between the central \textbf{BH} and the inner counterrotating accreting ring.
\emph{\textbf{(d)-}Panel}: sequence $\cc_{\times}^+<\cc^+\leq\cc^{-}$.
$(x, y)$ are Cartesian coordinates. Tori notations follow here the relative distance from the attractor in the \textbf{RAD}: thus $\pp_1$ is for the closest and $\pp_3$ for the farthest. Integration is stopped at the emerging of disks collision.}\label{Fig:but-s-sou}
\end{figure*}

The set of radii  $R^{\pm}$ can be decomposed, for $a\neq0$, into   $R^-$ for the corotating and   $R^{+}$ for  counterrotating matter.  
Specifically, for timelike particle circular geodetical orbits, $r_{\gamma}^{\pm}$ is  the \emph{marginal circular orbit}  or  the photon circular orbit, timelike  circular orbits  can fill  the spacetime region $r>r_{\gamma}^{\pm}$. The \emph{marginal stable circular orbit}  $r_{\mso}^{\pm}$: stable orbits are in $r>r_{\mso}^{\pm}$ for counterrotating and corotating particles respectively\footnote{ Given $r_i\in R$,  we adopt  the  following notation for any function $\mathbf{Q}(r):\;\mathbf{Q}_i\equiv\mathbf{Q}(r_i)$,  for example $\ell_{\mso}^+\equiv\ell^+(r_{\mso}^+)$. 
}.  The \emph{marginal  bounded circular  orbit}  is $r_{\mbo}^{\pm}$, where
 $E_{\pm}(r_{\mbo}^{\pm})=1$  \citep{Pu:Kerr,Pugliese:2013zma,Pugliese:2010ps,ergon,Stuchlik(1980),Stuchlik(1981a),
 Stuchlik(1981b),Stuchlik:2008fy,Stuchlik:2003dt}
 --see Figs\il\ref{Figs:GROUND-StA0}.
 { The  $\ell$counterrotating sequences are affected strongly by the fact that
  radii of the sets $R^{+}$ and $R^-$, curves $R^{\pm}(a)$,   cross  in the $(r-a)$ plane--Figs\il\ref{Figs:GROUND-StA0} and \ref{Figs:pranz}.}

 There is always  $r_{cent}>r_{\mso}^{\pm}$,  $r_{\max}\in]r_{\gamma}^{\pm}, r_{\mso}^{\pm}]$ and  $K_{\pm}\in [K^{\pm}_{\min}, K^{\pm}_{\max}[ \subset]K_{\mso}^{\pm},1[\equiv \mathbf{K0}$ with specific   momentum $\ell_{\pm}\lessgtr\ell_{\mso}^{\pm}\lessgtr0$  respectively.
 A one-dimensional ring of matter  located at  $r_{\min}^{\pm}$ is  the limiting case for  $K_{\pm}=K_{\min}^{\pm}$.

The specific angular momenta  $(\ell_{\gamma}^{\pm}, \ell_{\mbo}^{\pm},\ell_{\mso}^{\pm})$, related to  $r_{\gamma}^{\pm}, r_{\mbo}^{\pm}$,  $r_{\mso}^{\pm}$, define     three ranges $\mathbf{L1}^{\pm}$, $\mathbf{L2}^{\pm}$, $\mathbf{L3}^{\pm}$ respectively.
We   denote by the  label  $i\in\{1,2,3\}$, any  quantity $\mathbf{Q}$ related to the range  of specific angular momentum $\mathbf{Li}$ respectively;  for example,
$\cc_1^+$ indicates a closed  (regular) counterrotating configuration with specific angular momentum  $\ell_1^+\in\mathbf{L1}^+$.
No  maxima of the effective potential exist for $\pm\ell_{\mp}>\ell_{\gamma}^{\pm}$ ($\mathbf{L3}^{\mp}$) therefore, only equilibrium configurations, $\cc_3$, are possible.
An  accretion  overflow of matter from the  closed, cusped  configurations in   $\cc^{\pm}_{\times}$ (see Fig.\il\ref{Fig:but-s-sou}) towards the attractor  can occur from the instability point  $r^{\pm}_{\times}\equiv r_{\max}\in]r_{\mbo}^{\pm},r_{\mso}^{\pm}[$, if $K_{\max}\in \mathbf{K0}^{\pm}$  with fluid specific angular momentum  $\ell\in]\ell_{\mbo}^+,\ell_{\mso}^+[\equiv\mathbf{L1}^+$  or $\ell\in]\ell_{\mso}^-,\ell_{\mbo}^-[\equiv \mathbf{L1}^-$. Otherwise,  there can be  funnels of  material along an open configuration   $\oo^{\pm}_{\times}$,  proto-jet or for brevity jet,  representing limiting topologies for the  closed  surfaces \citep{KJA78,Sadowski:2015jaa,Lasota:2015bii,Lyutikov(2009),Madau(1988),Sikora(1981)}
 with   $K^{\pm}_{\max}\geq1$ ($\mathbf{K1}^{\pm}$), ``launched'' from the point $r^{\pm}_{\jj}\equiv r_{\max}\in]r_{\gamma}^{\pm},r_{{\mbo}}^{\pm}]$ with specific angular momentum $\ell\in ]\ell_{\gamma}^+,\ell_{\mbo}^+[\equiv \mathbf{L2}^+ $ or $]\ell_{\mbo}^-,
  \ell_{\gamma}^-[\equiv\mathbf{L2}^-$--Figs\il\ref{Figs:GROUND-StA0} and Figs\il\ref{Figs:pranz}.

However, we can  locate the center of each torus
  more precisely,   by
 introducing  the   ``\emph{complementary}'' geodesic structure $R_{\rho}$,  associated  to  the geodesic  structure $R$. \citep{dsystem}. This is  constituted by  the radii  $\rho_j\in R_{\rho}$, defined as $\rho_j>{r}_{j}$ solutions of $\bar{\ell}_{i}\equiv\ell(\rho_{i})=\ell(r_{i})\equiv\ell_{i}$--see Fig.\il\ref{Figs:pranz} and Figs\il\ref{Figs:GROUND-StA0}.
Radii of  $R_{\rho}$ satisfy the same equation as   the  notable radii $r_{i}\in R$ for corotating and counterrotating configurations,
 analogously  to the couples  $r_{\mathcal{M}}^{\pm}$  and  $\rho_{\mathcal{M}}^{\pm}$ satisfying relation  $r_{\mathcal{M}}^{\pm}>r_{\mso}^{\pm}$, where  $\ell_{\mathcal{M}}^{\pm}$ is associated to the   maximum of $\partial_r |\ell(r)|$-- \citep{open}.
The geodesic structure of the Kerr spacetime and the complementary geodesic structure are both significant in the analysis,  especially in the case of $\ell$counterrotating  couples.
There is  $r_{\gamma}^{\pm}<r_{\mbo}^{\pm}<r_{\mso}^{\pm}<\rho_{\mbo}^{\pm}
<\rho_{\gamma}^{\pm}$.
The location of the radii $r_{\mathcal{M}}$ and $\rho_{\mathcal{M}}$ depends   on the fluid  rotation with respect to the Kerr attractor.
  Thus the configurations $\pp_1$ are centered in $]r_{\mso},\rho_{\mbo}[$ (with accretion point in $r_{\times}\in]r_{\mbo},r_{\mso}[$), the  $\pp_2$ rings have centers in the range  $[\rho_{\mbo},\rho_{\gamma}[$  (with $r_{\jj}\in]r_{\gamma}, r_{\mbo}[$), finally the  $\cc_3$ toroids are centered at $r\geq \rho_{\gamma}$.

Figures \ref{Fig:Fig3D} show  three dimensional surfaces of special couples of tori obtained by a 3D-\textbf{GRHD}  integration.
\subsubsection{{Summary of main notation}}\label{Sec:notation-sec}
This section  represents a reference section and provides the summary of the main concepts introduced for the \textbf{RADs}  model, see also Table\il\ref{Table:pol-cy-multi}.
Quantities introduced and discussed here have been grouped according to the properties of the disks or attractors to which they refer.
\begin{table*}
\caption{{Lookup table with the main symbols and relevant notation  used throughout the article.}}
\label{Table:pol-cy-multi}
\centering
\begin{tabular}{ll}
 \hline \hline
$ \cc$&   cross sections of the closed Boyer surfaces (equilibrium torus)\\
$ \cc_{\times}$&   cross sections of the closed cusped  Boyer surfaces (accretion torus)\\
$ \oo_{\times}$&   cross sections of the open cusped  Boyer surfaces
 \\
 $\pp$& any of the  topologies $(\cc, \cc_{\times},\oo_{\times})$
 \\
 $(r_{in}, r_{out})$& inner and outer edge of $\cc_i$ torus
 \\
 $\pm$&counterrotating/ corotating
 \\
$r_{cent}$& location of torus center
\\
$r_{\times}$& accretion point ( inner edge of accreting  torus)
\\
$r_{\jj}$& unstable point in open configurations
\\
$r_{coll}$& contact point in collisions among two  quiescent tori
\\
$\lessgtr$& tori sequentiality according to the centers $r_{cent}$ (inner/outer disks)
\\
$\prec \succ$ &tori sequentiality according to the critical points $(r_{\times},r_{\jj})$
\\
$\widehat{\mathbf{C}}_{{m}}$& {mixed} $\ell$counterrotating sequences
\\
 $\widehat{\mathbf{C}}_{{s}}$& {isolated} $\ell$counterrotating sequences
 \\
$\mathfrak{r_{\times}}$ & {rank}  maximum number of unstable points $r_{\times}$ in a ringed accretion disk
\\
$\ell_{i/i+1}\equiv\ell_i/\ell_{i+1}$&
 ratio in  specific  angular momentum of $\cc_i$ and $\cc_{i+1}$
\\
$r^{\pm}_{\mathcal{M}}$& maximum point    of
derivative $\partial_r(\mp \ell^{\pm})$ for given  $a/M$
 \\
\hline
\end{tabular}
\end{table*}

\textbf{General notations:}
\begin{description}
\item[Given a radius $r_{\bullet}$],
  we adopt  the  notation $\mathbf{Q}(r):\;\mathbf{Q}_{\bullet}\equiv\mathbf{Q}(r_{\bullet})$ for any function $\mathbf{Q}(r)$.
\item[Notation $r_{\bullet} \in \pp$]  means the inclusion of a radius $r_{\bullet}$ in the configuration $\pp$ (location of $\pp$ with respect to $r_{\bullet}$) according to some conditions;
  $\non{\in}$  is  non inclusion
\item[Notation $\bowtie !$] is an \emph{intensifier},  a reinforcement of a relation $\bowtie$, indicating that this is  a necessary relation which is  \emph{always} satisfied;
    \item[Indeces $i$ and $o$]  in general define the inner  $(i)$ torus, the  closest to the central \textbf{BH}  and the outer one $(o)$ as there is $\pp_i<\pp_o$, according to notation in  Table\il\ref{Table:pol-cy-multi}.
\end{description}

\textbf{Expanded  geodesic structure}

Alongside the geodesic structure of the Kerr spacetime represented by the set of radii $R\equiv (r_{\mso}^{\pm}, r_{\mbo}^{\pm},r_{{\mathrm{\gamma}}}^{\pm},r_{\mathcal{M}}^{\pm})$, we associate the  following relations:
\bea&&\label{Eq:conveng-defini}
R_{\rho}\equiv (\rho_{\mbo}^{\pm}, \rho_{{\mathrm{\gamma}}}^{\pm},\rho_{\mathcal{M}}^{\pm}):
\\
&&\nonumber
r_{\mathrm{\gamma}}^{\pm}<r_{\mbo}^{\pm}<r_{\mso}^{\pm}<
 \rho_{\mathrm{mbo}}^{\pm}<
 \rho_{{\mathrm{\gamma}}}^{\pm}\quad\mbox{where}
\\&&\nonumber
 \rho_{\mathrm{mbo}}^{\pm}:\;\ell_{\pm}(r_{\mbo}^{\pm})=
 \ell_{\pm}(\rho_{\mathrm{mbo}}^{\pm})\equiv \mathbf{\ell_{\mbo}^{\pm}},\\
 &&\nonumber
  \rho_{{\mathrm{\gamma}}}^{\pm}: \ell_{\pm}(r_{{\mathrm{\gamma}}}^{\pm})=
  \ell_{\pm}(\rho_{{\mathrm{\gamma}}}^{\pm})\equiv \mathbf{\ell_{{\mathrm{\gamma}}}^{\pm}},\\
  &&\nonumber
\rho_{\mathcal{M}}^{\pm}: \ell_{\pm}(\rho_{\mathcal{M}}^{\pm})= \ell_{\mathcal{M}}^{\pm},
\eea
see Fig.\il\ref{Figs:GROUND-StA0} and Fig.\il\ref{Figs:pranz}-\emph{top panel}.
This expanded structure rules good  part of the geometrically  thick disk physics and multiple structures. The presence of  these radii stands as one of the main effects of the  presence of a strong curvature  of  the background geometry.

\textbf{Notation on the angular momentum $\ell$ and its ranges:}

Indices $i\in\{1,2,3\}$ refer to the following ranges of angular momentum $\ell\in \mathbf{Li}$;
\begin{description}
\item[Range- L1]

$
\mp \mathbf{L1}^{\pm}\equiv[\mp \ell_{\mso}^{\pm},\mp\ell_{\mbo}^{\pm}[$  where   topologies $(\cc_1, \cc_{\times})$ are possible,  with accretion point in  $r_{\times}\in]r_{\mbo},r_{\mso}]$ and center with maximum pressure  $r_{cent}\in]r_{\mso},\rho_{\mbo}]$;
\\
\item[Range- L2]
$\mp \mathbf{L2}^{\pm}\equiv[\mp \ell_{\mbo}^{\pm},\mp\ell_{\gamma}^{\pm}[ $ where   topologies    $(\cc_2, \oo_{\times})$ are possible,  with unstable point  $r_{j}\in]r_{\gamma},r_{\mbo}]$  and  center with maximum pressure $r_{cent}\in]\rho_{\mbo},\rho_{\gamma}]$;
\\
\item[Range- L3]
$\mp \mathbf{L3}^{\pm}\equiv\ \ell \geq\mp\ell_{\gamma}^{\pm} $    where only equilibrium torus  $\cc_3$  is possible with center $r_{cent}>\rho_{\gamma}$;
\end{description}
--see Figs\il\ref{Figs:GROUND-StA0}.

\textbf{Mixed and isolated subsequences}

We introduce also the following  definitions for the $\ell$counterrotating subsequences of a  decomposition of the  order $n=n_++n_-$, {of     \emph{isolated} $\ell$counterrotating sequences   if  $\cc_{n_-}<\cc_{1_+}$ or $\cc_{n_+}<\cc_{1_-}$,   and \emph{mixed} $\ell$counterrotating sequences, if $\exists\; {i_+}\in[1_+, n_+]:\; \cc_{1_-}<\cc_{i_+}<\cc_{n_-}$, or vice versa,   $\exists\; {i_-}\in[1_-, n_-]:\; \cc_{1_+}<\cc_{i^-}<\cc_{n_+}$, where for example $\cc_{i_-}$  is the $i$-torus of the corotating subsequence, or alternatively
$\widehat{\mathbf{C}}_{{m}}$ ($\widehat{\mathbf{C}}_{{s}}$)-{mixed} (isolated) $\ell$counterrotating {sequences}:  $[r_{cent}^{1_-},r_{cent}^{n_-}]\cap[r_{cent}^{1_+},r_{cent}^{n_+}]\neq(=)0$.
Examples of mixed sequences are in
Figs\il\ref{Fig:but-s-sou}\textbf{(a)} and \textbf{(c)} panels.
Examples of isolated subsequences are in
Figs\il\ref{Fig:but-s-sou}-\textbf{(b)} and \textbf{(d)} panel,
with an isolated inner corotating subsequence \textbf{(b)}  and inner counterrotating isolated sequence \textbf{(d)}.

\subsubsection{RAD seeds}

The study of the aggregate of  tori can be carried out  considering \emph{seed} couples. The adoption of this  method,  used in   \citet{ringed,open},    greatly simplifies  the characterization  of  multiple toroidal structures  of the  order  $n>2$.
The  $n$ order \textbf{RADs}   can be investigated starting with   analysis of its subsequences of  the order $n=2$.
Firstly, as   the  parameters  $\{\ell_i\}_{i}$ and  $\{K_i\}_i$ are fixed,  the \textbf{RAD} is uniquely fixed  and therefore it  has a unique state,  \citep{ringed}. A \emph{state} consists   in the  precise arrangement of the following characteristics of the couple: parameters   $(\ell_i, \ell_o)$ and $(K_i, K_o)$,  and relative location of the tori edges, topology  (if accreting or non accreting torus),  if in collision or not \footnote{
The notion of state  is    useful to clarify different aspects of the  macro-configuration structure and evolution. A  ringed disk of the  order  $n=2$, with fixed  critical topology   can be in $N=8$ different states according to the relative position of the centers and rotation: $N=4$ different states if  the rings are $\ell$corotating,  and $N=4$ for  $\ell$counterrotating rings. Considering also the relative location of points of minimum pressure, then the couple $(\mathbf{\pp_a-\pp_b})$  with different but fixed topology,  could  be in $N=16$ different states.}.

\subsubsection{Some notes on emergence of tori collision and tori correlation}

In a seed, tori collision can emerge  depending on fixed constraints on the tori.
 When the seed parameters $(\ell, K)$ are such  that these conditions may  occur at a certain point of tori  evolutions, then  we say that the two  tori are \emph{correlated}.
As a general result, the correlation is possible in all $\ell$corotating couples $\pp_{\pm}-\pp_{\pm}$, according\footnote{More precisely, this means that $\ell$corotating tori may generally evolve (change of $\ell$ and $K$ parameters starting from an initial couple) in order to reach a collision phase. Further discussion about this definition can be found specially in  \citet{open,dsystem,app}} to a proper choice of the density $K$-parameter and specific angular momentum $\ell$.
Because of the intrinsic rotation of the Kerr attractor,  the   correlation in a $\ell$counterrotating couple is disadvantaged and consequently collision may be generally more frequent in a $\ell$corotating couple.  This case  is in fact mostly constrained by particular  restrictions  on the specific angular momentum.  This situation is obviously  less clear in the geometries of the  slower \textbf{BHs}. The dynamics of each torus of an   $ \ell$counterrotating couple   is in this sense   more  ``independent'' from the other,  as  tori of these configurations are separated in some extent during  their evolutions.
A more detailed  look at the  Table\il\ref{Table:nature-Att}, {where main spin-classes and their properties are listed,} reveals that in the  $\ell$counterrotating case the angular momentum is not sufficient to uniquely fix the state and the correlation (indicated with  $({\mathrm{\mathbf{C}}})$). Indeed,  this  depends on the class of the Kerr attractor, and  the fluid density through
  the  $K$-parameter.
It is found  that the   \emph{proto-jet equilibrium}  correlation,  i.e. a $(\oo_{\times}-\cc)$ couple, is not possible  with a  corotating  proto-jet  $\oo_{\times}^-$, and
particularly restrictive conditions are required,   if the corotating   equilibrium torus is the outer one with respect to the proto-jet configuration.
The study of the ranges of   the torus edges $(r_{in},r_{out})$ location with respect to  the  geodesic structure of the  Kerr geometry  is essential to establish  the  conditions for   tori collision. We therefore studied   the inclusions  $r_{a}\in \pp_i$  for a  configuration $\pp_i$ and a radius $r_{a}\in R^{\pm}$.

Collision may arise, if the edges of tori in a couple are  $r_{{in}}^o=r_{{out}}^{i}= r_{coll}$,  for    ({non-accreting or accreting) $\ell$corotating or $\ell$counterrotating seed; we indicate this particular colliding seed with notation $\mathbf{\cc}_{coll}$. An example of such a case is shown in Fig.\il\ref{Fig:but-s-sou}-\textbf{ (b)}.   This process may     eventually lead   to tori merging of a kind of drying-feeding process, outlined by a loop evolution in  \citet{open}. For example, such a collision can occur as    the outer torus   grows  or loses   its angular momentum,  approaching the accretion phase.
Collisions due to \emph{inner} torus growing   is a particularly constrained case.
In general, the emergence of  tori  collision   can be affected by the evolution of the inner torus  of a \textbf{RAD}, especially    in the early stages  of  the  torus dynamics towards the accretion.
The second process  related to  a tori collision  is  featured in Fig.\il\ref{Fig:but-s-sou}-\textbf{(a)-(b)} . This process   implies   emergence of an  instability  phase for the  outer tori  of the \textbf{RAD}: the fluid accretion onto  the central \textbf{BH} necessarily impacts on the {(non-accreting or accreting)} inner torus of the couple.
 Finally, tori collision can emerge also  as a combination of the two processes considered before,  where  there is a contact point,  $r_{out}^{i}=r_{\times}^o=r_{coll}$ with an accreting outer torus.  Collision appears   in this case  together with an  hydro-gravitational destabilization   due to the Paczy{\'n}ski-Wiita  mechanism \citep{Pac-Wii,Paczynski:2000tz}.
\section{Relating  Kerr Black holes  to RAD systems}\label{Sec:RADs-sis}
In this section we  consider the properties of \textbf{RAD} aggregate of multiple tori orbiting around a Kerr attractor  in relation to its dimensionless spin. Our aim is to provide  most complete  description  of the \textbf{RAD} systems orbiting spinning attractors,    characterizing the central Kerr \textbf{BHs} in  classes uniquely identifiable through the properties of these objects rotating around   them.
Findings of our investigation  are listed in Table\il\ref{Table:nature-Att} providing   an overview of the  major features of the classes of spinning attractors,   on the basis of the properties of  the orbiting toroidal structures. In Table\il\ref{Table:nature-pics} we list some general features of the \textbf{RAD} seeds  independent from the \textbf{BHs} spin while in Appendix we discuss more details of the  \textbf{RAD} systems and their attractors.
Our analysis  is conducted by  numerical integrating   the  hydrodynamic equations for multiple systems with fixed boundary conditions for each torus of the  set--see Figs\il\ref{Fig:but-s-sou} and Figs\il\ref{Fig:Fig3D}. We carefully  explored the ranges of  fluid specific angular momentum and  the  tori $K$-parameters,  drawing  the constraints  for the  radii $(r_{cent},r_{\times})$ and  $(r_{in},r_{out},r_{coll})$ for each torus-see also
  \citet{open,dsystem,app}.

 Spin values of  Table\il\ref{Table:nature-Att}  are remarkably close, consequently ranges of \textbf{BH} attractors are very narrow.
  This reference table  could be therefore used to identify, through some of the \textbf{RAD}  features, the properties of the background geometries, placing a Kerr \textbf{BH},  around which a ringed structure can orbit, into one of the particular classes of  Table\il\ref{Table:nature-Att}.
  Therefore, we can effectively read Table\il\ref{Table:nature-Att},  looking for information on a particular class of attractors, narrowed on a limited dimensionless  spin range. Vice versa, for fixed properties of  the \textbf{RADs}, we  could  read the associated class   of attractors where these \textbf{RADs}  can be found according to Table\il\ref{Table:nature-Att} . This analysis   will eventually provide a guide for \textbf{RAD} observations and     provide constraints for a dynamical study  of \textbf{RADs}  evolution in more complex framework, where for example the contribution of magnetic fields is included.
 Many properties used  in  the classification actually refer to   each single \textbf{RAD}, torus (a limiting \textbf{RAD} of the order $n=1$), independently from the macrostructure. Nevertheless,
 in the development of the classification  in  Table\il\ref{Table:nature-Att} we also focused on  several    \textbf{BHs}-\textbf{RAD} features: the \textbf{BH} dimensionless spin (left column), the  tori
location (i.e. the specification of $(r_{in},r_{cent})$ and $r_{\times}$ or $r_{\jj}$)  and relative disposition in the \textbf{RAD},  the emergence of couple instabilities,    the fluid specific angular momentum (if $\ell$corotating or $\ell$counterrotating), chance of collision (seed correlation), maximum number of tori in a \textbf{RAD} (the \textbf{RAD}'s order), and general properties of the \textbf{RAD}  morphology.

{Due to the fact that our analysis includes several  properties of each torus of the aggregate and the \textbf{RAD} structure, it is  convenient  to first introduce   some of the main classes,   focusing on  general tori properties. Thus, below we  provide preliminary and general notes on the classification and  a description of relevant cases.   Specifically
we discuss the  occurrence of the double accretion phase  in the \textbf{RAD} and the presence of possible screening \textbf{RAD} tori. Concluding this part, we examine more closely the tori couples of  the  seed--schemes of Figs\il\ref{Figs:GROUND-scheme}, particularly  the case with an  inner counterrotating torus.
We then concentrate our discussion on specific notable \textbf{BH} spins, by describing the main \textbf{RADs} and attractors properties  in a defined range  of spins. }

\subsection{Double accretion phase in the \textbf{RAD} and appearance of  screening tori.}
 As a general result,  the \emph{maximum number} of critical points  in a \textbf{RAD} can be \emph{two}. This very special \textbf{RAD}, featuring a double accretion,   can be observed \emph{only}  in the spacetimes where  $a>0$, therefore it is a characteristic of all the Kerr spacetimes. These aggregates are  particularly favored by fast spinning \textbf{BHs}. In such configurations  the inner torus in accretion should  \emph{always} be corotating  and  it has to be the \emph{inner} torus of the accreting couple  as  also shown in Figs\il\ref{Fig:but-s-sou}.  Then, the outer  torus, which is not necessarily adjacent to the inner accreting one, \emph{must}  necessarily be the \emph{inner}  of the counterrotating subsequence of the \textbf{RAD}. The only  possible scheme for multi-tori  with  double accretion  is the following\footnote{We exclude   the open configurations $\oo_{\times}$ from this analysis. Further comments including the $\oo_{\times}$    formation can be found in  \citet{open}}:
 \bea&&\nonumber
\mbox{for}\quad 0<a\leq M:\\\label{Eq:spar-bay-cbc}
&&
 \underset{\textbf{(a)}}{\underbrace{\mathbf{\cc_{\times}^-}}}<\underset{\textbf{(b)}}{\underbrace{...<\cc^-<...}}
 <\underset{\textbf{(c)}}{\underbrace{\mathbf{\cc_{\times}^+}}}<...\underset{\textbf{(d)}}{\underbrace{<\cc^{\pm}<}}...
 \eea
 where $\mathbf{(a)}$ is the inner corotating  torus in accretion, $\mathbf{(b)}$ is the  inner subsequence of corotating tori in equilibrium, $\mathbf{(c)}$ is  the outer  counterrotating accreting torus  and  $\mathbf{(d)}$ is the outer (mixed or separated) subsequence composed of corotating or counterrotating  quiescent tori. In Fig.\il\ref{Fig:but-s-sou}-\textbf{ (a)}-panel is an example of   a \textbf{RAD}  of the order $n=3$, $\cc_{\times}^-<\cc_{\times}^+<\cc^-$, with     vanishing  $\mathbf{(b)}$  component of the corotating inner sequence of Eq.\il(\ref{Eq:spar-bay-cbc}).
 The \textbf{BHs} spin
  constrains  the specific angular momentum, elongation, and number $n$ of tori  in the subsequences  $\mathbf{(b)}$ and $\mathbf{(d)}$.

In the special \textbf{RAD} of Eq.\il(\ref{Eq:spar-bay-cbc}),  the subsequence $\mathbf{(b)}$ is made up by ``screening'' corotating, quiescent tori, between the two accreting tori of the agglomerate.
 We should also note that no torus, quiescent or in accretion, can form between the central \textbf{BH} and the inner accreting $\cc_{\times}^-$ configuration.}

\subsection{Preliminary notes on the counterrotating inner tori. }

{
We consider the schemes   in Figs\il\ref{Figs:GROUND-scheme} and the geodesic  structure in Figs\il\ref{Figs:GROUND-StA0} and \ref{Fig:Relevany}. The innermost
stable orbit in counter--rotation, $r_{mso}^+$, increases with the dimensionless \textbf{BH} spin, or   $\partial_{a} r_{mso}^+>0$ (dimensionless units). Whereas the corresponding  corotation radius
decreases with  $a$, or $\partial_{a} r_{mso}^-<0$, due to the corotation and the geometry frame dragging, which acts oppositely  on the   $\ell$counterrotating tori.  This implies that with increasing \textbf{BH} dimensionless spin, $a/M$,  the difference  $r_{\mso}^+-r_{\mso}^-$ increases. The maximal difference occurs  for attractors
with the   extreme maximal  spin value $a=M$. This fact has interesting implications on the \textbf{RAD} stability and the possibility to observe \textbf{RAD} couples. Increasing  $\Delta r_{\mso}\equiv r_{\mso}^{+}-r_{\mso}^-$ with the spin  implies
$\partial_a\Delta r_{\mso}>0$. This fact  has  consequences on the formation of the  $\ell$counterrotating  couples and on their dynamics. The radius $r_{mso}$  is an equilibrium    discriminant for the tori, fixing the regions of the  maximum and minimum of the hydrostatic pressure in  the tori.
Similar relations,   $\partial_{a} \Qa^{\pm}\gtrless0$ and
$\partial_{a} \Delta \Qa>0$, hold for  $\Delta \Qa$  defined analogously to $\Delta r_{\mso}$, where   $\Qa\in{R,R_{\rho}}$. As discussed  in Sec.\il\ref{Eq:tori} and  Sec.\il\ref{Sec:notation-sec}, these radii regulate the equilibrium and the extension of the torus on the equatorial plane as well as  other  morphological characteristics of the tori as   the geometrical thickness.
 Considering these relations and Figs\il\ref{Figs:GROUND-StA0}-\ref{Fig:Relevany}, a double accretion for the couple $\cc^-<\cc^+$,
 scheme \textbf{(4)} of Figs\il\ref{Figs:GROUND-scheme},
  is therefore favored by high  \textbf{BH} spins--Figs\il\ref{Fig:but-s-sou}.
This also has  consequences on the energetics of the processes associated to the  \textbf{RAD}, the  mass accretion rates, the  torus  cusp luminosity, as well as  other tori  characteristics depending on the location of the tori inner edge  and \textbf{BH}  parameters--\cite{app}.
On the other hand,
schemes\textbf{(1)} and \textbf{(3)}  of Figs\il\ref{Figs:GROUND-scheme} feature  an inner
corotating  torus  where the outer toroidal disk   is  corotating and quiescent,  scheme \textbf{(1)}, and
counterrotating in scheme \textbf{(3)}.
The Lense--Thirring \textbf{(L-T)} effect plays an
 essential role for  these couples. 
The inner torus, orbiting  fast spinning \textbf{BH} attractors, is  strongly dominated by the \textbf{L-T} effect and  can also form in the ergoregion being generally also rather small. Such accreting tori can also   influence heavily the \textbf{BH} attractor, establishing a runaway instability, or extracting energy through jet emission.
 The increase of  the \textbf{BH} spin   $a$,   favors  tori collision in  the  \textbf{RAD}  $\ell$corotating couples formed by corotating tori (see the   narrowing of the regions bounded by the radii in $R^-$ and $R_{\rho}^-$ as shown in Figs\il\ref{Figs:GROUND-StA0} and \ref{Fig:Relevany}).
Schemes \textbf{(2)} and \textbf{(4)}  of Figs\il\ref{Figs:GROUND-scheme} feature an inner counterrotating torus, which is part of the   $\ell$corotating couple, $\cc^+<\cc^+$, in scheme \textbf{(2)}, and  component of   the  $\ell$counterrotating  couple,  $\cc^+<\cc^-$,  as represented in  scheme \textbf{(4)}--see also Figs\il\ref{Fig:but-s-sou}.
 Increasing the spin  $a$, the inner  $\cc^+$ torus can be observed also far away  from  the central \textbf{BH}, but accretion is still possible because of the conditions on the radii $R$ and $R_{\rho}$.
Considering scheme $\textbf{(4)}$,  couples $\cc^+<\cc^-$, inner counterrotating torus and outer corotating torus are  expected to be observed especially in the geometries of the slow rotating attractors--Figs\il\ref{Fig:but-s-sou}. We shall detail this case below, within  the attractors classification.
Here we note that
the outer, corotating, quiescent  torus of this couple, can approach the instability only  in the spacetimes of  the  slower spinning attractors (accretion is always associated to  a decrease of the magnitude of the  fluid specific angular momentum  in the range $\textbf{L1}$, and to an increase of $K$-parameter where the torus inner edge moves inwardly towards the central attractor).
The presence of an outer corotating torus,  $\cc^-$, of the \textbf{RAD} couple in  scheme \textbf{(4)} implies, during the \textbf{RAD} evolution, and especially with  the emergence of the outer torus  instability,  the tori collision. This process can also end in a disrupting phenomena, leading  eventually to the tori merging. Therefore, this couple could be generally  considered as  a feature proper of the  slower spinning \textbf{BH}.
At fixed  spin, shifting the torus  center  outwardly,  the accreting torus will be larger than the accreting tori close to the  central \textbf{BH}. On the other hand, the increase of the \textbf{BH} spin has  similar effects, facilitating larger counterrotating tori  and smaller corotating tori very close to the \textbf{BH}, located eventually  in the ergoregion \cite{pugtot}.
 These effects are  investigated in  \cite{ringed,dsystem} and in  \cite{long}.}

\medskip
We  proceed now by considering  some  specific notable  spins of Table\il\ref{Table:coud-many}  in connection with the  principal \textbf{RAD} characteristics.
Firstly we discuss  the case of a static Schwarzschild  \textbf{BH} and then  we concentrate  on  the   Kerr \textbf{BHs}, by evaluating the influence of the spin of the central attractor  examining first  the  slower spinning \textbf{SMBHs}  singled  out  from Table\il\ref{Table:coud-many}.

\subsection{Schwarzschild attractors  $(a=0)$}
In the static geometry described by the Schwarzschild metric   any \textbf{RAD} sequence of tori behaves as an $\ell$corotating sequence, independently of the relative rotation of fluid in the ringed  disk, because of the unique geodesic structure of the spacetime\footnote{This situation  has been  described in  \citet{dsystem} by a \emph{monochromatic  evolutive graph.}}.
As consequence of this,  accretion around these attractors may occur only from the inner torus of the \textbf{RAD}, while  any further torus must be non-accreting (quiescent). Moreover, there can be no screening inner torus, comprised between the outer accreting torus and the central attractor.  This implies also that any further torus of the aggregate, shall be outer with the respect to the accreting inner one. The  tori sequences are characterized by the outer in the \textbf{RAD}  with respect to the inner torus in accretion.
For
$a=0$ a \textbf{RAD} seed can be only in the following configurations  $\pp_{\pm}<\cc_{\pm}$ or $\pp_{\pm}<\cc_{\mp}$.
Then an  emission spectrum from this \textbf{RAD}    should give track as  the single accretion inner edge only--in agreement  with    \citet{Schee:2008fc,Schee:2013asiposs,S11etal,KS10}.
Collision in this geometry is  possible according to specific conditions on the  fluid density and angular momentum and, in case of emergent  Paczy{\'n}ski  instability for one of the  outer tori of the sequence, collision is inevitable.
One torus may evolve from a non-accreting phase to accretion due to  decrease of the  angular momentum followed by an  increase of the  torus elongation on the equatorial plane,  approaching  the attractor. This phenomenon can  be detected as a very violent event with large energy release as shown in the first evaluation of energy collision in  \citet{app}.
\textbf{RAD}  aggregates, orbiting a static attractor, shall therefore characterize  the earlier  phases of evolution  of a \textbf{RAD}, where all the tori, but the inner one, are in equilibrium.
    It is clear than that these considerations for the static \textbf{BHs}, hold in some extent also for very slowly spinning Kerr \textbf{BHs}: in fact the study of the Schwarzschild  case can be seen as the limiting for a Kerr \textbf{BH} where we can consider a ``non-relevant'' influence of the frame-dragging.
    The classification   specifies the limits on the attractor spin  and the \textbf{RAD} features which are mainly affected.   As mentioned above  a   mixed sequences or $\ell$counterrrotating couples are  to be considered  as $\ell$corotating in the Schwarzschild spacetime  because of
    the equivalences  $R^{+}=R^{-}$ and $R_{\rho}^+=R_{\rho}^-$,  as clear also from  Figs\il\ref{Figs:GROUND-StA0}, and \ref{Figs:pranz}.

 In general, in the case of a  spinning Kerr  \textbf{BH}, the physics associated with the \textbf{RADs} is much more complex and  phenomenology is much more rich   as will be discussed  in the following section.
 {Following Table\il\ref{Table:nature-Att}, we explore the \textbf{RAD} in the Kerr spacetimes by starting with the small spins. Classes in the Table are indicated  by arrows, the  center column  to be read for decreasing spin, down-to-up direction, the  right column  to be read for increasing  spin, up-to-down direction.
Thus, for example, considering  spin $a_u$, Table\il\ref{Table:nature-Att}  provides information on   the ranges $a<a_u$  and $a>a_u$ respectively.}

\medskip

\subsection{Kerr Black holes $(a\neq0)$ }

In the case of axi-symmetric  attractors the situation is  complicated due to  the non-zero intrinsic spin $a$ of the attractor and the  relative rotation of the \textbf{RAD} tori.
The situation has been  addressed in details for ringed disks of the order $n=2$ in  \citet{dsystem}.
Below we describe the situation  considering some remarkable values of dimensionless spin with reference to Table\il\ref{Table:coud-many}. These notable  dimensionless spin ranges  identify  different classes of Kerr attractors, according to their dimensionless spin  $a/M$. The representation of the  main classes defined is  pictured schematically on  Fig.\il\ref{Fig:Relevany}-Appendix\il\ref{Appendix:sub-reference}-- note that there are many  intersected \textbf{BH}-classes. {Some general results, holding  irrespectively of the attractor spin-values  $a\neq0$,  are  listed in Table\il\ref{Table:coud-many}.}
In the following we focus our discussion on the special spin ranges defined by
Kerr \textbf{BHs} spins \textbf{(I)} $a_{\delta}=0.638285M$; \textbf{(II)} $a_{\iota}=0.3137M $;
 \textbf{(III)} $a_{\mathcal{M}}^{(3)}= 0.934313M$;
\textbf{(IV)} $a_{IV}=0.461854M$; \textbf{(V)} $a_u= 0.47403M$; \textbf{(VI)} $a_{VI}=0.73688M$;  \textbf{(VII)} $a_{K}=0.372583M$.

To simplify our discussion  on the seed couples,  in the following we make use of the  notation  $(\pp,\lessgtr,\prec \succ)$
of Table\il\ref{Table:pol-cy-multi}, for the relative location of the tori in the couples and their state of quiescence
or instability, as well as the range of variation of the magnitude of the specific angular momentum--see also
Sec.\il\ref{Sec:notation-sec}.

\medskip

%

\medskip

\textbf{I: }[\textbf{Kerr BHs spin $a_{\delta}$ }]

The first class of attractors we consider are the  fastest Kerr \textbf{BHs}  having spin      $a_{\delta}<a\leq M$.   Attractors with   values of $a/M$ in this range, have been singled out because couples $\cc_{\times}^-\prec \oo_{\times}^+$, formed by a corotating accreting  torus with an outer  counterrotating proto-jet, can  be observed  \emph{only} in these geometries.

\medskip

\textbf{II: } [\textbf{Kerr BHs spin $a_{\iota}$ }]

On the other hand, only in the spacetimes  where  $a_{\iota}<a\leq M$  it is possible to find the couple of proto-jets  $\oo_{\times}^-\prec \oo_{\times}$, where the inner configuration is corotating  while the outer can be corotating or counterrotating.
More specifically,
 \emph{jet-jet} correlation, with the production of this double-shell configuration of {proto-jets}, is always possible except for the fast attractors, class $\mathbf{A_{\iota}^>}$,  where  the case of  an outer  counterrotating proto-jet   is prohibited. Vice versa  for slower spinning Kerr \textbf{BHs}, class $\mathbf{A_{\iota}^<}$,   any combinations of  proto-jets are possible.

\medskip

\textbf{III: }[\textbf{Kerr BHs spin } $a_{\mathcal{M}}^{(3)}$]

Spin $a_{\mathcal{M}}^{(3)}$, introduced in Table\il\ref{Table:nature-Att},  discriminates the behavior of corotating  and counterrotating tori  located far  from the source-see Figs\il\ref{Figs:GROUND-StA0}.  Radii $r_{\mathcal{M}}^{\pm}$ and corresponding specific angular momenta $\ell_{\mathcal{M}}^{\pm}$  play an important role in the characterization of  the classes of attractors defined by this  special spin.
%
%
In fact, radii $(r_{\mathcal{M}},\rho_{\mathcal{M}})$ and the angular momenta $\ell_{\mathcal{M}}$ regulate the tori   angular momentum  also in region far from the source, i.e.,  $r>r_{\gamma}^{\pm}$.
Radii  $r_{\mathcal{M}}$ and $\rho_{\mathcal{M}}$ play a role in tori formation and, particularly, their collision  \citep{open}.
 %
Indeed, radii $r_{\mathcal{M}}^{\pm}(a)>r_{\mso}^{\pm}$  correspond to the maximum point  of the variation of the  magnitude of the tori fluid specific angular momenta   $\ell$ with respect to the orbital distance from the attractor;  they are the solutions of $\partial_r\partial_r\ell=0$, providing the  maximum point  of the functions  $\mp\partial_r\ell^{\pm}$, respectively--Table\il\ref{Table:pol-cy-multi}. This means that
increasing  of  the magnitude of the  specific angular momentum     with the radial distance  from the attractor is not permanent,  but  there is a  limiting radius,
 which is  different for the $\ell$counterrotating subsequences-see Fig.\il\ref{Figs:pranz}-\emph{bottom panel}. We specify the situation below.
The maximum $\ell_{\mathcal{M}}$ is associated to a torus centered at  $r_{\mathcal{M}}$, having critical point at  $\rho_{\mathcal{M}}$ (see definition \ref{Eq:conveng-defini}). This implies that  the tori of  $\ell$corotating couple  with fixed  angular momentum magnitude difference, $\ell_o-\ell_i=\epsilon$,  are increasingly closer to the central  \textbf{BH}  as their angular momentum approaches the limiting  value $\ell_{\mathcal{M}}$.
Noticeably this  is a relativistic effect that disappears in the Newtonian limit  when the orbital distance is large enough with respect to  the radius   $r^{\pm}_{\mathcal{M}}$.
 On the other hand,  in the case of static spacetimes,  the Schwarzschild \textbf{BH}  has  $r^{+}_{\mathcal{M}}=r^{-}_{\mathcal{M}}$.
Therefore, the existence of the  $(r_{\mathcal{M}},\ell_{\mathcal{M}})$ is a  relativistic effect, present also in the static case $a=0$, but for a rotating attractor this is  strongly differentiated from an   albeit minimal intrinsic rotation of the gravitational source.
Furthermore, this behavior of the specific angular momentum  strongly  distinguishes the two $\ell$corotating  subsequences   of corotating and    counterrotating tori.
As there is  $\partial_{|a|}r_{\mathcal{M}}^\mp\lessgtr0$,  for corotating and counterotating  tori respectively, the region where less additional specific angular momentum  is due increases with the spin-mass ratio of the central \textbf{BH} in the corotating   case, while it decreases with the spin in the other case--there is  $\partial_{r_{cent}}|\ell_{r_{cent}}|>0$ $\forall r_{cent}$,  but  $\partial^2_{r_{cent}}|\ell_{cent}|>0$  up to a $r_{\mathcal{M}}$,  which corresponds  to the ring with the center located on  $r_{\mathcal{M}}$ where
$\partial^2_{r_{cent}}|\ell_{cent}|=0$. Then  there is  $\partial^2_{r_{cent}}|\ell_{cent}|<0$ for $r_{cent}>r_{\mathcal{M}}$, or also:
$
\partial_{r_{cent}}\left.\delta \ell_{cent+j,cent}\right|_{\jj} \gtreqqless0$ {for} $r_{cent} \lesseqqgtr r_{\mathcal{M}}$, {and} $r_{cent}\in]r_{\mso},+\infty[
$.
These results apply in any Kerr spacetime, but depending on the attractor spin, the  maximum $r_{\mathcal{M}}$,  being  a function of $a$,  is located  in  different orbital regions\footnote{More specifically, there is  $
r_{\mathcal{M}}^-(a_{\mathcal{M}}^{(3)})=r_{\gamma}^+(a_{\mathcal{M}}^{(3)})$,  where  $ a_{\mathcal{M}}^{(1)}<a_{\mathcal{M}}^{(2)}<a_{\mathcal{M}}^{(3)} $ and spins $(a_{\mathcal{M}}^{(1)},a_{\mathcal{M}}^{(2)})$ verify the relations
$ r_{\mathcal{M}}^-(a_{\mathcal{M}}^{(1)})=r_{\mso}^+(a_{\mathcal{M}_1})$ and $
 r_{\mathcal{M}}^-(a_{\mathcal{M}}^{(2)})=r_{\mbo}^+(a_{\mathcal{M}}^{(2)})$ respectively--see Figs.\il\ref{Figs:pranz}-\emph{bottom panel}}--see Fig.\il\ref{Figs:GROUND-StA0}.
 For the case of corotating tori,  the region of increasing gap of specific angular momentum decreases with the spin; that is,  increasing of the  \textbf{BH} spin  corresponds to a decrease of the additional specific angular momentum to be supplied to locate the ring centers in an exterior region.  In general, one could say  that if the $\ell$corotating   tori are corotating with respect to the \textbf{BH} we need less additional specific angular momentum to locate outwardly ($\ell$corotating and corotating) tori with  respect to the counterrotating case. In this sense, one could also say   that the corotation with respect to the \textbf{BH}   has a stabilizing effect for the \textbf{RAD} structure.
Conversely, in terms of the  criticality indices  $r_{j}$ (location of critical points),   there is
$\partial_{r_{\jj}}\ell <0$, with
$\partial^2_{r_{\jj}}\ell=0$ for a configuration with specific angular  momentum magnitude   $\ell_{\mathcal{M}}>\ell_{\mbo}$,  centered in  $r_{\mathcal{M}}$.

Indeed, since
 $r_{\mathcal{M}}^{\pm}>r_{\mso}^{\pm}$, the radii $r_{\mathcal{M}}^{\pm}$ can be   minimum points of the effective  potential, but \emph{not} maximum points.  Then $\ell_{\mathcal{M}}$ is a maximum value for   the function $\partial_{r_{cent}}\ell_{r_{cent}}=\partial_{r_{cent}} r_{j}\partial_{r_{j}} \ell_{r_{j}}$.
 The $\ell$corotating    sequences  of corotating and counterrotating open configurations with specific angular  momentum  $\ell_{\mathcal{M}}^{\pm}$ have generally different topologies associated with their critical phase. In fact, for the counterrotating tori around attractors with  $0<a\leq M$, we have
$-\ell_{\mathcal{M}}\in\mathbf{L2}^+$; therefore,   critical configurations with  $\ell=\ell_{\mathcal{M}}^+$ always correspond to the proto-jets $\oo_{\times}^+$.
Conversely, this is not always the case for the corotating tori
 where, at higher spin  of the attractor, i.e.,   $a\geq a_{\mathcal{M}}^{(3)}$, there is  $\ell_{\mathcal{M}}^-\in \mathbf{L3}^-$, where  there are no critical topologies, while in the geometries with   $0\leq a<a_{\mathcal{M}}^{(3)}$ there is  $\ell_{\mathcal{M}}^-\in \mathbf{L2}^-$,  and only critical configurations $\oo_{\times}^-$ are possible.
  Summarizing, the tori centers are  more spaced in the region $r>{r}_{\mathcal{M}}$, and vice versa at  $r<{r}_{\mathcal{M}}$; the tori are closer each other   as they approach  ${r}_{\mathcal{M}}$.
In conclusion, in a ringed  model,   where the specific angular  momentum varies (almost) monotonically with a constant step $\kappa$, two remarkable points in the distribution of matter appear:  the  radius $r_{\mathcal{M}}$, where the density of stable configurations  reaches the  maximum (the maximum hydrostatic pressure), and  the corresponding point $\rho_{\mathcal{M}}$  where the density of jet launch is   at minimum.

\medskip

\textbf{IV: }[\textbf{Kerr BHs spin $a_{IV}$}]

Spin $a_{IV}$ identifies one of the  most remarkable  classes of Kerr \textbf{BHs}  in our classification.
We detail  the properties of the \textbf{RAD} orbiting attractors of these classes as follows.
For the faster spinning  \textbf{BHs} with    $a> a_{IV}$, there are \emph{no} $\pp^+<\pp_1^-$ couples (that is with an inner counterrotating configuration and outer corotating one with $\ell\in \textbf{L1}$),
 whereas  it is possible to find   a  $\pp_1^+<\pp_i^-$   couple with  $i\in\{2, 3\}$, in the \textbf{BH} geometries where  $a_{IV}<a<a_{VI}$. This is a strong constraint on the specific angular momenta and on the evolution of the couple tori, which we can read as follows: if the counterrotating inner torus  has specific  momentum  $\ell^+\in \textbf{L1}^+$,  thus in a  possible  phase of accretion, the corotating outer torus must have a sufficiently large momentum,  $\ell_i^-\neq\ell_1^-$, which also implies that torus must be sufficiently far from the central attractor and that this couple can only be observed by having a quiescent outer torus (or proto-jet); we  refer to  Figs\il\ref{Figs:pranz}.

Finally only  in the geometry of slower spinning \textbf{BHs}, where the spin is   $a< a_{IV}$ (including  also the Schwarzschild \textbf{BH} with $a=0$),  the  couple made up by an inner counterrotating configuration and outer corotating, quiescent or accreting one, with angular momentum $\ell^-\in \textbf{L1}^-$, that is the seed  $\pp^+<\pp_1^-$,  can  be found with initial state  having  tori $\pp_1^+<\pp_1^-$, which means an  inner  counterrotating, quiescent or accreting, torus with $\ell^+\in \textbf{L1}^+$. This fact  makes this class a very special case. In fact, Kerr \textbf{BHs}  belonging to this class  of attractors constitute the only   spacetimes  where couples of the kind  $\pp_1^+<\pp_i^-$ for $i\in\{1,2,3\}$, can  be observed (inner counterrotating, quiescent or accreting torus with $\ell^{+}\in \textbf{L1}^{+}$, with an outer corotating configuration)  and,  in particular, \textbf{RAD}   tori $\pp_1^+<\pp_1^-$ (inner counterrotating, quiescent or accreting torus, with an outer corotating torus with angular momenta $\ell^{\pm}\in \textbf{L1}^{\pm}$ respectively) can  orbit only  around these attractors.  As discussed above, these are the only couples where a double accretion phase in the \textbf{RAD} is possible.
 On the other hand,
\be
\mbox{for  }\quad  0<a<a_{IV},\quad \mbox{tori as }\quad \cc_{\times}^+< \cc^-,
\ee
can be observed,  i.e.,  an inner counterrotating accreting torus and an outer  corotating  (non-accreting) torus--see Fig.\il\ref{Fig:but-s-sou}.
A couple  of tori $\pp^+<\cc^-$ can be observed  around any  Kerr \textbf{BH}   attractor  with  $0\leq a\leq M$ (note that we did not specify the range of the specific angular momentum), but only if the central \textbf{BH} has spin in the range
   $0\leq a<a_{IV}$,  the corotating outer torus  of the couple $\cc^-$   approaches  the instability   phase, i.e. there  is  $r_{\times}\gtrapprox r_{\mso}^-$.
The  {faster}  spinning  is the Kerr \textbf{BH},   the {farther away} ($r_{cent}>\rho_{\gamma}^-)$
should be the outer torus to prevent { tori collision}
\footnote{ For  the corotating tori with specific angular momentum  $\ell^- \in \mathbf{L2}^-$, whose unstable mode is a proto-jet,  orbiting  the fast attractors of class   $\mathbf{A_{\varrho}^>}$,  the marginally stable orbit can be ``included'' in their equilibrium configurations i.e. $r_{in}<r_{mso}<r_{out}$ where $r_{in}\neq r_{\times}$. At lower spins, i.e. $a\in\mathbf{A_{\varrho}^<}$, this  is not possible.
For higher specific angular momentum magnitudes, $\mathbf{L3}$,  where there are no unstable modes of tori, the inner edge of a counterrotating equilibrium configuration is always outer to the $r_{\mso}^+$ orbit, $r_{in}^+>r_{mso}^+$, so as  a corotating torus orbiting attractors at  low spin, $\mathbf{A_{\varrho}^<}$. In the $\mathbf{A_{\varrho}^>}$  geometries,  the specific angular momentum has to be low enough for the equilibrium torus $\cc_3^-$. It is clear then that this discussion crosses  the problem of torus location and specifically the inner edge of a torus. This is indeed a relevant problem of the accretion disk theory, which has been variously faced in literature, we mention   \citet{open,long} for a deeper discussion in the \textbf{RAD} scenario.}.

\medskip


\textbf{V: }[\textbf{Kerr \textbf{BH} spin $a_u$}]

Spin  $a_u$  defines     two classes  of  \textbf{BHs} particularly significant  from  the point of view of the structure of ringed accretion disks.
We start by considering the  faster attractors with  spin $a>a_u$ which are  characterized by \textbf{RAD} couples
 $\pp_2^+<\pp^-$ (inner counterrotating configuration with $\ell^+\in \textbf{L2}^+$ and outer corotating one) that can only be observed  as  $\pp_2^+<\cc_3^- $, this means  that the outer  corotating configuration has to be quiescent and located far from the attractor according to the limits provided by $\ell\in \textbf{L3}^-$ (we note that particularly  there can never be a seed with $\pp_2^+<\pp_1^-$).
Couples of the kind  $\pp^+<\pp_2^-$  (inner counterrotating configuration and outer corotating one with $\ell^-\in \textbf{L2}^-$) can be observed only in a restriction of these geometries, i.e.,  they can  orbit  around attractors with  $a_u<a< a_{VI}$, and  this couple must  necessarily  have an inner counterrotating torus with specific angular momentum $\ell\in \mathbf{L1}^+$ (i.e. $\pp_1^+<\pp_2^-$). Indeed, this case is particularly relevant, being possible, for $\ell\in\mathbf{L1}^+$, either as a quiescent counterrotating torus or  an accreting torus.  Around  \textbf{BH}  attractors  of this class,
\textbf{RAD} couple $\cc_3^-<\pp_i^+, \forall i$ (inner corotating quiescent torus with $\ell^-\in \textbf{L3}^-$ and outer counterrotating one)  can orbit.
 This couple has an inner corotating quiescent torus  with very high specific angular  momentum ($\ell^-<\ell_{\gamma}^-$), therefore it could be located  also far from the central \textbf{BH} and, according to the  mutual location of the tori of the couple, the outer torus  would therefore  be also far away from the \textbf{BH}. Particularly, \emph{only} in these Kerr geometries,   couples $\cc_3^-<\pp_1^+$ (inner corotating quiescent torus with $\ell^-\in \textbf{L3}^-$ and outer, quiescent or accreting, counterrotating one  with $\ell^+\in \textbf{L1}^+$) can be observed.  However, we  note that the tori distance from the attractor depends actually  on the \textbf{BH} spin. We can see this considering   Figures \ref{Figs:pranz},  combining information on the specific  angular momenta  $\ell^{\pm}(a)$, and the relevant radii   $R(a)$ and  $R_{\rho}(a)$ with respect to the  \textbf{BH} spin that determine the location and conditions necessary for the instability of the couple. These  conditions inform us that, being $\partial_a\Qa^{\mp}\lessgtr0$  where
 $\Qa^{\mp}\in\{\ell^{\mp},R^{\mp},R_{\rho}^{\mp}\}$ with very large spins, the inner torus can also be very close to the \textbf{BH} hole while the outer one can be  very far away (depending on the angular momentum).   Nevertheless, despite the fact that the outer counterrotating torus can be accreting, the inner corotating one must be quiescent according to the constraint  $\ell\in \mathbf{L3}^-$. If the outer torus of such couple   is accreting, then the inner corotating, quiescent torus, acts as a ``screening'' inner torus, matter flows from the outer one impacting on the inner torus.  This fact plays a role in the tori evolution and  ringed disk evolution\footnote{For  convenience  of discussion, we may assume the  existence of a  phase in the formation of a seed,   in which both  tori are in an equilibrium state, $\cc_i<\cc_o$, and   the system    evolves towards an  instability  phase.
In a very simplified scenario, one can assume that the inner torus  may  be formed   even after or simultaneously with formation of the outer torus  from some local material.
 The torus evolution takes place following a possible decrease, due to  some dissipative processes, of its specific  angular momentum  magnitude  towards the  $\mathbf{L1}$ range where accretion is possible. Nevertheless, these states can be reached only in few cases, and under particular conditions collision  occurs--see also \citet{dsystem}.}. Note that this situation is made possible by the different behavior of the corotating and counterrotating tori with respect to  spin shift.
Then, couples  $\pp^-<\pp_1^+$ (inner corotating configuration and outer, quiescent or accreting, counterrotating one  with $\ell^+\in \textbf{L1}^+$) can orbit \emph{only} around these \textbf{BHs} as  $\cc_3^-<\pp_1^+$   tori, which means that  the inner torus is far enough from the central attractor.
 On the other hand,  \textbf{RAD} seeds of the kind  $\pp_2^+<\pp^{-}$  ($\ell$counterrotating couple  made up by  an inner counterrotating configuration with $\ell^+\in \textbf{L2}^+$, and an outer, quiescent or accreting, corotating  one), orbiting attractors with
$a<a_u$,  can only be observed   as $\pp_2^+<\pp_i^-$ with $ i\in(2,3)$, i.e.,  the outer corotating torus cannot have specific angular momentum $\ell\in \textbf{L1}$. Noticeably this is  the  only \textbf{BHs} class  where  $\pp_2^+<\pp_2^-$ is possible, this couple is made by an inner counterrotating torus and outer corotating torus with fluid specific angular momentum $\ell^{\pm}\in \textbf{L2}^{\pm}$ respectively,  tori can be quiescent or  we have proto-jets.
Tori couples  $\pp_i^+<\pp_2^-$ (inner counterrotating configuration, outer corotating configuration with $\ell^{-}\in \textbf{L2}^{-}$) can only have $i \in(1,2)$, therefore  the fluid specific  angular momentum  has to be  $\ell^+\non{\in}\textbf{L3}^{+}$ and the geometries of this class are the only   where $\pp_2^+<\pp_2^-$ can be observed  (there is $i\neq3$ for $a\neq0$)  in accord  with the result presented above.
 Couples $\cc_3^-<\pp^+$ (inner quiescent corotating torus with $\ell^-\in \textbf{L3}^-$ and an outer counterrotating toroidal configuration)   can be observed only as $\cc_3^-<\pp_i^+$ with  $i\in(2,3)$, or the couple is
 $\ell^+\non{\in}\textbf{L1}^+$; particularly this means the outer torus cannot be in accretion.
Similarly
 $\pp^-<\pp_1^+$ (the case of an inner corotating and outer counterrotating torus which can be quiescent or in accretion) can be observed only as $ \pp_i^-<\pp_1^+$ with $i\in (1,2)$; thus the inner corotating torus can be also in accretion and its specific angular momentum has to be $\ell_{mso}^-<\ell^-<\ell_{\gamma}^-$--see Figs.\il\ref{Figs:pranz}. Finally, the \emph{accretion-accretion} correlation is possible but  with an inner counterrotating accretion point in the spacetimes with $a<a_{IV}$ only {(although
 the accreting couple $\cc_\times^-<\cc_\times^+$, according to (\ref{Eq:spar-bay-cbc}) can be observed in all the  spacetimes where $a\neq0$, but  they are largely more expected orbiting around the faster spinning  attractors \cite{open})}.
 \medskip

\textbf{VI: }[\textbf{Kerr BHs spin $a_{VI}$}]
 A \textbf{RAD} seed  of the kind  $\pp^+<\pp_2^-$ (inner counterrotating torus with an outer corotating configuration having $\ell^-\in \textbf{L2}^-$) cannot be observed  around attractors with spin  $ a> a_{VI}$,
while the  couple $\pp_1^+<\pp^-$ (inner counterrotating torus with angular momentum $\ell^+\in \textbf{L1}^+$, which can be quiescent or in accreting, and outer torus corotating) is constrained as $\pp_1^+<\cc_3^-$, that is the outer corotating torus must have fluid specific angular momentum $\ell^-\in \textbf{L3}^-$.
    Properties of the spacetimes with $a<a_{VI}$  are discussed  in reference to the limits $a_u$ and $a_{IV}$.
    \medskip

\textbf{VII: }[\textbf{Kerr BHs spin $a_{K}$}]

The main properties of the classes of attractors defined by  spin $a_{K}$ refer  mainly to the location of the inner and outer  edges  of the tori  with respect to radii  $R$ in (\ref{Eq:conveng-defini}).
As we discussed earlier, this information is relevant both for determining whether collisional effects between adjacent tori of the sequence emerge,  and   for a more deeper understanding of the instability emergence  for a single torus of the aggregate-- \citep{open}. Therefore, the fast attractors defined by this spin class  $\mathbf{A_{K}^>}:\,a_{K} \leq a\leq M$ and the slow spinning attractors belonging to the class $\mathbf{A_{K}^<}: 0\leq a<a_{K}$
are distinguished by the location of  tori centers $(r_{cent})$ and critical points with respect to the radii in the set $R$ and $ R_{\rho}$ respectively. Since the study of these cases is  quite technical, we particularize the  results of this specific analysis  in the   of Appendix (\ref{Appendix:sub-reference}).

\medskip

Finally, we conclude this section with some notes on the role of the Kerr equatorial frame  dragging  in the regulation of  the multiple accreting periods of  the  \textbf{BHs}. We will consider directly attractor classes  delimited by the spins $a_0$, $a_b$ and $a_2$,  defined by considering the crossing  of radii $R^-$  with the static limit $r_{\epsilon}^+$-- see Fig.\il\ref{Figs:GROUND-StA0}.
In fact, an important  and intriguing possibility relies  in the situation where  the inner edge of the inner torus   of the \textbf{RAD} aggregate is located on   $\Sigma_{\epsilon}^+$, which is the region closest to the central \textbf{BH} horizon.
The possibility that  corotating tori may penetrate or even form inside  the region $\Sigma_{\epsilon}^+$  has been recently  explored in details in  \citet{ergon,pugtot}, and particularly in relation to \textbf{RAD} model in \citet{ringed,open,dsystem}.
  In  axi-symmetric spacetimes, the static limit $r_{\epsilon}^+$ separates the two $\ell$counterrotating subsequence. In general,  the  inner edge, and in some cases  also the  torus center $r_{cent}$, of $\pp_{}^-$ or $\oo_{\times}^-$ configurations may cross the static limit--\citep{pugtot,ergon}\footnote{
Noticeably, the occurrence of these cases  for the solutions of hydrostatic equations of the tori, depends in fact  directly on the  $ \ell/a$  ratio-see  \citet{pugtot}.}.
An interesting complementary study of the velocity profiles representing peculiarity of the frame dragging influence  on the  toroidal structure  can be found in  \citet{Stuchlik:2004wk}.

Matter of the toroids will  penetrate the ergoregion $\Sigma_{\epsilon}^+$ in the equatorial plane for sufficiently fast  Kerr attractors.  The funnels of material from the tori will eventually cross outwardly the static limit with an initial velocity  $\dot{\phi}>0$,  following a possible energy extraction process \citep{pugtot,ergon}. The maximum elongation of such toroids decreases with the \textbf{BH} spin, being  closer  to the central attractor; we may say  that the \textbf{BH}  spin acts to squeeze  the tori because of the  frame dragging,  and the faster is the attractor, the smaller are these peculiar    corotating tori.
We start from the class  of the slowest attractors with   dimensionless spin  ${a<a_1}$,  where no configuration (no part  of any $\pp^-$ torus) could be in  $\Sigma_{\epsilon}^+$.
The geometries where   ${a_1<a< a_b}$,  presents a  rather limited region in which the spin varies approximately of $\Delta a\approx 0.1 M$. In these spacetimes,
non-equilibrium  $\oo_{\times}^-$ configurations  can have the instability point   $r_{\jj}^-\in \Sigma_{\epsilon}^+$.
However, the center  $r_{cent}$ of any $\pp^-$ torus  must be external to this region, and the inner edge of any equilibrium configuration  could be included in this region for $\oo_{\times}^-$ (more detailed analysis of this inclusion with the restrictions caused by specified range of angular momentum is provided in  \citet{open,long}).
This  possibility    is particularly relevant  considering that  the  $\oo_{\times}$ accreting tori have been differently associated to the possibility of jet launch (\emph{proto-jet} configurations \citep{open,2010A&A...521A..15A,long}), while for the case of  lower  spin  of the  \textbf{BH}, $a<a_1$, no such instability points  can be allowed in  $\Sigma_{\epsilon}^+$.
The scenario in the geometries of the fast Kerr attractors is more diversified, as also tori  with lower specific angular momentum $\ell$,  non-accreting $\cc_1^-$ and accreting $\cc_{\times}$ tori,  can be, and in some cases must be, in  $\Sigma_{\epsilon}^+$.
In general, with the increase of $ a/M $, equilibrium for axi-symmetric tori is  possible also for lower specific  angular momenta. In $\Sigma_{\epsilon}^+$, the frame dragging is  able to compensate for  the centrifugal force in the disk forces balances,  as evident from  the effective potential function $V_{eff}$.
Then we consider  \textbf{SMBHs} with spin in  ${a_b< a<a_2}$,  with  spin range  having   extension  $\Delta a\leqslant0.1 M$.  In these geometries, the \textbf{RAD} $\cc^-$ tori   center $r_{cent}$ can be placed in $\Sigma_{\epsilon}^+$, and particularly,  the accreting point  $r_{\times}^-$ can be included;  i.e.,  the accretion can take place within the ergoregion $\Sigma_{\epsilon}^+$. {We should also note how    the Table\il\ref{Table:nature-Att},  with additional  combination of the information provided in Table\il\ref{Table:nature-pics} and Table\il\ref{Table:nature-Atcol},  shows a    remarkable closeness of   the  notable spins.}
For fast attractors, where spin is  ${a>a_2}$,  the accretion point $r_{\times}$ \emph{must} be included  in $\Sigma_{\epsilon}^+$.
Then a torus may be also entirely contained  in  the ergoregion \citep{pugtot}.
However, as  $\Sigma_{\epsilon}^+$    is bounded below by  the \textbf{BH}  horizon $r_+$, thus we can say it acts as a ``contact region'' between the accretion torus (corotating) and the  \textbf{BH}  horizon, representing a ``transition region'' where \textbf{BHs} interaction with environment matter in accretion  is essentially regulated by the   Lense-Thirring  effects.
 It is, particularly, the  region firstly affected by any   variation of the spacetime structure,  induced  especially by a change of the  \textbf{BH} parameters due to, for example,  a back-reaction  of the accretion process itself (as the  runaway instability).
A remarkable possibility  in this sense resides in the transition  of spin values     in close proximity of the limiting spins   which should be reflected as quite huge changes     in the \textbf{RAD} configurations we are considering. These processes may give rise to  transient phenomena of positive or negative feedback \textbf{RAD}-\textbf{BH}  analogue, for example, to the runaway instability.
%
\newcommand{\mc}[2]{\multicolumn{#1}{c}{#2}}
\definecolor{Gray}{rgb}{0.75,0.75,0.75}
\definecolor{LightCyan}{rgb}{0.88,1,1}

\newcolumntype{a}{>{\columncolor{Gray}}c}
\newcolumntype{b}{>{\columncolor{white}}c}
\begin{turnpage}
\begin{table}[h!]

\centering

\caption{\label{Table:nature-Att}Classes of the attractors. {Values of the Kerr \textbf{BH}  spin considered for the characterization of the ringed structures in the Kerr  geometries and the properties of the states of the \textbf{RAD} are listed.  Further details on the classes of attractors are enlightened throughout the  text. Notation  $(\non{\mathbf{\mathcal{C}}})$ indicates  non-correlated configurations,  $(\non{\mathbf{\mathcal{C}}}_*)$    stands for correlated tori with particularly restrictive conditions to be satisfied. $({\mathbf{\mathcal{C}}})$ stands for the possibility of tori correlation.
 The  $(*)$ sign in general means   particularly restrictive conditions to be satisfied for the property to be realized.
 Arrows    are in accordance to the decreasing spin (center column)  or, vice versa, increasing spin (right column). In this way we can characterized the classes  of the Kerr  attractors   through the  correspondent  \textbf{RAD} properties.  For a fixed spin  $\bar{a}$, center column, ``decreasing spin '', collects all the  \textbf{RAD} properties  holding in the geometries with  $a<\bar{a}$ (decreasing spin), down-up reading. Vice versa,  right  column, ``increasing spin '', collects all the  \textbf{RAD} properties  holding in the geometries with  $a>\bar{a}$ (increasing spin), up-down reading.  For a bounded spin range, say $\bar{a}<a<{\bar{\bar{a}}}$},  information of the center and right columns  have  to be combining by reading the center column from ${\bar{\bar{a}}}$ to ${\bar{a}} $, and right column from  ${\bar{a}} $ to  ${\bar{\bar{a}}}$. Angular momentum definitions $({\ell}_*,{\ell}_{\varrho}^{\pm},{\ell}_{\beta}^-,\ell_{\Gamma}^{-},\ell_{\mu}^-,\ell_{q}^-)$ are in Table\il\ref{Table:def-flour}--see also Figs\il\ref{Figs:GROUND-StA1} and \ref{Figs:GROUND-StA0}.
General review of definition of notations and symbols can be found in the summary Table\il\ref{Table:pol-cy-multi}, Table\il\ref{Table:def-flour} and Sec.\il\ref{Sec:notation-sec}. More details on specific values of the spin may be found in Sec.\il\ref{Sec:RADs-sis} and  in \citet{dsystem}. }
\centering
\resizebox{1.25\textwidth}{!}{%
\begin{tabular}{b|a|b|a|b}
\hline
\textbf{{\large{{\upshape Classes of attractors}}}
}&\textmd{\scriptsize\textbf{down-to-top}}& \textbf{{\large{\upshape{Decreasing BH spins}}}}& \textmd{\scriptsize\textbf{top-to-dow}}& \textbf{{\large{\upshape Increasing BH spins}}}
\\\hline\hline
  ${a_{\iota}}\equiv0.3137M:r_{\mbo}^-=r_{\gamma}^+$&$\uparrow$&${\mathbf{A}}_{\iota}^<:$ $r_{\mbo}^-\non{\in} \cc_2^+,\;r_{\mbo}^-{\in} \oo_{\times}^{+}(*)$& $\downarrow$&${\mathbf{A}}_{\iota}^>:$ $r_{\mbo}^-\non{\in} \pp_2^+$
  \\
  & &\;\;\;\;$\widehat{\mathbf{C}}_{{m}}:$ $\oo_{\times}^+\succ \oo_{\times}^-$ or $\oo_{\times}^+\prec \oo_{\times}^-$& & \;\; $\widehat{\mathbf{C}}_{{s}}:$ $ \oo_{\times}^+\succ! \oo_{\times}^-$-${\mathrm{\mathbf{(\non{C})}}}$;
 \\
 \hline
  ${a_{K}}\equiv0.372583M:r_{\mso}^-= r_{\mbo}^+ $ &$\uparrow$&$  \mathbf{A}_{K}^<$:   $\cc_{\times}^-<!\cc^+$, $r_{\times}^{-}\leq r_{in}^{+}$-$(\mathcal{C})$, $\cc_{\times}^-\non{\prec} \oo_{\times}^+-(\mathcal{C})$&$\downarrow$&$ \mathbf{A}_{K}^>$:  $\cc_{\times}^-<!\cc*^+$,-$\non{
 (\mathcal{C})^*}$, $\cc_{\times}^-\prec! \oo_{\times}^+-(\non{\mathcal{C}}^*)$
   \\
 &&\;\;\;\; $r_{\mso}^-\in! \oo_{\times}^{+}$, $r_{\mso}^-\in \cc_2^{+}*$, $r_{\mbo}^+\non{\in}\cc_3^-$&&\;\;  $r_{\mso}^-\non{\in}\{ \pp_1^+,\cc_2^+\}$, $r_{\mso}^-\in \oo_{\times}^{+}*$
    \\
      \hline
    ${a_*}\equiv 0.401642 M:{\ell}_*=\ell_{\gamma}^-$&$\uparrow$&$ \mathbf{{A}^<_*}:$ $
r_{\mso}^+\in \cc_1^-$,  $r_{\mso}^+\non{\in} \cc_3^-$; $r_{\mso}^+\in \cc_2^-(*)$,&$\downarrow$&$ \mathbf{{A}^>_*}:$ $r_{\mso}^+\in \{\cc_{1}^-,\cc_2^-\}$, $r_{\mso}^+\in \cc_3^- (*)$
    \\
    \hline
    $ {a_{IV}}\approx 0.461854M:\ell^-(r_{\mso}^+)=\ell_{\mbo}^-$&$\uparrow$&$ r_{\mso}^+=r_{cent}^-\in\{\cc_1^-, \cc_{\times}^{1^-}\}$& $\downarrow$&$ r_{\mso}^+=r_{cent}^-\in\{\cc_2^-, \oo_{\times}^{-}\}$
    \\
    \hline
\rowcolor{lightgray}
   $a_{u}=0.474033M:\; \rho_{\mbo}^+=\rho_{\gamma}^-$
&$\uparrow$&$$& $\downarrow$&
        \\
  \hline
     $a_{\beta}\equiv0.628201 M:\ell_{\beta}^-=\ell_{\gamma}^-$&
     $\uparrow$&$\,r_{\mbo}^+\non{\in} \cc^-_3$, $r_{\mbo}^+\in \{\pp^-_{1},\pp^-_{2}\}(*) $ &$\downarrow$&$ r_{\mbo}^+{\in} \cc^-_3(*)$, $r_{\mbo}^+\in \{\pp^-_{1},\pp^-_{2}\}$
     \\\hline
      ${a_{\delta}}\equiv 0.638285 M:r_{\gamma}^+=r_{\mso}^-$&$\uparrow$&$ r_{\mso}^-{\in}!\oo_{\times}^{+}(*)$ $ r_{\mso}^-\non{\in} \cc_2^{+} (*)$&$\downarrow$&$  r_{\mso}^-\non{\in}\pp_2^{+}$
      \\\hline
${{a}_1}\approx0.707107M:r_{\gamma}^-=r_{\epsilon}^+$&$\uparrow$&$  r_{\jj}^-\non{\in } \Sigma_{\epsilon}^+$&$\downarrow$&$  r_{\jj}^-\in \Sigma_{\epsilon}^+$
\\\hline
 ${a_{VI}}=0.73688M:\ell^-(r_{\mso}^+)=\ell_{\gamma}^-$&$\uparrow$&$ r_{\mso}^+=r_{cent}^-\in\{\cc_2^-, \oo_{\times}^{2^-}\}$& $\downarrow$&$ r_{\mso}^+=r_{cent}^-\in \cc_3^-$
 \\\hline
  $a_{\Gamma}\equiv0.777271M: \ell_{\Gamma}^-=\ell_{\gamma}^-$&
   $\uparrow$&$ r_{\gamma}^+\in \pp^-_1$,$r_{\gamma}^+\non{\in}\cc_3^-$&$
\downarrow$&$ r_{\gamma}^+\in \{\pp^-_{1},\pp^-_{2}\}$,
$ r_{\gamma}^+\in \{\pp_{2}^-,\pp_{3}^-\};(*)$
\\\hline
   ${a_b}\approx0.828427M:r_{\mbo}^-=r_{\epsilon}^+$&$\uparrow$&$  r_{\jj}^-\in \Sigma_{\epsilon}^+$, $ r_{\times}\non{\in}\Sigma_{\epsilon}^+$& $\downarrow$&$  r_{\jj}^-\in! \Sigma_{\epsilon}^+$, $ r^-_{\times}{\in}\Sigma_{\epsilon}^+$
 \\\hline
    ${a_{\mathcal{M}}^{(3)}}\equiv 0.934313M:\ell_{\gamma}^-=\ell_{\mathcal{M}}^-$&$\uparrow$&$ \exists\,  \pp_{\mathcal{M}}^{2^-}$& $\downarrow$&$ \exists\, \cc_{\mathcal{M}}^{3^-}$
    \\\hline
    ${a_2}\approx0.942809M: r_{\mso}^-=r_{\epsilon}^+$& $ \uparrow$&$ r^-_{\times}{\in}\Sigma_{\epsilon}^+$& $\downarrow$&$  r^-_{\times}{\in!}\Sigma_{\epsilon}^+$
    \\\hline
${a_{\varrho}}\equiv 0.969174M{\ell}_{\varrho}^-=r_{\gamma}^-$&$\uparrow$&$ \mathbf{A_{\varrho}^<}: r_{\mso}^-\non{\in}\cc_3^- $, $r_{\mso}^-{\in}\cc_2^-(*)$&
$\downarrow$&$ \mathbf{A_{\varrho}^>}:\, r_{\mso}^-{\in}\cc_{2}^-$, $r_{\mso}^+\non{\in}\cc_{3}^+$, $ r_{\mso}^-{\in}\cc_{3}^-(*)$
\\
\hline
\end{tabular}}
\end{table}
\end{turnpage}
%
%
\begin{figure}[h!]
\centering
\includegraphics[width=1\columnwidth]{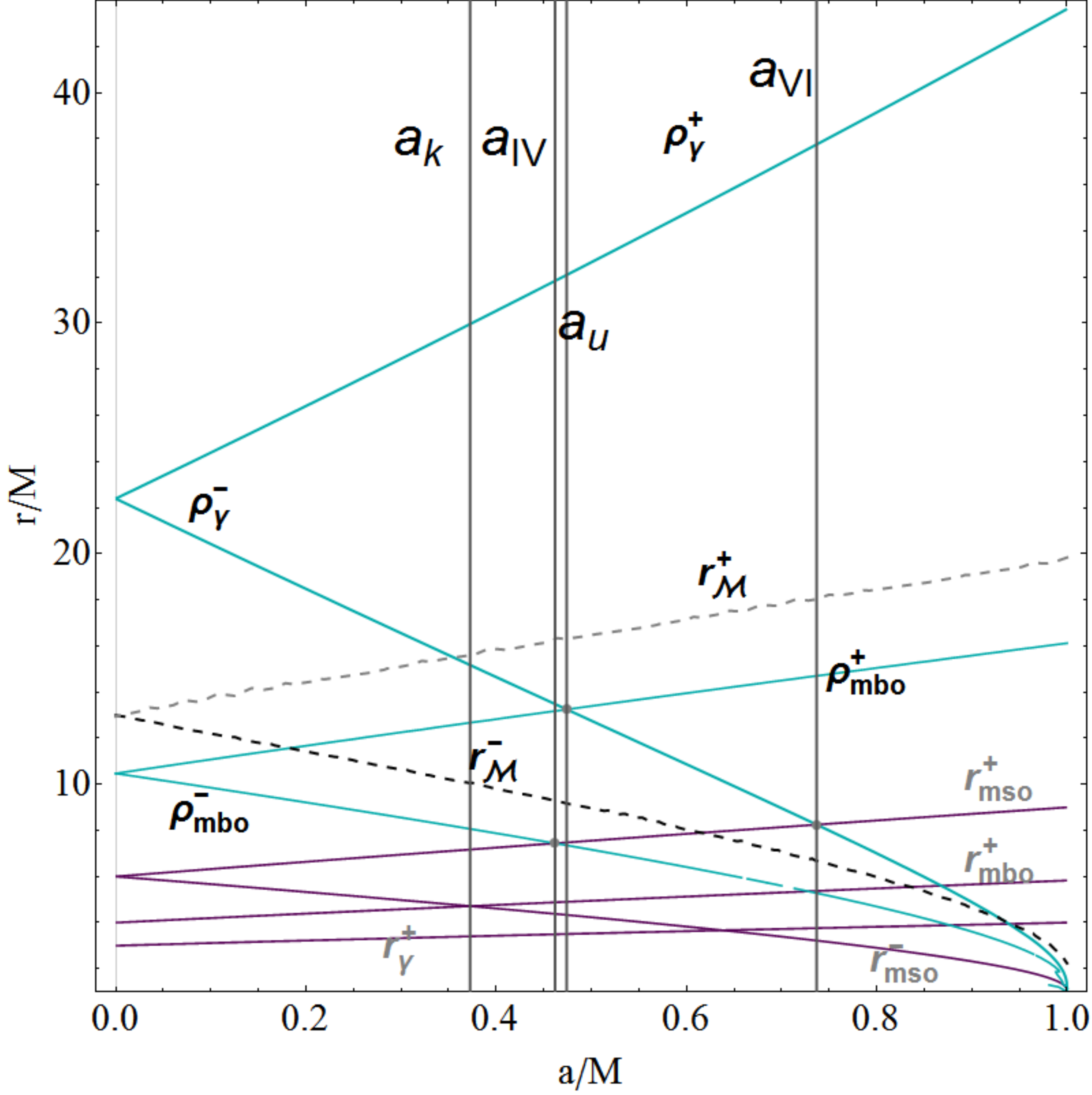}\\
\includegraphics[width=\columnwidth]{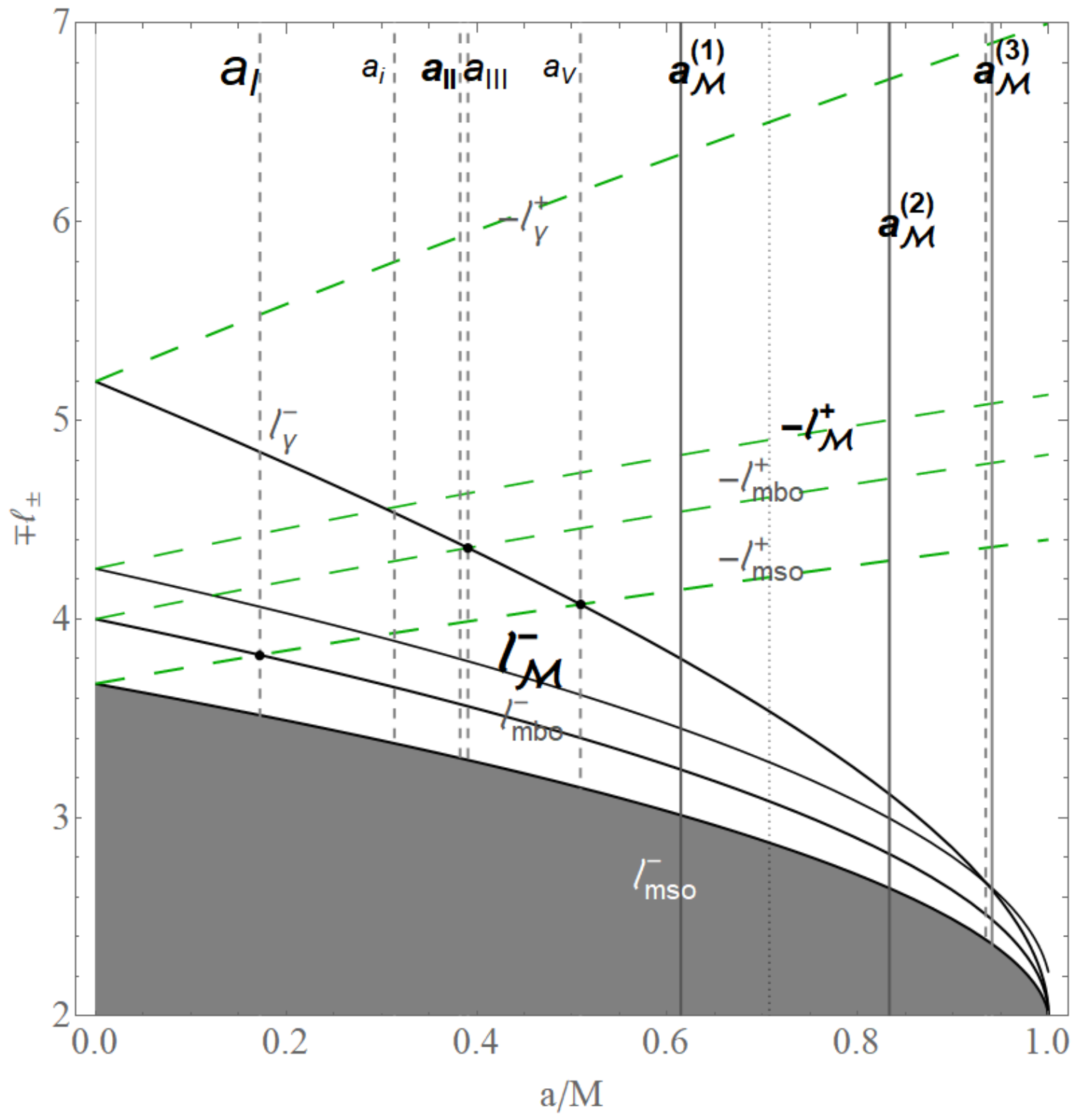}
\caption{Upper: Complementary geodesic structure of the Kerr geometry: notable radii  $\rho_{i}\equiv (\rho_{\mbo}^{\pm}, \rho_{{\mathrm{\gamma}}}^{\pm})$ and $r_{\mathcal{M}}^{\pm}$.
Bottom:fluid specific angular momentum  $ (\ell_{\mbo}^{\pm}, \ell_{{\mathrm{\gamma}}}^{\pm},\ell_{\mathcal{M}}^{\pm})$.
}\label{Figs:pranz}
\end{figure}
\begin{figure}[h!]
\includegraphics[width=1\columnwidth]{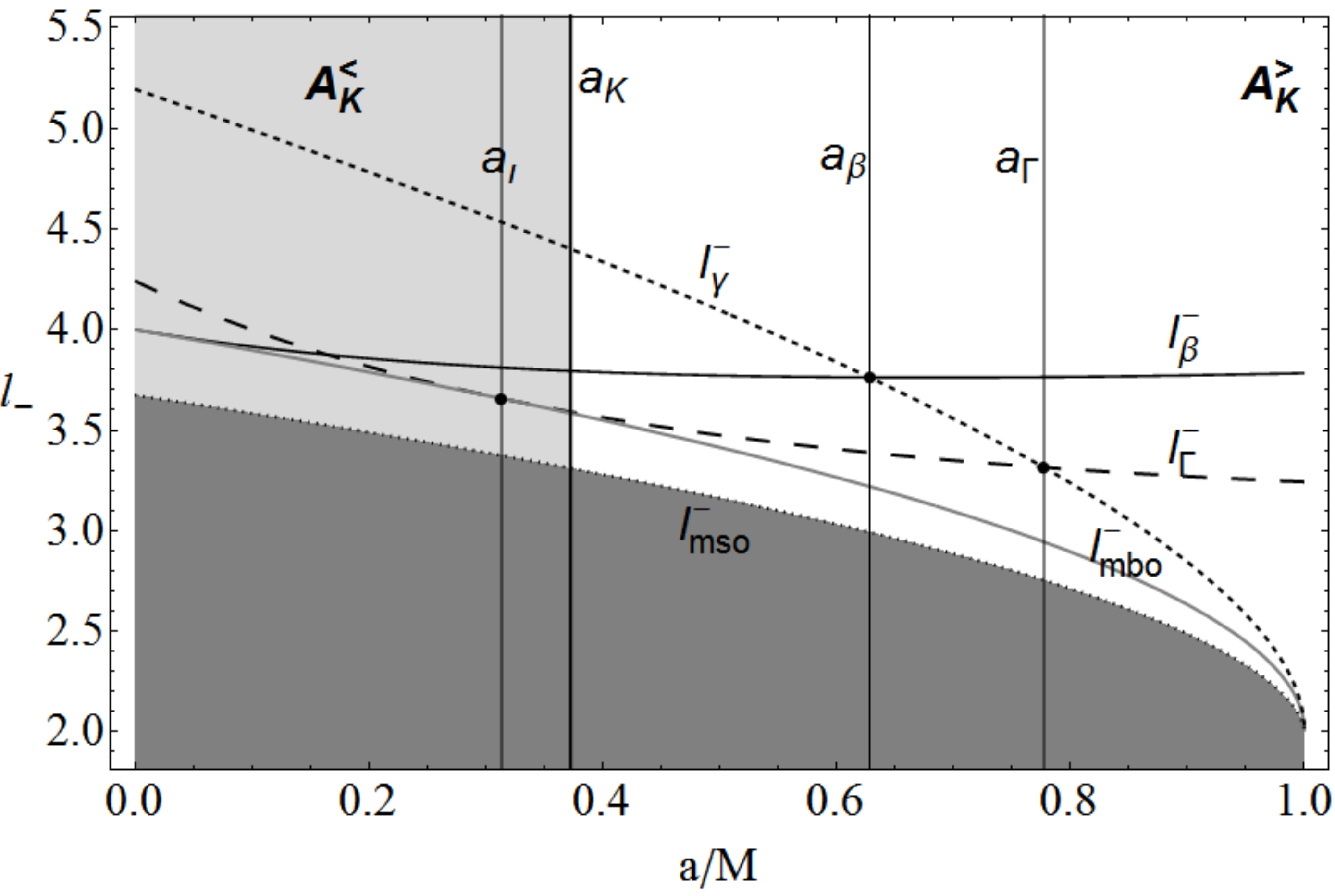}\\
\includegraphics[width=1\columnwidth]{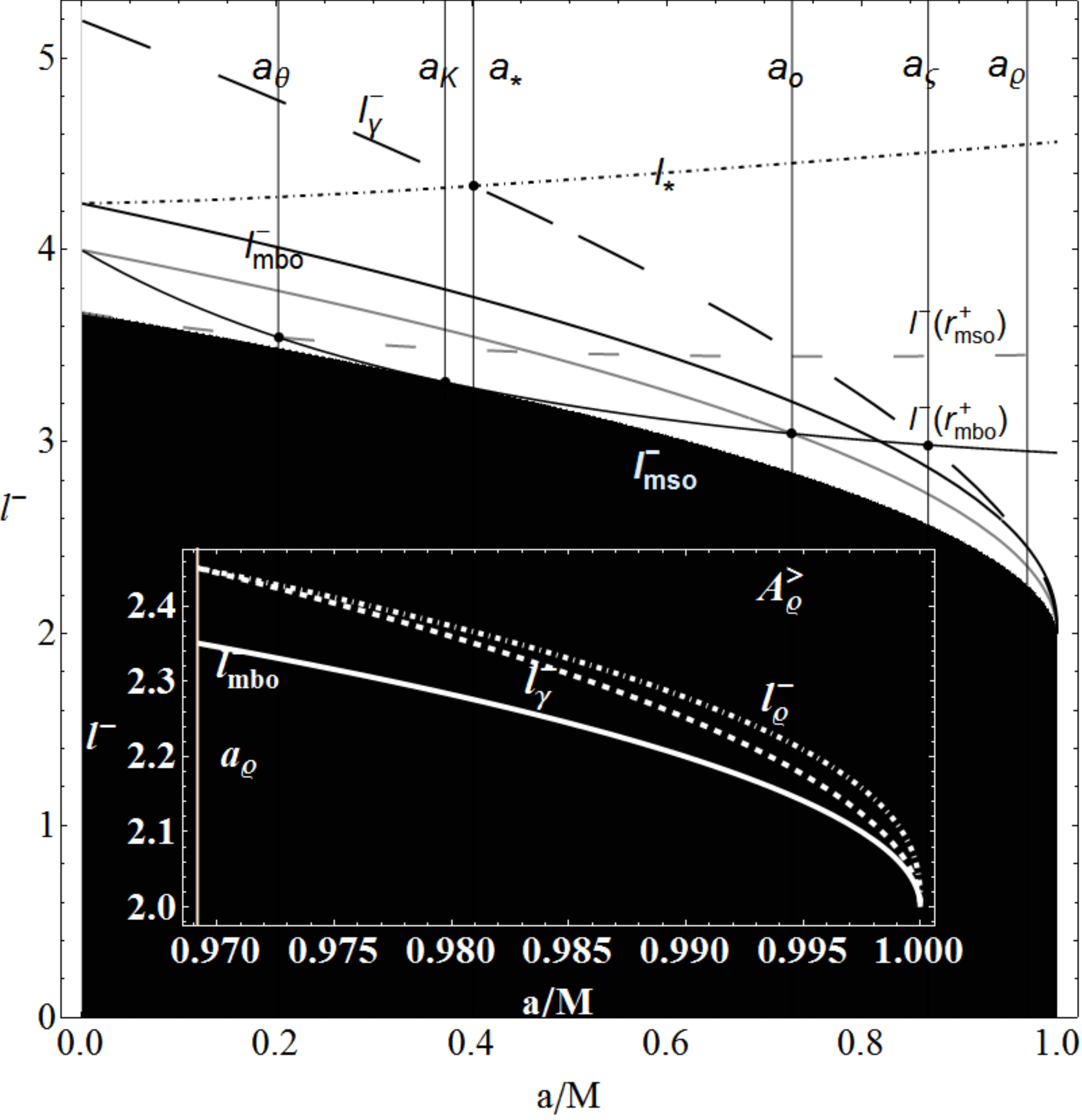}
\caption{Some notable spacetime spin-mass ratios are also plotted, with remarkable specific anglar momentum with reference to  Table\il\ref{Table:nature-Att}, Fig.\il\ref{Fig:Relevany} and Table\il\ref{Table:def-flour}}\label{Figs:GROUND-StA1}
\end{figure}
%
\section{{On the energetics of   $\ell$counterrotating  accreting tori}}\label{Sec:nw}
We conclude the analysis  with some notes on the  energetics of  the  processes  involving   $\ell$counterrotating  accreting tori.
We concentrate on the  specific example of the \textbf{RAD} accreting tori $\cc_\times^-<\cc_{\times}^+$ represented in  Fig.\il\ref{Fig:but-s-sou} \textbf{(a)}; this is a special  case of the \textbf{RADs},  where the outer torus is counterrotating and the inner torus is corotating  with respect to the central Kerr \textbf{BH}, and both  tori are accreting.
A further interesting aspect    singling out  this special \textbf{RAD} is  the  occurrence of a  jet emission associated with each  of the accreting    toroid. Jet launching  is associated to proto-jets in the \textbf{RADs}, studied in   \cite{open,app} and to accretion,  because of   the  correlation  of the inner accretion edge,  $r_\times$, and the jet emission launch,  $r_J$.
{In this regard we  mention
\cite{Fender(2001),
Fender:2009re,
Fender:2004aw,
1998NewAR..42..593F,
1999MNRAS.304..865F,
1998MNRAS.300..573F}
for a detailed perspective of the possibility that highlights oscillation between the inner  disk and jet
 in GRS 1915+105.}

In the specific case considered in this  section, where a double accretion occurs and  double jets can be expected, it is also possible that the   two jets associated to the \textbf{RAD} $\cc_{\times}^-<\cc^-< \cc_{\times}^+$ are separated by   a screening, corotating,  and quiescent torus,  located between the two lunching  points: $\cc_{\times}^-\leq r_{J}^-<\cc^-<r_{J}^+\leq \cc_{\times}^+$. The jets are correlated  with the  corotating inner shell, and the counterrotating outer shell.
In this case, we can evaluate the separation between the two jet emission points and its  variation  with  the \textbf{SMBH} spin $a/M$ as in    Figs\il\ref{Fig:corso} and \ref{Fig:Raccon},
identifying   $r_x\approx r_J$.  The maximal    separation of the  shell emission points is
$(r_{mso}^+ - r_{mbo}^-)$, which is a spin $a/M$ depended quantity.
We could also  consider the variation of the cusp luminosity   with the spin parameter $a/M$,  quantifying this distance in relation to the
accretion rates. We examine the tori and \textbf{BH}  accretion rates, tori cusp luminosity, and  enthalpy-flux,
mass-flux,
and
 thermal-energy    carried at cusp,
 for a specific case where the \textbf{RAD} fluid angular momentum distribution, the tori masses and the \textbf{BH} spin $a/M$ are fixed. Finally, we comment  the variation of these quantities in dependence  on the \textbf{BHs} spin  $a/M$.

In many  models of \textbf{SMBHs}  at high redshift,    $z>6$,  alternate accretion phases of the   \textbf{SMBHs} evolution are considered as a
 succession of accretion episodes  from accreting tori  causing, eventually, a    random seed-\textbf{BH} spinning-up or spinning-down. 
 A sequence of interrupted  super-Eddington accretion phases,  combined   with  sub-Eddington phases can appear.
 (Note  that a super-Eddington  accretion  can also imply a low efficiency
 mass-radiation conversion  \citep{abrafra}.)
 In the \textbf{RAD} frame the sub-Eddington phase may be associated to the presence of a screening, non accreting,  corotating torus $\cc^-$  in the system  $\cc_{\times}^-<\cc^-<\cc_{\times}^+$, the accretion from the outer torus would be absorbed partially by  the torus $\cc^-$. As a consequence of this,  the efficiency of the \textbf{RAD} and its luminosity are not  determined  uniquely by the inner accreting torus, in fact  our  analysis falsifies this assumption because of  the  possibility  of the screening effects.
 We limit here  to consider   a simple version of this problem, examining the accretion rates of  tori and \textbf{BHs} for
 $\cc_{\times}^-$  and $\cc_{\times}^+$,  in dependence on the parameter $a/M$ variation.
 \subsection{Accretion conditions}
We consider an inner corotating accreting torus and an outer counterrotating accreting torus  
where   ($\mathbf{A}^{\pm}$, $\mathbf{B}^{\pm}, \mathbf{S}^{\pm})$ accreting  models are defined in the following way:
$\mathbf{S}:\, \{r_s=r_{mbo},\, r_{\times}=r_{mso}\} $,
$\mathbf{A}:\, \{r_s=r_{mbo},\, r_{\times}=\widehat{r}_x\} $,
$\mathbf{B}:\, \{r_s=\widehat{r}_s,\, r_{\times}=\widehat{r}_x\} $.
In the example  where the inner edge of  the accreting tori  $r_{\times}^\pm\approx r_{mbo}^{\pm}$, there is   $W_x\equiv W(r_{\times})=\ln K_\times\approx0$ (coincident with the limiting asymptotic value for very large  $r/M$ of the P-W potential $W$). We define the accretion point
$\widehat{r}_x(a)\equiv r_{mbo} + (r_{mso} - r_{mbo})/\alpha_x$
where  the constant
{$\alpha^-_x\approx3.900$, for corotating tori, and $\alpha^+_x\approx6.76029$ for counterrotating tori}, while $\widehat{r}_s(a)<\widehat{r}_x(a)=r_{mbo} + (\widehat{r}_x -r_{mbo})/2$.
We  illustrate spin dependence of these radii
 and the corresponding fluid specific angular momentum and the density parameter $K$ in Figures\il\ref{Fig:Raccon}.
For the double-accretion couple, the situation   where $r_J\approx r_{\times}\approx r_{mbo}$ is a limit case  occurring when  the centrifugal component of the disk force balance  tends to dominate the gravity and pressure force components in the torus.
We  give a preliminary evaluation of the center of maximum hydrostatic pressure in the tori as  $r_{mso}^{\pm}<r_{cent}^\pm<\rho_{mbo}^{\pm}$, in Fig.\il\ref{Fig:corso}.
\begin{figure}
  \includegraphics[width=\columnwidth]{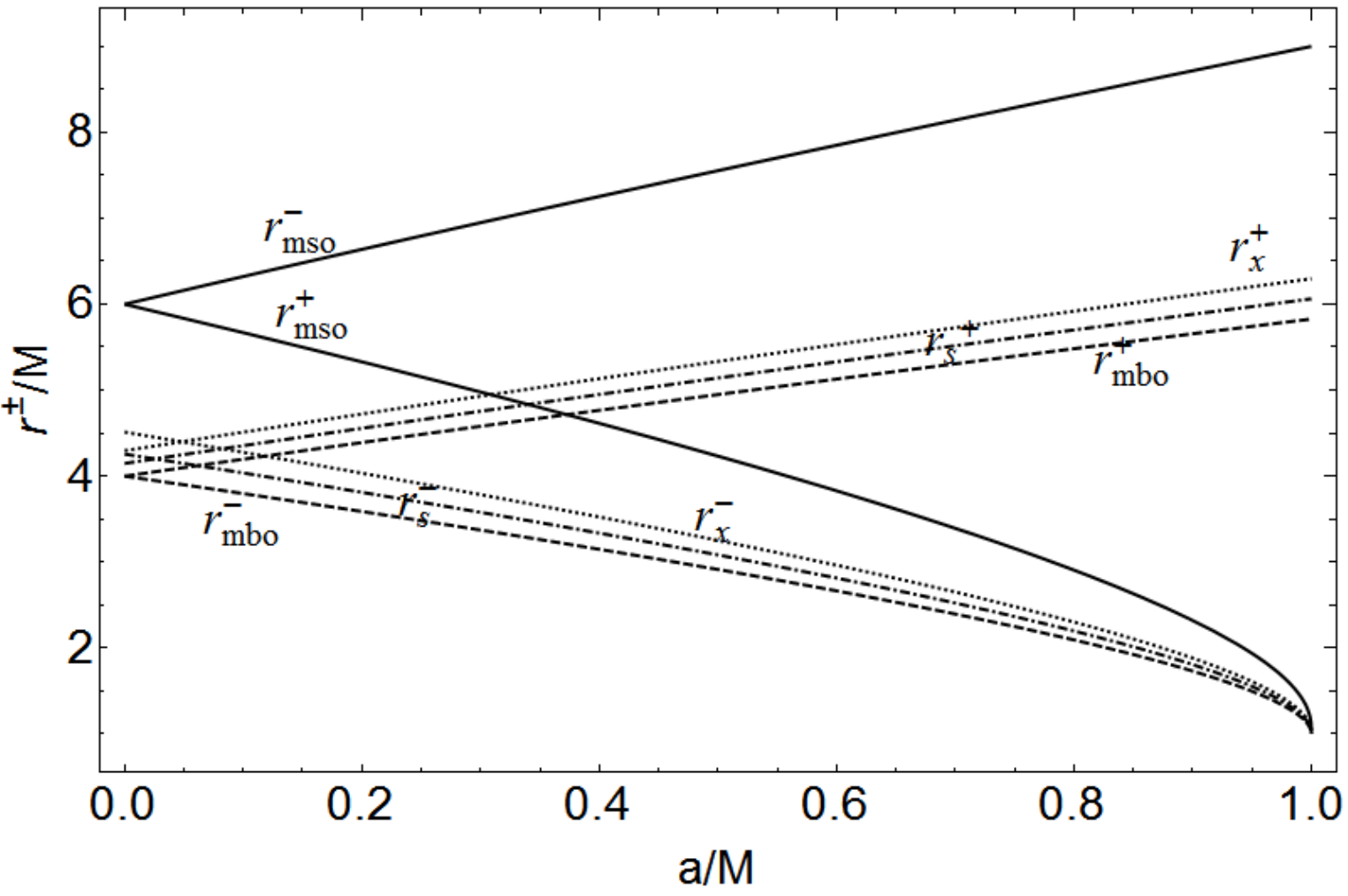}
 \includegraphics[width=\columnwidth]{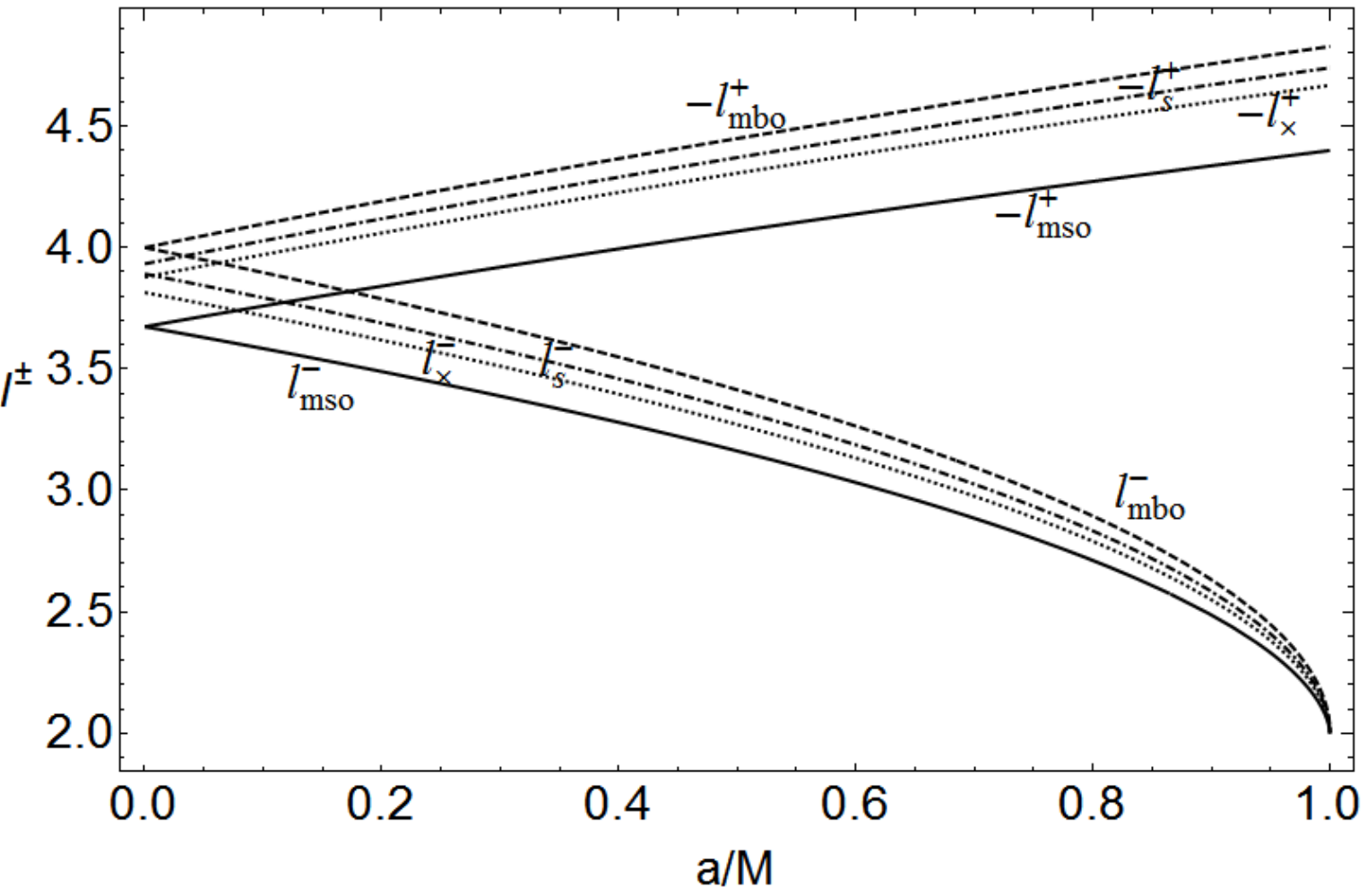}
  \includegraphics[width=\columnwidth]{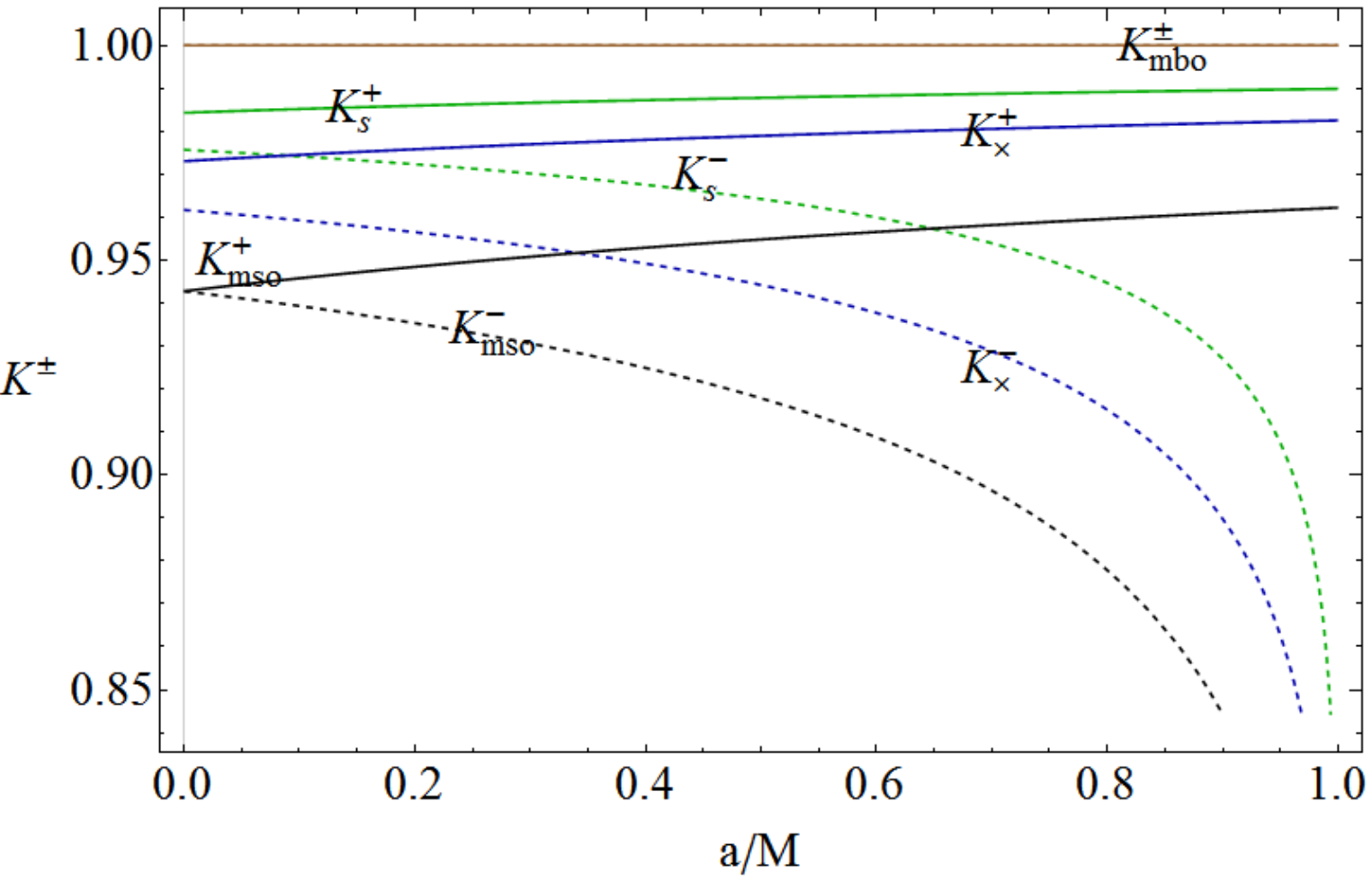}
  \caption{Radii $\{r_{mbo}^{\pm},r_{\times}^{\pm},r_{s}^{\pm},r_{mbo}^{\pm}\}$ (\emph{upper panel}), fluid specific angular momenta $\{\ell_{mbo}^{\pm},\ell_{\times}^{\pm},\ell_{s}^{\pm},\ell_{mbo}^{\pm}\}$ (\emph{center panel}),  $K$-parameters $\{K_{mbo}^{\pm},K_{\times}^{\pm},K_{s}^{\pm},K_{mbo}^{\pm}\}$ (\emph{bottom panel})  for  models ($\mathbf{A}^{\pm}$, $\mathbf{B}^{\pm}, \mathbf{S}^{\pm})$). }\label{Fig:Raccon}
\end{figure}
\begin{figure}
  \includegraphics[width=\columnwidth]{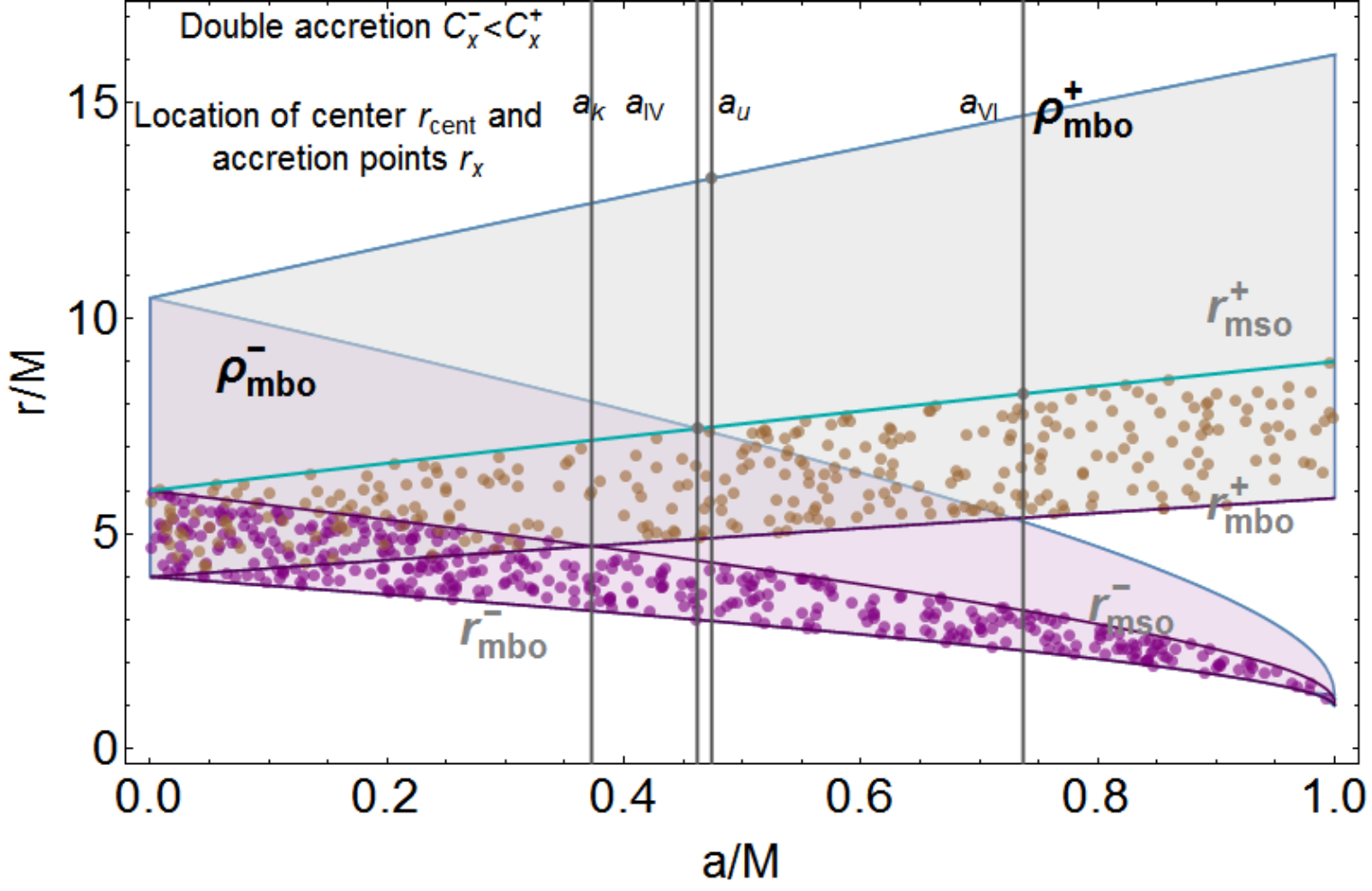}
  \caption{Double accretion $\cc_{\times}^-<\cc_{\times}^+$, location of tori centers $r_{cent}^{\pm}$ (dotted regions) and accretion points  $r_{\times}^{\pm}$ (shaded regions) for corotating $(-)$ and counterrotating $(+)$ fluids. Location is according to  Eqs\il(\ref{Eq:conveng-defini}). Constrains are used in the analysis of the models ($\mathbf{A}^{\pm}$, $\mathbf{B}^{\pm}, \mathbf{S}^{\pm})$ in Figs\il\ref{Fig:ostagh}. Crossing of the regions provides limits of  the \textbf{BH} classification.  }\label{Fig:corso}
\end{figure}
\begin{figure}
 \includegraphics[width=\columnwidth]{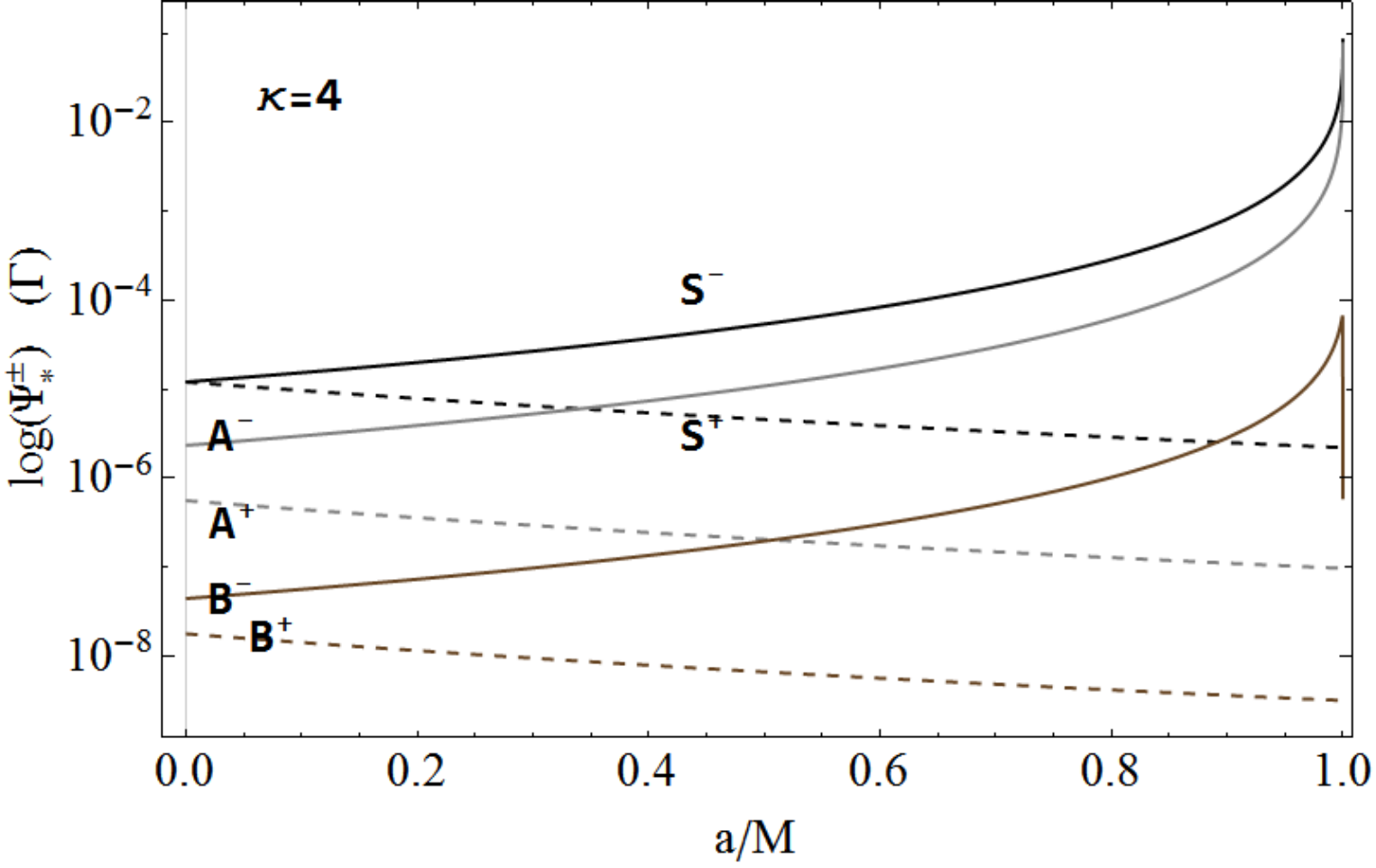}
 \includegraphics[width=\columnwidth]{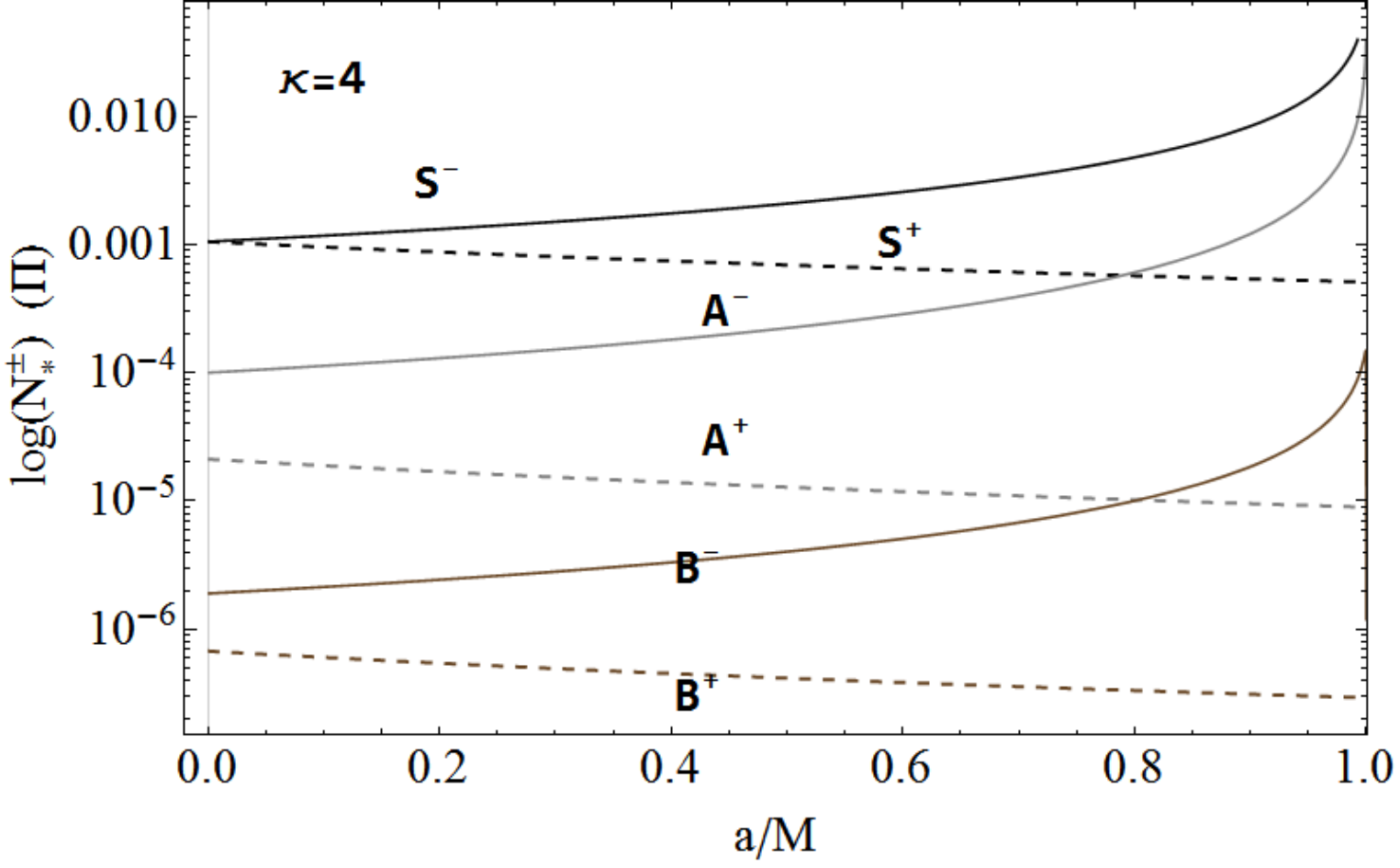}
  \caption{Evaluation of $\Psi$  and $\mathrm{N}$ in models  ($\mathbf{A}^{\pm}$, $\mathbf{B}^{\pm}, \mathbf{S}^{\pm})$ and dependence of $\Psi$  and $\Gamma$ on the \textbf{SMBH} spins $a/M$.
Models  ($\mathbf{A}^{\pm}$, $\mathbf{B}^{\pm}, \mathbf{S}^{\pm})$ are constructed on the data  on $r_s,r_{\times}$  provided in Figs\il\ref{Fig:Raccon}. This analysis refers to Sec.\il(\ref{Sec:nw}), {where  $\kappa\equiv n+1$}.}\label{Fig:ostagh}
\end{figure}
We  start by considering   two \textbf{RAD} tori constituted   of polytropic fluids
with pressure
 $p=K \varrho^{1+1/n}$. Within the formalism introduced in
\citep{abrafra},
we can   estimate   the
mass-flux, the  enthalpy-flux (evaluating also the temperature parameter),
and  the flux thickness.
All these quantities   can be written in general form $\Gamma (r_x,r_s,n)=q(n,K)(W_s-W_{x})^{d(n)}$, where $\{q(n,K),d(n)\}$, different for each torus, are  functions of the polytropic index,  $r_x$ is the cusp location and the inner edge of accreting disk while   $r_s<r_x$ is related to thickness
 of the accreting matter flow and the P-W potential
  $W=\ln V_{eff}$. thus
$W_s(W_{x})$ denotes, for a torus with fixed specific angular momentum $\ell$,  the (constant) value of the  P-W potential of  the $p=$constant surface corresponding to radius $r_s$ ($r_x$).

Specifically the $\Gamma (r_x,r_s,n)$-quantities read:

$(\diamond)$ $\mathrm{{{Enthalpy-flux}}}=\mathcal{D}(n,K) (W_s-W)^{n+3/2}$

$(\diamond)$ $\mathrm{{{Mass-Flux}}}= \mathcal{C}(n,K) (W_s-W)^{n+1/2}$

$(\diamond)$ $\mathrm{E-L}=\mathcal{L}_{x}/\mathcal{L}= \mathcal{B}/\mathcal{A} (W_s-W_{x})/(\eta c^2)$
  which is  the  fraction of energy produced inside the flow and not radiated through the surface but swallowed by central \textbf{SMBH}.

$(\diamond)$   Efficiency
$\eta\equiv \mathcal{L}/\dot{M}c^2$,    $\mathcal{L}$ is  the total luminosity, $\dot{M}$  is the total accretion rate,  and  for a \emph{stationary flow}, $\dot{M}=\dot{M}_x$.

 We  examine also $\Pi$-quantities having  general form  $\Pi={\Gamma}(r_x,r_s,n) r_x/\Omega_K(r_x)$; $\Omega_K(r_x)$ is the Keplerian frequency  of the accreting  tori cusp  $r_x$, where the pressure vanishes.
 Making explicit  the polytropic index, there is

$(\bullet)$ the cusp luminosity
\be\mathcal{L}_{x}={\mathcal{B}(n,K) r_{x} (W_s-W_{x})^{n+2}}/{\Omega_K(r_{x})}\ee,
measuring the
rate of the thermal-energy    carried at the  cusp;

 $(\bullet)$ the \emph{disk accretion rate }    $\dot{m}= \dot{M}/\dot{M}_{Edd}$,

 $(\bullet)$ the  mass flow rate through the cusp (i.e., mass loss accretion rate)
 \be\dot{M}_{x}={\mathcal{A}(n,K) r_{x} (W_s-W_{x})^{n+1}}/{\Omega_K(r_{x})}\ee.
In Figures \ref{Fig:ostagh} $\Psi _*^{\pm }\equiv \Gamma (r_x,r_s,n)/q(n,{K})$  for $\Gamma $-quantities  and
 $\mathrm{N}_*^{\pm}={\Gamma}(r_x,r_s,n) r_x/q(n,K)\Omega_K(r_x)$  for $\Pi$-quantities, for $\kappa\equiv n+1=4 (n=3)$. (For more general discussion see also \cite{1951ApJ...114..165V,1974MNRAS.168..603L})
\begin{figure}
  \includegraphics[width=\columnwidth]{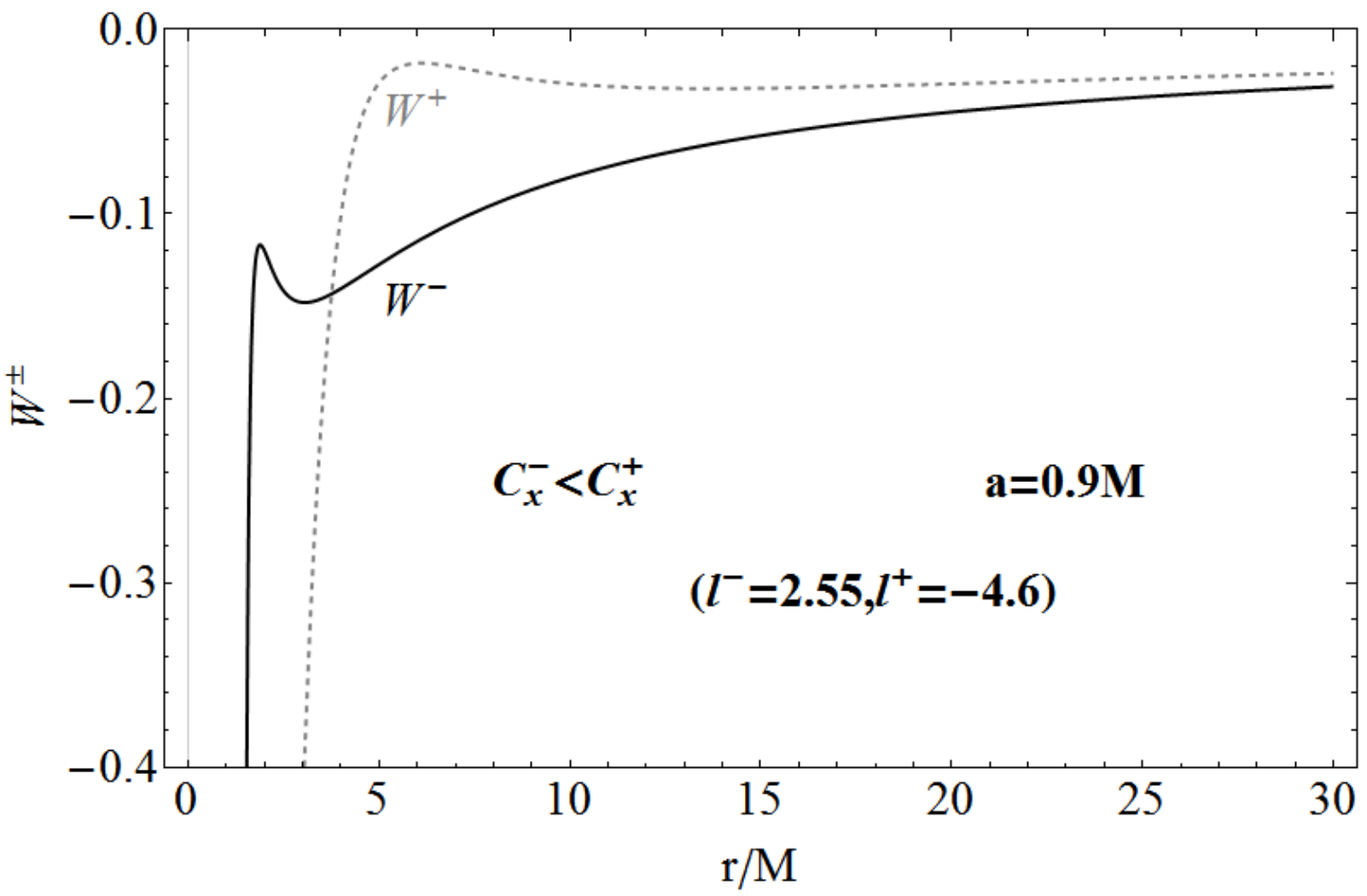}
  \caption{$W^{\pm}$-functions for models  ($\mathbf{A}^{\pm}$, $\mathbf{B}^{\pm})$ for $\mathbf{BH}$ spin $a=0.9M$. $W^+$  ($W^-$) is the P-W potential for the counterrotating outer (corotating inner) torus of the \textbf{RAD} couple with specific angular momentum $\ell^+=-4.6$ ($\ell^-=2.55$), with  cusp $\hat{r}_{x}^{+}$ ($\hat{r}_{x}^{-}$) used in models $\mathbf{A}^{+}$ and $\mathbf{B}^{+}$ ($\mathbf{A}^{-}$, $\mathbf{B}^{-}$). }\label{Fig:Premark}
\end{figure}
\subsection{Time-dependent accretion }
The stationarity of each \textbf{RAD} component, and therefore of  the \textbf{RAD} agglomerate, has several  relevant advantages but also some limitations. One significant   advantage  is that such set-up is  able to provide very simple \textbf{RAD} configurations  as initial data in \textbf{GRHD} and \textbf{GRMHD} dynamical analysis. However, stationarity also implies that for the dynamical analysis associated to the instability processes and, for example the emergence of the deterministic chaos occurring  possibly  after tori interactions, collision or double accretion,
 we need to extend our considerations beyond  the stationary  model,  introducing the time-dependence and, eventually, explicit  tori interaction terms.
 An interesting study  would consist in  the analysis of the  \textbf{BHs}  runaway instability to understand how the double accretion can affect this phenomenon.
Here, we examine    more  simple general   examples  of situations where deterministic chaos might emerge, considering   modified  accretion rate laws \citep{May}. For this purpose we focus on the accretion rates of Figs\il\ref{Fig:ostagh} and explore the possibility that tori collision or screening effects,  as described above, would provide a mechanism for interrupted phases of super-Eddington accretion as often considered in models of \textbf{SMBHs} evolution, altering $\dot{m}$ and $\dot{M}$ to include a time dependence. We explore the conditions under  which the   chaotic systems can emerge, starting from  the initial configurations considered in Fig.\il\ref{Fig:Premark}, as in the  models $\{\mathbf{A}, \mathbf{B}, \mathbf{S}\}$.
In these  models, we considered   the relations between the accretion rates and the black hole spin $a/M$;
we could assume this relation  persists also in the time-dependent generalization.
Accretion rates in the  stationary systems investigated here  are constant in time  $t$, thus  in general there is
$\dot{m} = \dot{m}(0)\equiv  \dot{m}_0$  and $\dot{M} = \dot{M}(0)\equiv  \dot{M}_0$ for each torus
at  any  $t\geq0$.
Depending on different constraints on evolutionary processes (analyzed at $t =$constant) in
\cite{dsystem}, several scenarios had been envisaged for  the double-accretion systems where,
\textbf{\texttt{(1)}} the flow from the external torus decreases  to stabilize, while the inner torus rate increases during the accretion process;
\textbf{\texttt{(2)}} by changing initial conditions, the flow from the outer torus increases,  the inner torus  increases accretion rate and tori collision occurs;
\textbf{\texttt{(3)}} in the third scenario, the  accretion rate   for  the outer counterrotating torus increases, the inner torus  decreases until accretion  stops, and the torus   stabilizes in a quiescent state.
Finally, these \textbf{\texttt{(1)}} and \textbf{\texttt{(3)}}  paths can be combined to give rise to some drying-feeding effects.
These situations are widely discussed in  \cite{dsystem}.
The identification of a specific scenario asks to fix fluid conditions (GRHD or GRMHD dynamical models), depending on the initial model  parameters   $(\ell, K, a/M, r/M)$, where   $ r/M $  stands for the tori separation. In  the stationary systems these  are time-independent  quantities.

  In general,  for small $t/M$, there is
$\mathbb{\mathbf{(g)}}: \mathrm{{G}}[t;0]=\frac{1}{2} t^2 G''(0)+t G'(0)+G(0)$,  in a proper range $\Delta t=t$, while
$G(0)$ is   evaluated  and represented in  Figs\il\ref{Fig:ostagh} where  we explore the case $a=0.9M$.

The first case  we consider  in $\mathbb{\mathbf{(g)}}$ is the exponential growth  with  time of the tori and \textbf{BH}  accretion  rates\footnote{Hypothesis of exponential growing  or exponential decay  is  considered  in astrophysics  in general situations related e.g. to
presence of viscous effects affecting the establishing of the accretion mechanism,  or in some binary system processes.}.
 Fixing then accordingly $G''(0),G'(0))$ and $G(0)$, we examine the    law
$\mathrm{\mathbf{(l_1)}}: \mathrm{G}^{\pm}[t+1] = b^{\pm} \mathrm{G}^\pm[t] + h^\pm[t]$, thus
 $\mathrm{G}^\pm[t]=(e^{(b^{\pm}-1) t} [(b^{\pm}-1) \mathrm{G}^\pm[0]+h^{\pm}]-h^{\pm})({b^{\pm}-1})^{-1}$.  The parameters $b^{\pm}$  give the  rate of growing  (decay) of the accretion rates. Parameter $h^{\pm}$ can be  fixed considering  the initial data $\mathrm{G}^\pm[0]$, thus $\mathrm{G}^\pm[t]= \mathrm{G}^\pm[0] e^{(b^{\pm}-1) t}$ for $h=0$.
The relation, $\mathrm{G}^\pm_{t+1}=f(b^\pm) \mathrm{G}^\pm_t$, where $f(b^\pm) $  is a function of $b^\pm$ is however linear--as consequence of this  there is no  chaotic behavior.

Chaotic behavior  appears in non-linear relations, as  in the
 second case  we consider, namely the  polynomial form $\mathrm{\mathbf{(q_1)}}:\mathrm{G}^{\pm}[t+1] = -b^{\pm} \mathrm{G}^\pm[t]^2+\mathrm{G}^\pm[t]+ h^\pm$,  leading to different tori evolutionary paths, including accretion rates decreasing in time.
 We constrain $b$ parameter (related to the rate of decreasing (or increasing) of the accretion rates) to the initial data. By setting  $h=0$, we obtain a one-parameter model, with parameter $b$ related to the  rapidity  of the flow evolution and  $\mathrm{G}^{\pm}[0]$.
By considering the differential $\partial_t \mathrm{G}^{\pm}\approx \mathrm{G}^{\pm}[t+1]-\mathrm{G}^{\pm}[t]$ with $\mathrm{\mathbf{(q_1)}}$, we obtain the general solution
{\small
\bea\mathrm{G}=\frac{b-1-\beta \tan \left[\frac{\left(c_1+t\right)\beta}{2}\right]}{2 b},\;
\beta\equiv\sqrt{b(2-4 h-b)-1},
\eea}
where $c_1$ is fixed by  $\mathrm{G}_0$ for each torus.
We simplify the analysis with condition
$h=0$, the general solution  has then a simpler exponential behavior 
\bea\mathrm{G}=(1-b)\left[\frac{e^{(1-b) t} [b (\mathrm{G}_0-1)+1]}{\mathrm{G}_0}-b\right]^{-1},
\eea
where  the rapidity depends on the  parameter $(1-b)$; $b=1$ is then a limiting value we investigate below.

 For small $t$, there is  \bea\nonumber\mathrm{G}\approx \mathrm{G}_0+t [1 + b (\mathrm{G}_0-1)] \mathrm{G}_0 [1 + b (2 \mathrm{G}_0-1)]+
 \\
 \frac{1}{2} t^2 \mathrm{G}_0 [b (\mathrm{G}_0-1)+1] [b (2 \mathrm{G}_0-1)+1]+O\left(t^3\right).\eea
It is easy to see that the rate is constant in time for  $\mathrm{G}=\mathrm{G}_0=\frac{b-1}{b}$
or $b=(1-\mathrm{G}_0)^{-1}$.

On the other hand, this situation occurs only when $b<0$ or $b>1$, which is satisfied for  $a=0.9M$, while it is likely not satisfied for larger  values of $a/M$. Eventually, for very large spin, a phase where $b<0$ could occur. Evolution  in time is represented in  Fig.\il\ref{Fig:widepro} with data  from  Fig.\il\ref{Fig:ostagh}.
As expected, the  study of the fixed points for the first cycle leads to
$\mathrm{G}=(b-1)/b$
 (for $b<0$ and  $b>1$), with first derivative of $(\mathbf{\mathrm{q_1}})$  in this point being   $(2-b)$, distinguishing   the range of $b$  in $b>3$ or  $b<3$ ($|2-b|>1$ or $|2-b|<1$ for stable equilibrium point).
 As clear from Figs\il\ref{Fig:widepro}, $\mathrm{G}$    decreases or increases in time  depending on $b$: in  one possibility we  consider, the inner corotating torus  decreases  to $\mathrm{G} =0$  being thus a quiescent torus, while for the  outer counterrotating torus the rate increases with time.
 Then we proceed with  the analysis of the successive steps (cycles) for the  neighborhood  of the unstable points, namely the first cycle $G[t+1]$ as function of    $G[t]$ (we assumed $t_0=0$), the second cycle for $G[t+2]$ as function of    $G[t]$ up to $\upsilon$-cycle, for $G[t+\upsilon]$ as function of    $G[t]$.
 The stability analysis for the  second, third and fourth cycle is in  Figs.\il\ref{Fig:widepro} where  the emergence of chaos for the outer counterrotating torus is demonstrated.
\begin{figure}
  \includegraphics[width=\columnwidth]{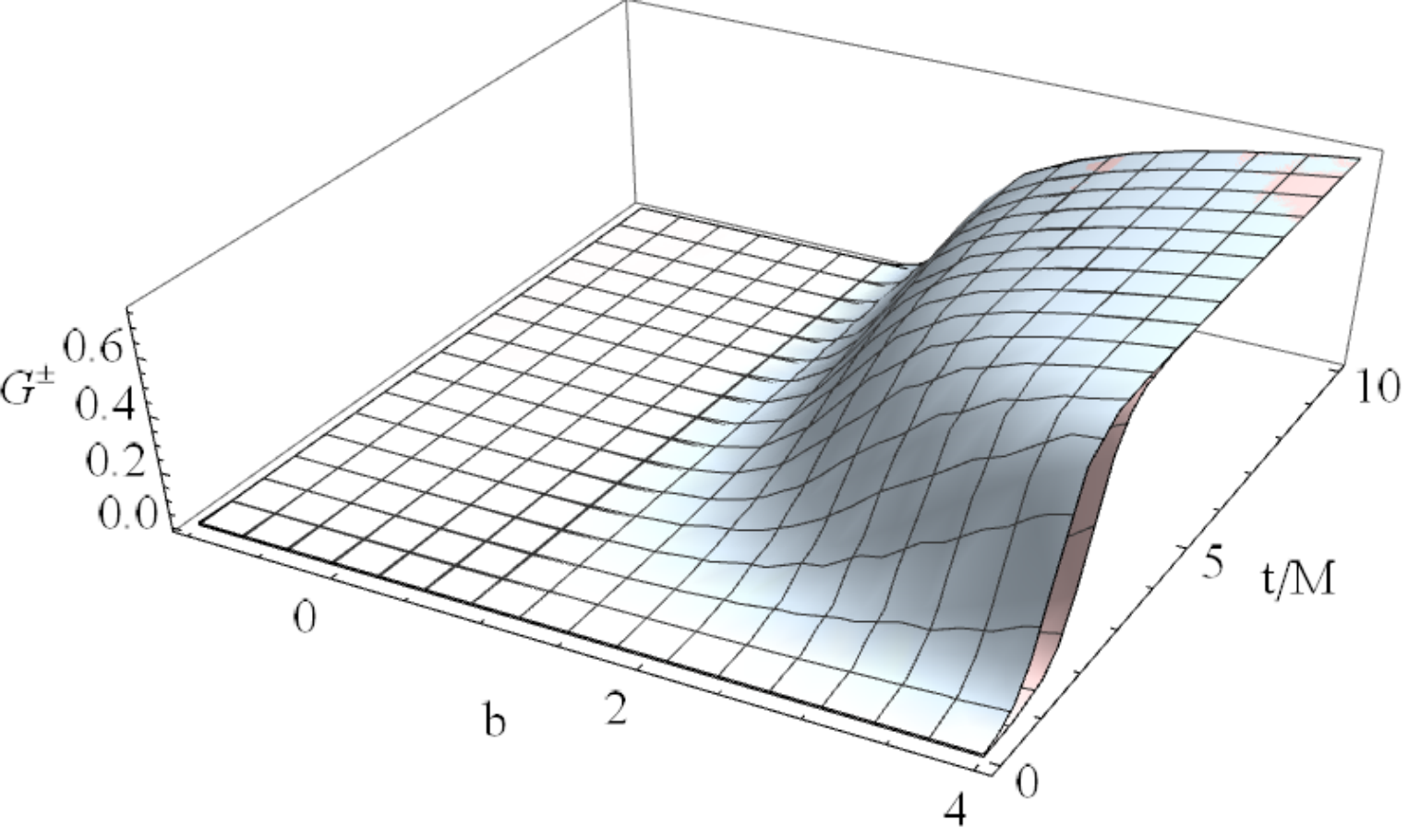}
  \includegraphics[width=\columnwidth]{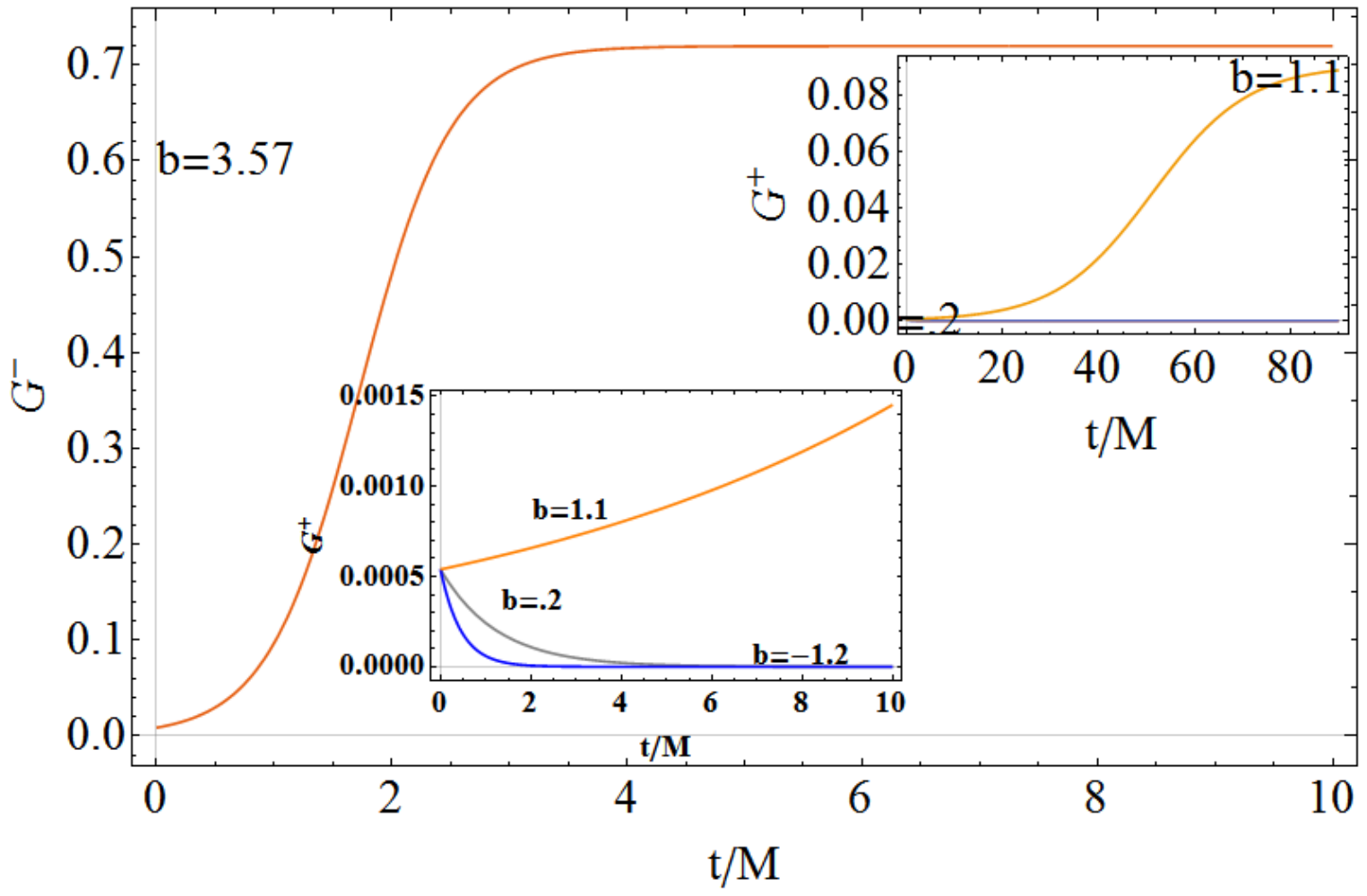}
  \caption{Upper Panel: $\mathrm{G}^{\pm}$ as function of $t/M$ and $b$ parameter for the  dynamical solution   of Sec.\il(\ref{Sec:nw}).
Initial data are for  the  accretion rates  $\dot{M}^+[0]=0.00054$ for the outer counterrotating  torus
(red surface), and  $\dot{M}^-[0]=0.00845$  for the inner  corotating   torus
(blue surface), related to function  $\Pi$   in Figs\il\ref{Fig:ostagh} for a=$0.9M$.
For  $b>2$ the rate generally increases  with time  towards a stable value
(stationary hypothesis),
for  $b<1$ the rate may decrease with time according to the initial data.
Bottom panel: $\mathrm{G}^{\pm}$ as function of $t/M$ for fixed  values of  $b$.
Inside panels show the situation for $\mathrm{G}^{+}$ for different values of $b$. The  increasing time profile of $\mathrm{G}^{-}$ is shown for
$b=3.57$.  The function $G^-$ is fast increasing, giving    emergence of  chaotic analysis  studied  in Figs\il\ref{Fig:widepro1}.
 }\label{Fig:widepro}
\end{figure}
\begin{figure}
  \includegraphics[width=\columnwidth]{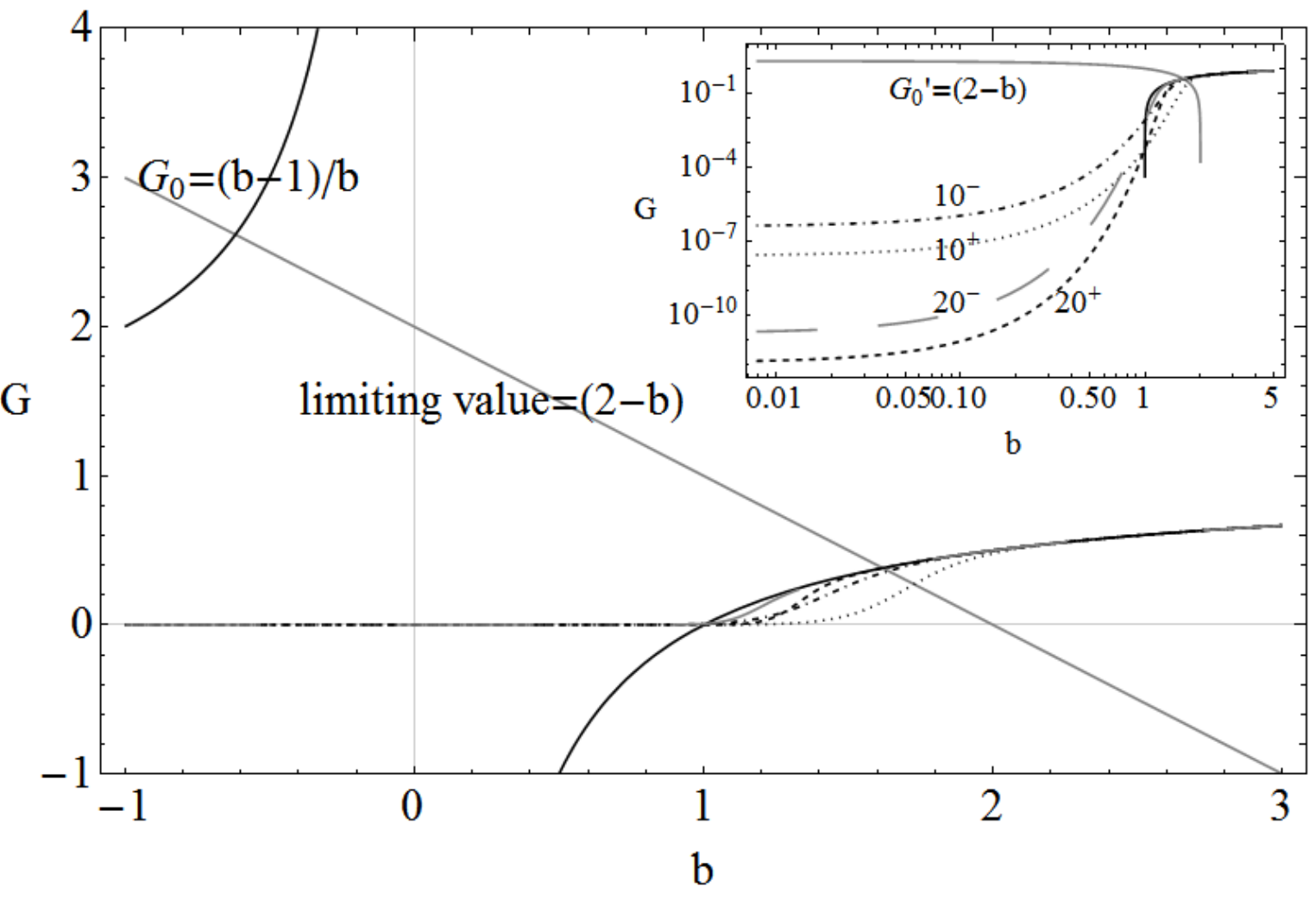}
  \includegraphics[width=\columnwidth]{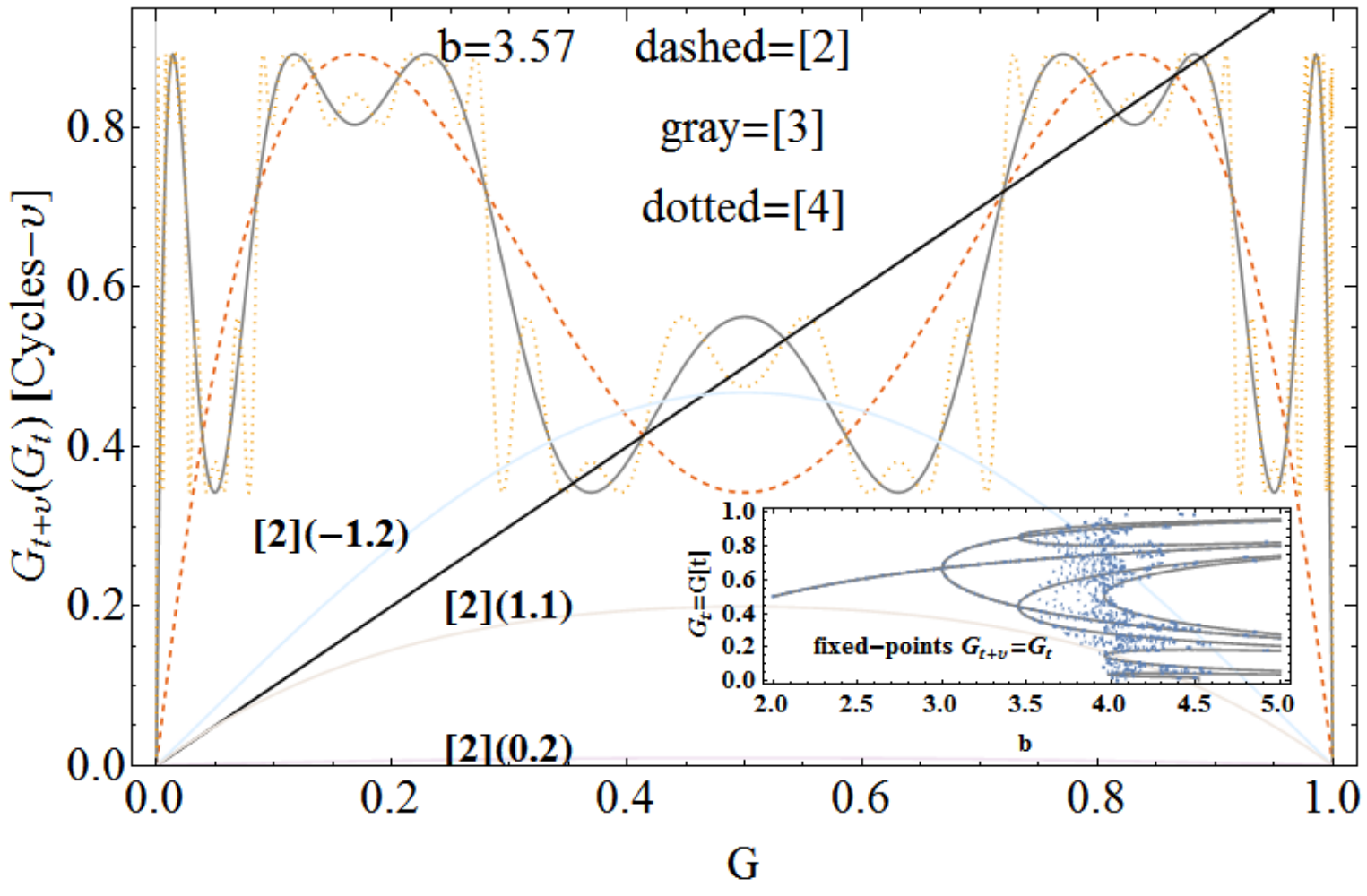}
  \caption{Upper Panel: $\mathrm{G}^{\pm}$ as functions of the parameter $b$. Plots shows fixed point $\mathrm{G}_0=(b-1)/b $ and  $(2-b)$-limiting function for the stability analysis.
Initial data  for  the accretion rates are   $\dot{M}^+[0]=0.00054$ for the outer counterrotating  torus and
$\dot{M}^-[0]=0.00845$  for the inner  corotating   torus related to functions  $\Pi$ in  Figs\il\ref{Fig:ostagh}, taken for a=$0.9M$.
Inside panel is a zoom for restricted range of $b$, numbers close to curves are  different times $t/M$,
signs $\mp$  correspond  to the  corotating/counterrotating  fluids.
Bottom panel: Stability analysis, emergent chaos,
 cycles  $\mathrm{G}_{t+\upsilon}(\mathrm{G}_t)$.
Different cycles $\upsilon$ are given in  square brackets,  numbers close to the curves are for $b$-parameter for the  corotating torus.
Inner panel:  analysis of the fixed points  $\mathrm{G}(t+\upsilon)=\mathrm{G}(t)$ versus parameter $b$ for different cycles $(\upsilon)$; some bifurcations appear, for example  at $b\approx 3$, $b\approx 3.5$--see also Figs\il\ref{Fig:widepro}.}\label{Fig:widepro1}
\end{figure}

\section{On the \textbf{RAD} instabilities and  generalization of the \textbf{RAD} model}\label{Sec:gothc}
The \textbf{RAD}  macrostructure    is relevant for  different  aspects of the    \textbf{BH} attractors  and their host  galactic environments, for example  as explanation of  the accretion
 rates of \textbf{SMBHs} at high redshift, or as model of screening tori, or also into the description of the \textbf{QPO} emission. The \textbf{RAD} hypothesis could play an important  role also   with  regard to   thick (outer) torus of gas and
dust encircling the   \textbf{AGNs} accretion {disk}  (obscuring tori).
   On the other hand, current analysis of astrophysical \textbf{BHs} interacting with their environments suggests   that the formation of multiple tori (from different accretion phases or fragmentation from an original disk and also galaxy interaction)  is likely to occur  \cite{2018ApJ...852..131B}.
Concerning \textbf{RAD} in their hosts,   \textbf{RAD}   proto-jet launch could be connected with
 \textbf{AGNs} very fast (almost speed of
light) jets. (\textbf{RAD} accretion  in a galaxy could emit
radiative power  outshining  the
host galaxy). Further analysis also
suggests  the possibility to detect
\emph{{structured proto-jets}} as   sequences of jet-like configurations (or \emph{jet-bundles}), constrained in spacings and relative fluids rotation.
Below we  outline  some of most intriguing aspects of the \textbf{RAD} hypothesis which can have  interesting implications

$\mathbf{(\bullet) }$ Our investigation  suggests   that,  as expected, the \textbf{BH} spin   distinguishes corotating from  counterrotating  tori by   favoring generally the formation of corotating configurations, proto-jets and accreting tori, with respect to the counterrotating ones, according to the constraints provided on tori dimension ($K$-parameter), the specific angular momentum ($\ell$-parameter), and the \textbf{BH} dimensionless spin. This frame  enters     the   wide  field of investigations on the relevance of   the  counterrotating accreting tori, binding the situations where such tori can collide  or double accretion can occur.

$\mathbf{(\bullet) }$ A further phenomenological application  of the \textbf{RAD}  is the connection  between \textbf{RADs} seismology and the \textbf{QPOs}, the low
and high frequency peaks in the
power density spectra\footnote{These are analyzed for example  in missions like \textbf{XMM-Newton} (X-ray Multi-Mirror Mission) {http://sci.esa.int/science-e/www/area/index.cfm?fareaid=23}  or \textbf{RXTE} (Rossi X-ray Timing
Explorer) {http://heasarc.gsfc.nasa.gov/docs/xte/xtegof.html}}, which could be interpreted  in the \textbf{RAD}  frame as derived  from the
 tori aggregate oscillation modes:
the axis-symmetric and incompressible mode  of each  torus of the aggregate,
variously associated  in several studies with the \textbf{QPO} emergence  \citep{Montero:2007tc}, has to be combined with the peculiar
modes arising   by  stratified structure of the \textbf{RADs} \citep{ringed}. The \textbf{RAD}  oscillations can also lead, eventually,  to an alteration of the
macrostructure elongations and spacings.

$\mathbf{(\bullet) }$ From the observational  view--point,  a \textbf{RAD} could be  ``disguised'' as a single  disk combing a  complex mix of several properties.
The \textbf{RAD} macrostructure can disrupt the usual  ``disk-model''--``disk geometry'' correlation,  especially as regards   the assessment of the  accretion rates.
In fact, \textbf{RAD} matches  the geometry of a geometrically thin disk with the  specific characteristics of a  geometrically thick disk,  as   the  high accretion rates,  in the  frame of   a   stratified  inner \textbf{RAD}  structure, a differential relational law and  a  \emph{knobby},
but  axial-symmetric, \textbf{RAD} disk (envelope) surface  \citep{ringed}.
\textbf{RAD} can      therefore play an important part  in the debate on the  \textbf{SMBHs}  origin.
The  episodic  \textbf{RAD}  accretion phases    with super-Eddington accretion rates    might be considered to  explain  the  \textbf{SMBHs}   origin   from    (intermediate   or low mass) \textbf{BH} ``seeds''  of  $10^4-10^2 M_{\odot}$--{\citep{1984A&A...137L..12V,1985A&A...153...99S}}. This issue applies to
\textbf{SMBHs} formation and evolution   at cosmological distances  \citep{Oka2017,Kawa,Li:2012ts,apite1,apite2,apite3,Allen:2006mh}.
Long and continuous accretion episodes  have been considered as origin of the   \textbf{BH}   masses,  involving  also   a relevant spin-shift (particularly in elliptic galaxy hosts) or, \emph{vice versa}, a series  of  small and random accretion
episodes  due to  cloud accretion or also   tidal disruption of a  star   (spiral galaxies), {see  for example \cite{Bonnell}}.
This situation can be  clearly reflected by the \textbf{RAD} macrostructure.  The internal structure of the \textbf{RAD}   induces a peculiar internal dynamics, made up of  tori interaction, the    drying--feeding processes and screening effects \cite{dsystem}.
 Indeed, the  \textbf{RAD} would be able to describe   screening  effects of  X--ray emission \citep{Marchesi,Gilli:2006zi,Marchesi:2017did,Masini:2016yhl,DeGraf:2014hna,Storchi-Bergmann} describing the screening tori as the inner torus of the \textbf{RAD} or the inter--torus located between two tori where the outer one of the couple is accreting towards the central attractor--see Eq.\il(\ref{Eq:spar-bay-cbc}). This situation was restricted  in Sec.\il\ref{Sec:RADs-sis} to some specific cases.

In this section we explore  possible generalizations of the \textbf{RAD}, where different models are considered for the aggregate components.
We have already pointed out  that a part of the \textbf{RAD}   impact  in its galactic environment derives from the  ``macrostructure-level'', i.e.,   from the tori interaction in the \textbf{RAD}, being therefore  in these aspects quite independent from the  accretion disk model adopted for each component.
In the following we consider the \textbf{RAD} symmetries  exploring  the relevance of the misaligned and warped tori within an agglomeration.
Then we discuss the tori self- gravity influence in the emergence of the    \textbf{RAD} instabilities, and the relevance of the self-gravity component of the toroidal disks with respect to the different contributions of the disk force balance.
Finally we  conclude this section   examining  briefly the main aspects of the  magnetic field contributions in \textbf{RAD} instabilities.

\subsection{Symmetries in RADs}

This  first realization or  \textbf{RAD} model  has been  constructed considering agglomerated     coplanar and axi-symmetric  toroidal disks centered on a spinning \textbf{BH}. This situation could be seen  as the final steady state  of an original,  more complex,  environment  constituted by an originally misaligned disk orbiting a central Kerr \textbf{BH}.
There are indications to assume,  at least in a first phase of   formation, a torus  misalignment with respect  to the \textbf{BH} attractors arises, and a warped and tilted accreting torus occurs--see for example \citep{Nelson:2000mw,1996MNRAS.282..291S,2005MNRAS.363...49K,liska,2018arXiv180110180S,
2015MNRAS.449.1251D,2006MNRAS.368.1196L,1999ApJ...525..909A,2017ApJ...845...10F,
2009MNRAS.399.2249P,2009MNRAS.400..383M}.
The disk  misalignment   depends,  however, significantly on the central \textbf{BH} spin.
The
dragging of  frames, Lense-Thirring (\textbf{L-T}) effect,  has a crucial  influence on
the morphology and equilibrium   of an accretion disk and this has also a notable importance in the misalignment.

More generally, the description of this  situation involves  several aspects of the accretion disk physics. \textbf{\textrm{(1)}} Misaligned disks  exploration faces aspects of the    accretion disk formation   after  different processes: from a  binary system, made up by a \textbf{BH} with a star companion or  as  formation from local matter  of  the \textbf{BHs} environments. Misaligned disks are   especially expected orbiting \textbf{BHs} in binary systems  where
the \textbf{BH} spin axis is generally misaligned with the
orbital plane of the  companion star.
Consequently, in the early ages of disk formation, the angular momentum of the infalling material from the \textbf{BH} companion  to the forming  torus, can be misaligned with respect to the binary orbital plane   and the  Kerr  \textbf{BH} equatorial
 plane.
Therefore, one may expect that the tori misalignment can
provide  description of the early phases of formation of the \textbf{RAD} multiple tori.
 As we  describe below,  this special situation can provide probable means to justify the origin of counterrotating tori of the \textbf{RAD}. \textmd{\textbf{(2)}} The warping and twisting of the disk   depends  on the disk  composition, the viscosity, the presence of magnetic fields, and  the frame-dragging (Bardeen--Petterson  effect \cite{BP75})  and,   in a binary system, the disk inclination with respect  to the  orbital plane of the star companion.  \textmd{\textbf{(3)}} The  probable   final steady stage of an initially misaligned  torus  is expected as a   coplanar, near--\textbf{BH}, accreting disk.  This inner torus  can be relatively
 small compared to the outer  disk  and the original  torus. The outer part of such  disk
is  aligned  more slowly on the longer timescales.    \textmd{\textbf{(4)}} Misalignment can also  provide a ringed structure by  creation of several  tori from an original warped disk which splits after the warping.
The initially inclined disk can  twist because of the  tidal
effect of the  star companion. The torus  has therefore two distinct regions: an \emph{inner} region, close to the central \textbf{BH} and an \emph{outer}  region, the farthest   from the central attractor. The inner
regions of the disks being   located on the equatorial plane, while the
outer regions are inclined because of the accretion from the companion.

For the  \textbf{RAD}  model, it is important to assess the  time periods of the accretion from the companion   star (companion accretion rate).
The  tori outer regions remain influenced by the
 differential precession of  the  tidal torques, following therefore the
binary  orbital plane.
The  disk region
precesses around  the  axis  parallel to the binary  orbital axis.
  The torus  inner regions, on the other hand,  are affected by the  differential
precession  due to the \textbf{BH} frame-dragging,  and in turn this  effects  aligns  the inner parts of the torus
with the \textbf{BH} spin (Bardeen--Petterson effect).
Consequently the torus finally becomes warped and twisted with an inner part aligned with the spin of the central Black hole.
The disk will eventually reach a steady state, which is  expected also  from
 the process of continual  matter feeding    from the star companion.
Noticeably  it has been proved that  when  the disk outer regions   are almost  counter--enlightened, the
 disk can form a   counterrotating torus with respect to the central \textbf{BH}.

Although probable in an early, transient, period of the attractor--accretion disk life, misaligned disks   can have a  considerable importance  also in  the \textbf{RAD} framework, in the jet--emission scheme related to the inner   part of the accreting disk.
Considering the  torus regions partition, due to the binary precession and \textbf{L-T} effects, jets  emitted in the inner torus  region (inner-disk---jet correlation), would be  aligned with the \textbf{BH} spin. A warped
disk finally could be considered as the  explanation for the correlation between  \textbf{AGN} radio jets  and the galaxy disk  plane.
This fact can have an  influence  also on the \textbf{QPO} emission. More precisely,
the torque derived from the \textbf{L-T}  effect has been   related to the \textbf{QPO} emission in  X-ray light--curves.

Magnetic fields are  also able to condition the first phases of the disk formation, the galactic magnetic field could alter
the structure of the accretion torus, influencing  the disk initial inclination.
(We note that the large--scale vertical magnetic field influences  also the jets orientation). Furthermore, in presence of strong large-scale magnetic field  there is a bending of  the inner part of the  disk and jets into partial alignment with
the \textbf{BH} spin.
 However some studies have shown  that the increase of disks masses (in \textbf{RAD} parametrization, this means a $K$-parameter increase)  such  inclination would  disappear thus, under certain conditions, the disk inclination does not arise or turns finally to vanish.

Viscosity is a further aspect affecting the viscous disk misalignment,
     leading   the  inner disk  regions   aligned  in the \textbf{BH} equatorial plane   \citep{BP75}.
We can asses the  impact  of the viscosity in the viscous disk dynamics  distinguishing a   (warping) wave--like regime and a  diffusive
regime,  depending on the Shakura \& Sunyaev dimensionless viscosity parameter  and the disk  thickness.
In fact, warps in the misaligned disks can  diffuse  on a
viscous time--scale, and  can  also  decay extremely rapidly
 as the
differential precession combines with the  viscosity.
We can specify this mechanism,
identifying in the disk  a transition radius defining the point  at  which the
disk specific angular momentum  becomes   aligned with the \textbf{BH} spin.  This radius,  depending on several factors  of the disk model,  has a role in the eventual fragmentation of the original disk into a  multi-toroidal structure.
 Therefore, more generally, considering the balance of the tidal forces, the frame-.dragging and viscous effects, we can define  the following  three relevant radii:
\textmd{\textbf{(i)}}The Lense--Thirring radius, $\mathrm{R}_{\mathbf{LT}}$, where
 Lense--Thirring
precession and the  effects due to the viscosity balances;
\textbf{\textmd{(ii)}} The tidal radius, $\mathrm{R}_{\mathbf{tid}}$, where the tidal
term balances the  viscous term;
\textmd{\textbf{(iii)}}The warp radius, $\mathrm{R}_{\mathbf{warp}}$, related to the
rate of disk warping.
Some models are expected to  be  warped  in a region between
 $\mathrm{R}_{\mathbf{LT}}$ and $\mathrm{R}_{\mathbf{tid}}$, having  a largest  rate of
warping at  $\mathrm{R}_{\mathbf{warp}}$.

\subsection{BHs and misaligned disks}
In this article  we have shown the importance of the spin--to--mass ratio of \textbf{BHs} in shaping  the main morphological and stability characteristics of \textbf{RADs}.
The impact  of the \textbf{BHs} spin on the misaligned disks  reflects in  the Bardeen \& Petterson effect.
In the context of the misaligned  disks, a \textbf{BH} warped torus evolves \emph{together} with its attractor, changing its mass, the magnitude of  its spin,  and the spin  orientation. It has been shown that a \textbf{BH}  can  align  with  the accretion disk  outer part   within a time--scale shorter than the accretion
time-scale. However, for a  geometrically thick disk  considered here as model for a torus of the \textbf{RAD} agglomerate and associated to
super-Eddington rates, alignment should occur  at the  mass accretion time--scale.
On the other hand, the  twisting  of the accreting disk  can enforce the internal  dissipation
as a mechanism which leads to increase of  the   accretion
rate.
We also note  that
 jets can extract   large amounts of \textbf{BH} rotational  energy affecting also the disks structure; this holds  particularly  for a tilted torus, where  jets can   contribute  to the alignment of the torus inner parts with the  \textbf{BH} spin.
Concerning the spin magnitude, it is  generally   expected that
corotating  disk accretion  is associated to a \textbf{BH}  spin--up, however, accreted material from
 counter-aligned accretion disks
   could induce a \textbf{BH} spin--down rather than a spin--up.
There are  further aspects to be considered to  trace a more complete  picture  of the \textbf{RAD} instability process.
 For example, the presence of wind (particularly  in the mass loss from accretion flow)  is a further  aspect  of the tori physics which could be  considered  in relation to the effects of the double accretion phase typical of the \textbf{RAD}.

\subsection{Tori self--gravity and    \textbf{RAD} instabilities}
The  \textbf{RAD}  stability has been  discussed  in Sec.\il\ref{Eq:tori} and Sec.\il\ref{Sec:RADs-sis}.
Then, as each torus of the aggregate is  modeled by a geometrically thick disk, torus \emph{self-ravity} may be relevant. However, we have also stressed   that in some circumstances a torus component,
  especially the inner corotating \textbf{RAD} tori, can also be  rather small, i.e., $K\gtrsim K_{\min}\equiv V_{eff}(r_{cent})$.
 In the \textbf{RADs},
tori self--gravity  should have a great impact with respect to the  concurrent effects in  the  force balance, such as the background curvature, the  centrifugal effects, and  the  magnetic field contributions, especially in   the largest tori,  with large magnitude of the specific angular momentum $\ell$, located far from the central \textbf{BH},   and with  large $K$-parameter.  Among these  tori  we mention the counterrotating tori orbiting in the spacetimes of the faster spinning \textbf{BH} attractors.

 It  has be  shown in other contexts that tori are unstable, if
considered as self-gravitating systems; formation of $\ell$corotating  pairs     after  fragmentation of  an original seed disk can be one of the outcomes of this effect,  similarly to the  misaligned disks where however the torsion, typical of the binary systems, is  at the origin of the original disk split.
However,  the newly formed outer torus always has, if it is part of a $\ell$corotating couple, a specific angular momentum greater (in magnitude) than  the inner torus and at fixed angular momentum, and  a disk in accretion reaches its maximum elongation (and dimension) with respect to the quiescent disk.
Whether  self--gravitating disks fragment, and the determination of the   number
 and  dimensions  of the  disk debris, depend on several parameters--tori   masses with  respect to attractor,  and
the rate of  energy loss and   cooling.
Fragmentation, on the other hand, can   provide a context  in the  galactic environment  for stars and
regular spiral structure waves formation \citep{ShlosmanBegelman(1987),Lodato:2008mu,Khajenab2006,2016A&A...588A..22K,MJ18,
Ghasemnezhad:2015yiz,2013ApJ...765...96A,1988ApJ...332..637A,1963MNRAS.126..299H,200120012001,BLodato99,
1994ApJ...420..247W,1990ApJ...361..394T,1989ApJ...341..685S}.
Also, rotation of disks in the  central part of various   galaxies, for example
\textbf{NGC4258}, shows  a divergence from the  Keplerian  rotation law--\citep{1998ApJ...508..243H,1998ApJ...497L..69H}.
In the \textbf{RAD} scenario, one might consider the internal differential rotation due to the ringed structure, but other interpretations   can involve the  misalignment or  the self--gravity of the disks.
Nevertheless, especially in \textbf{AGN} disks,
the presence of viscosity  combines with   self--gravity,
    affecting eventually the torus turbulence, modifying both accretion rates and the   disk  force  balance,   together with the angular momentum distributions.
  These factors  reshape the angular momentum profile, obviously conditioning the stability of the disk.
 Furthermore,  the induced  internal waves would contribute to the transport  of angular momentum in the accretion phase together with the eventual turbulence generated by the Balbus--Hawley   mechanism \citep{BH91}.

It should also be said that if the self-gravity is combined with the \textbf{L-T} effect for the region of the  disk closest to the central attractor,  then this  situation  may lead to  counter-rotation of the external part of the disk. Furthermore, runaway instability, typical of thick tori, can combine     with the torus self-gravity, affecting the torus
radial and vertical structure.
Note that  torus  self-gravity  implies a modification of the gravitational background that obviously  influences  the adjacent tori in the agglomerate.
The oscillations of each  toroid   of the \textbf{RAD} will be a further factor of instability for the entire aggregate which combines with  the disk self-gravity.

\subsection{Magnetic field  contribution in RAD instabilities}
Magnetic fields are extremely relevant  in the physics of accretion, altering   the accretion phase. We specify that, although not considered here, magnetic fields have a significant role  also  in the   jet emission  and in the  jet collimation.
Above, we  discussed  the impact of the magnetic fields in  several aspects of the formation of non--aligned disks.
However, the influence of magnetic fields is varied and we     report here some considerations in connection with the \textbf{RAD} stability.

\textbf{RAD} constituted by tori endowed with toroidal magnetic fields (relevant for geometrically thick disks and involved in dynamos processes) have been  treated in  \cite{Fi-Ringed},
On the other hand, the case of a poloidal magnetic field in  a more complex \textbf{GRMHD}-\textbf{RAD}  model is planned  for future investigation. We mention, concluding this section, some important aspects of the  magnetic field in \textbf{RAD} tori in regards to the    \textbf{RAD} equilibrium  and the classification of the \textbf{BH} attractors.
 The analysis of the  toroidal magnetic field case has shown the divergence with the purely \textbf{HD} case in the tori rotation law.  However, the introduction of the magnetic field does not alter many  of the main  characteristics  of   \textbf{RAD} disks as the constraints on main seeds couples;  the magnetic field rather combines  with the constraints derived by the \textbf{GRHD} model.
Moreover, the \textbf{GRHD} tori considered  for the \textbf{RAD} are often used an initial configuration  in the current \textbf{GRMHD} models \citep{Luci,abrafra}.
Therefore, we expect that a numerical integration of \textbf{GRMHD-RAD} should be set on the tori- attractor constraints provided by  our investigation.
 Concerning the influence of magnetic fields in the tori and \textbf{RAD} stability,  there  are several recent  studies focusing  on  the  combined effects of the \textbf{PP} instability, the  global non--axi--symmetric hydrodynamic Papaloizou--Pringle (PP)  instability, and the   magneto--rotational instability  (\textbf{MRI}) due to the  magnetic field and  fluid differential rotation.
Firstly, in general, \textbf{MRI}  is most effective and  fast in transport of  angular momentum across
the disk, and higher  accretion rates were proved to occur
in the magnetized models. (We remind that disks can be   locally \textbf{HD} stable, according to  Rayleigh criterion, but  unstable for  \textbf{MHD} local  instability.)
An extensive general discussion of the situation can be found in \cite{Fi-Ringed,Del-Zanna,Wielgus,Das:2017zkl,Bugli}.
\textbf{PP} instability
 is associated to the   formation of long--lasting,
large--scale structures  which have been also recently connected to the
 gravitational wave emission--see for example \cite{PRL}.
 A series of recent  analyses shows that
 inclusion of a toroidal magnetic field could strongly  affect, even with a
sub--thermal  magnetic field, the  \textbf{PPI}. \textbf{MRI}  can trigger  predominant larger modes of oscillation (i.e. smaller length scales) with respect to  \textbf{PPI} modes, altering therefore the equilibrium properties.
More precisely, geometrically thick tori  with a toroidal  magnetic field     develop the (\textbf{3D}) non-axi-symmetric \textbf{MRI}  affecting the configuration on the  dynamical timescales see \cite{Del-Zanna,Wielgus,Das:2017zkl,Bugli}.
One of the possibilities is the suppression of the  \textbf{PPI}  by the \textbf{MRI}, or vice versa
 \textbf{MRI} and \textbf{PPI} can  coexist depending on several parameters.
Considering the \textbf{RAD} framework, the  emergence   of the \textbf{MRI}  suggests an
accentuation of the effects of the    \textbf{RAD} own   instabilities, discussed in Sec.\il\ref{Eq:tori}  to accent   phenomena connected with energy release and matter impact.
\section{Discussion and Concluding  remarks}\label{Sec:Open-Concl}
We investigated agglomerations of  accreting coplanar  tori  (\textbf{RADs}) orbiting in the equatorial plane of a  Kerr \textbf{BH} attractor as  tracers of the  Kerr \textbf{SMBHs}. The limit of the \textbf{RAD} constituted by one torus in accretion is also considered.
This analysis eventually proves  the existence of  a strict correlation   between  the \textbf{BH}  spin-mass ratio $a/M$, and the accretion tori featured by the \textbf{RADs}. Our findings resulted in the     complete  description  of the \textbf{RAD} systems  characterizing the central Kerr \textbf{BHs} in  classes which are uniquely identifiable through the properties of the rotating tori of the agglomerate.
Overview of the  major features of the classes of spinning attractors has been   discussed,    according to the  properties of  the orbiting toroidal structures. Table\il\ref{Table:nature-Att} collects the main classes characteristics.
More specifically,  we developed  a general classification of attractors and orbiting tori gathering together information on the Kerr \textbf{BHs} spin and their orbiting accreting tori in the \textbf{RADs} framework  introduced in  \citet{pugtot,ringed,open,dsystem,long}.
As consequence of this analysis,  different   issues  related to the \textbf{SMBH-RAD} systems were   addressed such as    the  identification of observable features of the toroidal  agglomerates  the indication of the  associated \textbf{BH} classes according to their dimensionless spin. These classes associate a \textbf{RAD} to its   central attractor, therefore they provide  indication on  the \textbf{BH} environments where it  would be possible to observe  a \textbf{RAD}.   Eventually  we addressed more explicitly the question whether    there is a way to unambiguously   identify the \textbf{RAD} aggregates  through the specification of the \textbf{RADs} order, angular momentum distributions, the relative rotation of the toroids, toroid rotations  with respect to the attractor,  and internal dynamics of the \textbf{SMBHs}  (the spin-mass ratio).
This investigation  serves to different  purposes:  to envisage  an accretion disk--\textbf{SMBH} correlation   in the \textbf{RADs} scenario, or in the limiting case of one orbiting torus, and to provide  an indication of the possible  attractor-disk candidates to look for observational evidences of \textbf{RADs}.
  This work supplies   the most comprehensive reference template of \textbf{BH}-accretion tori correlation, which can be also used an a  strictly constrained  set of the initial configurations for the development of  a fully general relativistic  dynamical \textbf{GRMHD} simulation for the \textbf{SMBH}-\textbf{RAD} system--see for example \citet{Luci,abrafra}.

Our analysis is conducted by  numerical integration of  the  hydrodynamic equations for multiple tori with fixed boundary conditions for each configuration of the  set--see Figs\il\ref{Fig:but-s-sou}. Parallel  to this  analysis we carefully  explored  the ranges of  fluid specific angular momentum and $K$-parameters,  drawing then the constraints for   the  radii $(r_{cent},r_{\times})$ and  $(r_{in},r_{out},r_{coll})$--see also
  \citet{open,dsystem,app}. It is clear that the construction of the aggregates relies   essentially  on the boundary conditions imposed on the function (\ref{Eq:def-partialeK}) (the Heaviside functions). We used the studies  of  \citet{open,long}, which  fix the location of the accretion torus edges in the
spacetime regions confined by marginally bound, marginally
stable and marginally circular (photon) orbits.
 This setup turns to be  very important for the analysis of the \textbf{RAD}  oscillations emerging as   perturbations of tori and the boundary conditions \citep{ringed}. The \textbf{RAD} model  encloses a huge amount of   possibilities to be investigated, having a  large number   of cases even in the   simple three parameters  model  considered here (the specific angular momentum  $\ell$, the $K$ parameter and the attractor spin-mass ratio). Interestingly, this  investigation   enlightened also the importance of the dimensionless quantities   $\ell/a$,  $r_{cent}/a$  and $r_{in}/a$, while  a deeper analysis  of the \textbf{BHs}-accretion tori connection in this special parametrization  is  reserved for a planned future work.
We consider these configurations using a seed  to generate
a general  sequence of the \textbf{RAD} tori, starting from one of the four  germs $\pp_{\pm}> \pp_{\pm}$ (the $\ell$corotating couples) or
$\pp_{\pm}> \pp_{\mp}$ (the $\ell$counterrotating couples) respectively--Figs\il\ref{Figs:GROUND-scheme}.

Focusing here   on the central objects    around which these   configurations may orbit,  and considering different properties of the configurations, we  drawn  a  detailed classification of the  attractors in \textbf{$17$} general classes  singled by the their \textbf{BHs} spin-mass ratios  in the entire range $0\leq a\leq M$,  including the Schwarzschild static solution and the extreme Kerr solution.
As  expected, this classification tends to strongly differentiate  fast spinning attractors and slow spinning ones. Major differences are highlighted for   two classes of attractors,  with spin  $0< a\leq 0.47M$  and  $a>0.47 M$ respectively.  The static case, represented by the Schwarzschild solution, was considered separately. 
 A further relevant aspect of this analysis  is that the \textbf{BH} classes  often intersect--see Fig.\il\ref{Fig:Relevany}.
Importantly   we have taken into account the  \textbf{RAD} stability properties  in the construction  of the  classes of attractors. The main aggregate instabilities are mainly driven by  two families of processes: {\textbf{(1)}} Collisions between accreting or non-accreting tori, and \textbf{(2)} \textbf{RAD} instability following the accretion phase of one torus  of  the aggregate  \citep{ringed,dsystem}.

A double accretion  can be observed  \emph{only} in a     couple  $\cc_{\times}^-<\cc_{\times}^+$ (es Fig.\il\ref{Fig:but-s-sou}),  around all Kerr \textbf{BHs} ($a\neq0$).
Moreover,  in the \textbf{RAD} scenario, the maximum number of  accreting tori orbiting  around one  central Kerr \textbf{SMBH} is $n=2$.   This opens up important potential observations   encouraging also  a review of  the current interpretation of the accretion data  by considering  a \textbf{RAD} framework.
 In fact,  the ringed structure can be effectively disguised  as a geometrically thin, axi-symmetric  disk centered  on the equatorial plane of the Kerr \textbf{SMBH} with  interrupted  phases of super-Eddington  accretions and a very rich inner dynamics    with jet emission featuring  also inter-tori, proto-shell, jet emission.
The presence of an inner torus can  also enter as a new unexpected ingredient in the accretion-jet puzzle--see also \citet{Hamb1}. Therefore, in  Table\il\ref{Table:nature-Att} we also consider the possibility of launch of proto-jet configurations in the \textbf{RADs}. This situation however,  has been deeply analyzed   in   \citet{open,long}.
We have shown  an important restriction on any  screening effect from an inner torus of an aggregate. Screened X-ray emission by some ``bubbles'' of material are currently studied in many processes-\citep{Gilli:2006zi,Masini:2016yhl,DeGraf:2014hna,Storchi-Bergmann,
Ricci:2017wmr,Ricci,Marchesi,Marchesi:2017did}.
Then, particularly some \textbf{AGNs} have been proved to be
 obscured according to the X-ray spectral emission \citep{Marchesi,Marchesi:2017did}; several analyses also  suggest that  obscuration in optical and   X-ray emission profile may be due  to
 different phenomena,  caused by dust materials surrounding the
 inner part of the galactic nuclei.
So far these (free) dust materials were almost always supposed to be  randomly distributed around the central \textbf{BH} and, depending on  the gas  density, the light emitted during the growth could  be absorbed in the optic as in the X-ray electromagnetic band, distinguishing  \textbf{AGN}  as obscured, not obscured, or much
obscured (or \emph{Compton thick}).
We here explicitly claim for an analysis of this  obscuration  assuming   a \textbf{RAD} scenario, therefore considering the   inner corotating, accreting or quiescent torus  in accordance with the analysis in Sec.\il\ref{Sec:RADs-sis}, and with the constraints imposed by the spin of the central \textbf{SMBHs}. Such \textbf{RADs} are  very much constrained so that they cannot be randomly distributed neither assumed to exist independently on the evolution of the \textbf{SMBH} itself--see also  \citet{dsystem,long}. We believe this change of paradigm may have a huge impact on the current scheme adopted to explain the obscuration.
To be more precise,  screening tori may only exist as corotating fluids in particular \textbf{BH} classes and under specific restrictions on the specific angular momentum. These tori can be  accreting onto the central \textbf{BH} or quiescent-- see also  \citet{dsystem,long}.  A ``screening'' \emph{corotating}, non-accreting, torus  between the  two accreting tori   can be observed only as $\,\cc_{\times}^-<\cc^-<...<\cc_{\times}^+$ for any Kerr attractor having spin $a\neq0$. These special tori  are expected to be relatively small compared to the outer tori of the \textbf{RAD} agglomerate.
A further case is where the inner corotating or counterrotating (for \textbf{BHs} with   $a<0.46M$) accreting torus of the  \textbf{RAD} is ``obscured'' by an \emph{outer} screening and quiescent  torus.  As seen in   Table\il\ref{Table:nature-Att}, a counterrotating accreting torus  with an outer corotating  torus towards the accretion (i.e. a $\cc^-_1$ torus  having specific angular momentum  $\ell\in \mathbf{L1}^-$), can be observed only as a
$\cc_{\times}^+<\cc_1^-
$ aggregate, and orbiting around  slow spinning \textbf{SMBHs} with spin  $a<0.46M$.
In the \textbf{BH}  classification,  we have also included    very fast spinning \textbf{SMBHs} with   $a>a_1$,  which are characterized by the possibility that  corotating (screening, accreting) tori  can be orbiting  in  the ergoregion.
This aspect, also mentioned in  \citet{pugtot,ergon,ringed,dsystem,open},  is an interesting consequence of the frame dragging.

{More generally, we expect \textbf{RAD}    to be  a  relevant   feature of  the  faster-spinning  \textbf{SMBHs}. Specifically the  $\ell$corotating \textbf{RADs} are  favored features of
 slow attractors $(a\lesssim0.45M)$,  while around  faster spinning attractors, $\ell$counterrotating \textbf{RADs} would be most expected. Generally,  in the spacetimes of the slower rotating  \textbf{BHs}, tori collision appear more probably.  This episode may be followed by the collisional energy release and, eventually, tori merging and accreting onto the central \textbf{BH}--\cite{app}.
For the slow spinning  \textbf{BHs}, with  dimensionless spins $a/M$   close to the lower limit of the static attractors,  any \textbf{RAD} tori couple  has to be considered as an   $\ell$corotating couple, independently of the relative rotation of the tori, i.e., although the tori of the aggregate can   have alternate spin, eg.  $(0,+,-)$ or $(0,-,+)$, according to the notation introduced  in  Fig.\il\ref{Figs:GROUND-scheme},  their  properties  considered for \textbf{RAD} structure,  are entirely analogue to the $\ell$corotating case in a  Kerr spacetime. Moreover, in the Schwarzschild spacetime,  double accretion or obscuration is \emph{not} possible.
On the other hand,  very  fast-spinning \textbf{SMBHs}  $(a\approx M)$,  would favor  the  formation of $\ell$counterrotating  tori   which are  also largely separated in the \textbf{RAD}. This holds particularly for the  couples  $\pp^-<\pp^+$, where a double accretion phase occurs. Instead, the seed having an  inner counterrotating torus in accretion   with  an  outer corotating torus  is favored only in the early phases of formation of  the outer torus, for fast-spinning \textbf{SMBH}, while in  Schwarzschild spacetime such a couple can be always observed--\citep{dsystem}.

Kerr \textbf{SMBH} attractors are significant environments for the   \textbf{RAD}  observation. In fact,  \textbf{SMBHs} determine certainly strong curvature effects having  ultimately  a major influence in the determination of  the constraints in  (\ref{Eq:def-partialeK}). Then, as shown in  \citet{dsystem}  and then discussed in  \citet{long}, very massive \textbf{BHs}  would prevent the  emergence of  tori collision. Note also  that the space-scales are here  in
units of the \textbf{SMBH} mass; for example the maximum spacing $\bar{\lambda}$   considered  for two accreting tori is  $\bar{\lambda}\approx8 M$  in the case of  nearly extreme \textbf{BHs}.
To realize the significance of this  we  should note that the spacing
parameter $\bar{\lambda}$ between two tori regulates in fact the possibility of  tori collision.
}
{ The unstable phases of the aggregate,
on  one hand,  could undermine the survival of the ringed structure, eventually leading to the formation of a single disk. On the other hand, the \textbf{RAD} unstable phases constitute   environment for  set of  interesting  phenomena-- see \cite{dsystem}.
It is possible that a  \textbf{RAD} represents  a final steady state of life of an  attractor-disk system originating as extended disk misaligned with the \textbf{BH} axis; this  hypothesis stands as  particularly promising  for the very massive \textbf{BH} attractors in \textbf{AGNs}, as demonstrated  in Sec.\il\ref{Sec:gothc}--see also \cite{Nelson:2000mw,1996MNRAS.282..291S,2005MNRAS.363...49K,liska,2018arXiv180110180S,
2015MNRAS.449.1251D,2006MNRAS.368.1196L,1999ApJ...525..909A,2017ApJ...845...10F,
2009MNRAS.399.2249P,2009MNRAS.400..383M}. }

\subsection{Relevance of the \textbf{RADs } in the \textbf{AGN} environments }

It is clear  that a \textbf{SMBH}  is rarely an isolated \textbf{BH}, but rather a living object  in the host galaxy cores interacting with its environment and subjected to several evolutionary phases, where its mass and spin will often change during  different stages of the \textbf{BH} life. The \textbf{SMBHs} intercept the stellar population and  dust of their  galactic host and vice versa \textbf{SMBHs} remix, the  matter and radiation content of  the galactic environment.  Suffering  from galaxy  collisions,  the interaction between the \textbf{BH}  and the galaxy environment     ends in  changes of the    \textbf{BH} spacetime which is  initially considered as  ``frozen background''.  A non-isolated \textbf{BH} background can change following a spin-down or a spin-up process \citep{Abra83,Abramowicz:1997sg,Rez-Zan-Fon:2003:ASTRA:,Font:2002bi,Hamersky:2013cza,Adamek:2013dza,Lot2013}. This clearly should lead to a \textbf{BH} shift from one class of our classification  to another, and to a change of the equilibrium conditions of the \textbf{RADs}.
  Toroidal structures might be   formed as remnants of several accretion regimes  occurred in various phases of the \textbf{BH} life { \citep{Aligetal(2013),Lovelace:1996kx,Romanova,Volonteri:2002vz,Carmona-Loaiza:2015fqa,Dyda:2014pia}}
Therefore, the  analysis  considered here can be  significant  for  the resolution of the  so called mass problem for \textbf{SMBHs} in \textbf{AGNs}.
We can consider  accretion   in  long and continuous accretion episodes, arising due to    merging    involving  a large spin-shift or,  \emph{vice versa},  sequences of  small and random accretion
episodes,   being advocated     for the formation  of the   \textbf{SMBHs}.
The \textbf{RAD}  internal dynamics has  several inter-disk effects including double  accretion processes,   screening effects,   and tori collisions. These  situations  clearly represent mechanisms
for the mass growth of the \textbf{SMBHs}.
It is clear that  many aspects of  the physics of \textbf{SMBHs} and their  host galaxies would be altered  by the relevance of the \textbf{RAD} model  in support of the  hypothesis of a more complex \textbf{BH}-accretion disk system,  than is commonly considered.

In Sec.\il\ref{Sec:gothc} we  also examined  the  impact    of the  \textbf{RAD} scenario, regarding, more widely,   future generalizations of the tori aggregate model    to consider  misaligned tori  and, more generally, diverse  accretion disk models.
  It should be also  stressed that, although the current theoretical analysis  conceives on a large variety of  accretion models,  with a diversified  parametrization and disk shape  (depending  on disk optically depth,  the geometric thickness,   luminosity...),  there are some general constraints that  extended  matter orbiting configurations  must abide on the curved spacetimes---\citet{abrafra}. This is an important feature of \textbf{SMBH} accretion disks  validating also  the relevance  of the  \textbf{RAD} model.
However,   accreting  disk models are generally associated to specific  characteristics of their \textbf{BH} attractors  and especially their mass range\citep{abrafra}.
The geometrically thick disks, which were adopted here as  aggregate components, are  governed by   the gravitational forces  predominant with respect to other possible components of the force balance, and they are therefore  associated to \textbf{SMBHs} spacetimes where  the  curvature effects  and the fluid rotation are significant  in  the  determination  of the toroidal equilibrium and morphology.

\subsection{Phenomenology}

\textbf{RAD}  phenomenology is   significant for the high energy phenomena related to accretion onto \textbf{SMBHs} in \textbf{AGNs}  which
could be observed in their X-ray emission.
\textbf{RAD} framework also includes the   jet launch    with an interesting and intricate shell  structure. These  structures then fit into the more broad discussion on the role and significance of open surfaces in relation to  (matter) jets emission and collimation, as well as jet-accretion correlation--see  \cite{KJA78,Sadowski:2015jaa,Lasota:2015bii,Lyutikov(2009),Madau(1988),Sikora(1981)}.
Among the other  phenomenological application of these studies, there is a possible
connection between \textbf{RADs} seismology and \textbf{QPOs}--low
and high frequency peaks in the
power density spectra--see also \citet{Ingram:2016tbq}.
The pattern of the
possible oscillation modes of the tori aggregate has been provided
and related to the evolution of instabilities in \textbf{RAD}
in \citet{ringed,2013A&A...552A..10S}. Missions like \textbf{XMM-Newton}  or \textbf{RXTE}, \textbf{NuSTAR} (\textbf{Nu}clear \textbf{S}pectroscopic \textbf{T}elescope \textbf{Ar}ray)\footnote{https://www.nustar.caltech.edu/}  are on the verge of these studies--see also \citet{Mossoux:2014fqa,Gilli:2006zi}  and \citet{Gandhi:2017dix,Masini:2016yhl,Harrison:2013md}.
The  current status of observational mission and the   potentialities in the next future provide certainly  an increasingly high degree of details with which we can look at \textbf{BHs} and accretion disk morphology.   The results outlined here thus are directly comparable with the current data and  encouraging us to apply already these results  for  a re-interpretation of the current data analysis set up which is almost unanimously based on the scenario of \textbf{BH} one-accretion disk system -- see for example  \citet{Hamb,Hamb1}.
     Possible evidence of the existence of the ringed accretion disk can be inferred from the study of the optical properties of the ringed-like structures   \citep{KS10,S11etal,made,made0}.
{Concerning the  \textbf{RAD} optical appearance
we expect that  tori emanating radiation would be
distorted into a belt-like configuration because of the curvature  effects   of the Kerr geometry. (Then the Lense--Thirring effect becomes clearly more important for corotating tori approaching the \textbf{BH}).  This aspect  will be considered in planned  future investigation.
 We expect that   optical phenomena  could reflect the \textbf{BH} classification  as they  differentiate  between corotating and counter-rotating tori in dependence  of the \textbf{BH} dimensionless spin.
Optical properties depend also on the geometrical thickness of the disk and  the presence of a disk atmosphere (some
accretion disks atmospheres can show   similarity with  the   upper main
sequence stars, where  magnetic fields and winds appear \citep{1967ApJ...148..217W}.
The analysis of the optical appearance of the thick accretion disk   is also based on the model of  hot coronae above
the surfaces of accretion disks and  a hot coronal layer implied in the emission profile--
  \citep{dance,FW07,made0,made,Schee:2008fc}.}
\textbf{BH} \textbf{RAD} model may be  revealed by future X-ray spectroscopy, from the
 study of  excesses
on the shape of the relativistically broadened spectral line  profile,
related to  a sort of rings model which may be adapted as a special case  of the  \textbf{RADs}.
Specifically, in  \citet{KS10}
extremal energy shifts of radiation from a ring near a rotating \textbf{BH} were particularly  studied:
radiation from a narrow circular ring shows a  double-horn profile with photons having
energy around the maximum or minimum of the  range (see also  \citet{Schee:2008fc})\footnote{
Notice that significant influence of the self-occultation  effect  on the profiled spectral lines was for the first time demonstrated in \cite{1992ApJ...400..163B}.}.
This energy span of  spectral lines is a function of the observer's viewing angle, the \textbf{BH} spin and the  ring radius. %
 Accordingly,
the ringed disks  may be revealed thought
detailed spectroscopy of the spectral line wings,  claiming for observation and data re-analysis in this framework, especially in relation with the \textbf{BH} classes considered here.
}

\subsection{RAD phenomenology and future perspectives  for    the RAD extensions}

On one side, the possibility of more orbiting tori opens  new perspectives of enriched accretion phenomenology connected to the \textbf{RAD}  structure and   its own internal   dynamics. \textbf{RAD} typical   effects  as the double accretion phase   or the presence of screening tori which are located between the  \textbf{BH} and the accreting  torus,  or  tori between two accreting \textbf{RAD} tori,    the presence of \textbf{AGN} obscuring torus, are proved to be  expected in specific contexts and under precise conditions on the tori and the central \textbf{BH} attractor. In this analysis we present   limits indicating  the parameter ranges where these  situations occur.
On the other side,  in the \textbf{RAD}  frame, we pose rather narrow constraints for  different ad hoc models of several accretion processes which are currently considered,   ruling out many of the assumptions considered so far  on screening and obscuring phenomena. Those limits and some main indications on  \textbf{RAD} observation are given below.
This work provides a new  scenario for  the data  interpretation  which has been until now  essentially framed into the single axi-symmetric, corotating, accretion   disk model. Screening and obscuring tori in accretion processes are essentially  positioned ad hoc in order to fit the observations and models.
It is very likely  that at least in the initial phases of the \textbf{RAD} formation, ringed structures may appear,   and we explicitly call for a review of the analyses  carried out so far in the simplest scenario of one torus model, shifting this classic setup to the   model of  orbiting \textbf{RAD} axi-symmetric structures including  the counterrotating  tori. Indeed, the predictions implied
 by the adoption of a \textbf{RAD} system fit some features of the accretion disk physics   which puzzle the theoretical grounds  of  one class of accreting model, for example the geometrically thick tori with some features proper  of geometrically thin disks.

  A further important aspect in this regard stands in the fact that a \textbf{RAD}  could have been   disguised so far as one  geometrically thin disk; in   \cite{ringed} it has been proved that \textbf{RAD} can be treated as  a geometrically thin disk blending with  distinctive  features of geometrically thick torus  as  super-Eddington luminosity (high  accretion rates). This situation can fit a particular  hypothesis made to justify the masses of \textbf{SMBHs} located in  high redshift galaxies.
Concerning the possibility of
screening tori we proved that   such tori  \emph{must} be corotating and  screening effects could be possible \emph{only} if the outer accreting disk is counterrotating, but this in fact rules out several of the models currently considered in the screened  accretion.  According to our analysis   screening effects are likely  to be observed  around  high spin \textbf{BHs}--precise limits and constraints  are  in Table\il(\ref{Table:nature-Att}).
Further constraints   on thickness  and extension on the equatorial plane of  the \textbf{RAD},   relevant to set power spectra   obscuration follow, from more detailed analysis  in   Table (\ref{Table:nature-Att}).
Those constraints will emerge  as distinctive features of the emission spectra.

Most interestingly a screening effect may happen in the occurrence of  a  double accretion phase as in Fig.\il(\ref{Fig:but-s-sou})-(a).
 The possibility of a double accretion from an $\ell$counterrotating couple is a new feature firstly presented in the  \textbf{RAD}, {we clamor for attention on the scientific community to
focus on observations on high masses and high spin Kerr black hole where
pieces of evidence of this phenomenon should be found}. Besides we proved that no more then two accreting tori can be  observed.
 The detection of a double accretion phase  (associated to a double shell of jet emission) would imply the presence of an  outer counterrotating accreting  torus and an  inner corotating one,
 the central \textbf{BH} is not screened. Vice versa, a  quiescent corotating torus, screening the outer retrograde accretion, and the inner prograde one are  possible. Such inert screening tori must be   relatively small configurations.
The analysis of  the  \textbf{RAD} characteristics which could be extracted from the spectrum emission or accretion rates/luminosity analysis, \cite{long}, may be used to locate the \textbf{RAD} in one of the attractor class of  Table\il(\ref{Table:nature-Att}), thus identifying the central \textbf{BH} attractor.  Spectra, for example from X-ray emission,  would  show evidence of  the  \textbf{RAD} spacing.
 Note that the distance in spin boundaries of the \textbf{SMBHs} classes  can be  very close,
 also up to $10^{-2}M$, this means that we can distinguish, considering \textbf{RAD} features, a  spin range
 $\Delta a\approx 10^{-2}M$, depending on the constraints provided on the  accretions properties.
It should be noted that a large part of the accretion disk analysis is actually dealing mainly with  corotating disks, our \textbf{RAD} analysis provides  more precise insight on the physics connected with counterrotating tori.

Screening effects are possible only in the case of accreting  counterrotating  tori, this implies that the current analysis using a screening torus must take into account the fact that the outer  torus, in the \textbf{BH}-screening torus-accreting torus system, is counterrotating.
This is a strong restriction  in the analysis of retrograde disks.
 On the other hand, if there is a system with a screening torus and an obscuring one,  this situation implies that  the \textbf{BH} spin has to be relatively small   $a\lessapprox 0.46M$.
Most importantly a  corotating  torus  \emph{cannot} be screened to the \textbf{BH}, i.e., no screening effect from axi-symmetric inner torus can be set--this means that many screening  models are effectively ruled out.

A further interesting aspect is constituted by jet emission in \textbf{RADs}.
 \textbf{RADs} are related to jet emission in two different ways. Firstly, each toroid considered as \textbf{RAD} component allows  open funnels, proto-jets, occurring when the centrifugal component in the disk force balance is strong  enough to push matter along the vertical direction along the \textbf{BH} axis. Jet for retrograde tori is another interesting feature.  We considered  these open solutions too in the set-up of the attractor classifications, although a  focused investigation of these are  especially in \cite{open,long,app}.
Secondly,  as   jets are expected to be correlated to accretion and especially to the inner part of an accreting  disk, the dynamical picture  featuring  more accreting tori, up to maximum of two $\ell$counterrotating accreting tori, including the possibility of screening and  obscuring tori, is certainly an interesting perspective  in the context of jet emission.
We proved that there can be up to two jets (no proto-jets), i.e.,  double shell of jets. According to constraints on the double accreting phase, the outer jet being correlated to the outer retrograde \textbf{RAD} torus, and the inner jet from the inner corotating, accreting  torus. Constraints on spacing and tori dimensions are set  according to \textbf{SMBH} spin and details on these constraints   are presented   in  Table\il(\ref{Table:nature-Att}).
Spacing between the tori will reflect in the jets separations.

The change of \textbf{BH} spins will reflect in change of the spacetime and other features of jet emission.
In this context,  it is proper to stress that characteristic \textbf{RAD} features  in accreting processes, such as screening and obscuring tori act also in regard of  the jet emission, the  presence of  inert materials will affect the energy release, the jet launching and collimation. Small corotating tori will be located eventually between two accreting tori and inside the double jet shall, while in the case of one jet correlated to a counterrotating accreting torus, the \textbf{BH}-jet system, can be possibly  characterized by an inner, screening, corotating torus.
It is then crucial to establish  possible \textbf{RAD} evolutions following the occurrence of the instability processes in the structure.

Constraints on \textbf{RAD} evolution have been discussed more extensively  in \cite{dsystem,long}.
Generally, the situation after tori collision emergence, or  accretion  between  two tori,  with matter  impacting from one torus to another,  can result also   in the destruction of the \textbf{RAD}, i.e. formation of one torus after tori merging.
Evolution paths in these models were discussed in \cite{dsystem}, energy release associated to the collision and accretion phases including the evaluation of the  torus cusp luminosity and  accretion rates  are examined in \cite{long}, where  evolution is also discussed in context of energy release in \textbf{RAD}.

We can say  that one possibility is the  merging after  accretion of the outer counterrotating torus of a couple or  collision with one inert torus or during the double accretion  phase; in all these cases there will be  a modification of the stability properties of each involved  torus,  according to a variation of the parameters range, for  matter supply  from the outer accreting  counterrotating torus, followed by an alteration of the  fluid   specific  angular momentum and an  increase of the $K$-parameter,  reflecting change in density and  tori dimension.
Such  phenomenon has  to be considered, focusing on   a more accurate prescription of the torus inner dynamics.

 The possibility  that the accretion or collision may in fact not lead  to the destruction of the \textbf{RAD} inner structure, but rather  to a sequence of instability processes (accretion phases) involving an inner torus  has also been discussed.
 Concerning \textbf{RAD} evolution, it must be said that the   oscillation modes  of  each toroidal component,
  will be combined together in the ringed structures to reflect the QPOs phenomenon--see  discussion in \cite{ringed} for the \textbf{RADs} perturbation.
  One can assume oscillations  of an individual torus of the \textbf{RAD} system, and possible excitation of tori oscillation by external influence, e.g, due to original oscillation of the first oscillating torus.
The resonance models of the epicyclic frequencies could give relevant explanation of double peak \textbf{QPOs}. Such double \textbf{QPOs} are recently directly observed in \textbf{AGNs} (see \cite{Carpano}) and they  might be directly linked to the  \textbf{RAD} systems, giving independent estimates of the mass and spin of the Kerr \textbf{SMBH}.

{Finally, there are    possible  relevant  extensions of the \textbf{RAD} model to be considered, according to the specific frame where the aggregate hypothesis may be applied. There are two main  classes of model generalizations, the first involving a change in  each aggregate  component, for example in \cite{Fi-Ringed} magnetized \textbf{RADs} have been considered. Secondly, assumptions of \textbf{RAD} entire structure may be modified, particularly on regards of the aggregate symmetries. We  discussed the relevance of tori misalignment in \textbf{RADs} in several parts of this  work. Considering the different properties of in-falling matter  from several  companions during the accretion processes into the Kerr black hole,  a  tori misalignment   is highly expected at least in the early phases of formation of the \textbf{RADs}. The presence of initial misalignment  will lead also to a change of the  entire \textbf{BH}-disk system  inducing  a variation  of the \textbf{BHs} properties  due to modifications of the  spin magnitude  and orientation.

The \textbf{BH}-\textbf{RAD} tori  considered here might  be seen as the final state of this complex scenario.
In fact in several cases   the final ending  of the  \textbf{BH}- misaligned tori   dynamics    results in  equatorial disks.
{Wind and jet emission can be present in \textbf{RADs}, showing a very complicated behavior.
This might possibly lead to non linear dynamics, and  as a consequence  of this, the \textbf{RAD} system may be subjected to deterministic  chaos (e.g. \cite{May}).  Moreover, toroidal components of the \textbf{RAD} can exchange matter and angular momentum  leading  to different effects, this situation has been discussed in an evolutionary scenario in \cite{dsystem}: a possibility  includes ``drying-feeding'' effects or { evolutionary loops among tori}  of a \textbf{RAD} couple as described in  \cite{dsystem}.
Another possibility in this two-steps  exchange of angular momentum  between tori, is   the excitation of   oscillations  which   could eventually  be observed  as it occurs in the matter and angular momentum exchanges in    certain class of binary stars.
This   behavior in general  could be connected with the  quasi-periodic or
chaotic oscillations of  the  \textbf{RADs}. We shall deepen this  aspect of the \textbf{RAD} dynamics and the model generalizations  in  future works.}
Particularly inclined tori will be set first in the simplest case of  the spherically symmetric background of the Schwarzschild black hole   and  then slow rotating spacetimes will be analyzed. The ranges of spin variation in the slow regime will be located  considering also the  limits provided in the \textbf{SMBHs} classes provided of  Table\il(\ref{Table:nature-Att}). A further extension  on the model affecting tori symmetries is for the   warped disk analysis where the tori  warping will be  included.}
{Then in  binary \textbf{BHs}  systems (\textbf{BBHs}), \textbf{BHs}  spins are  usually not aligned \citep{Moran:2008dv}, a complex \textbf{RAD} related  problem  is considering \textbf{BBHs} with two mini-disks and a circumbinary disks, with not aligned \textbf{BHs} spins.}

\begin{acknowledgements}
D. P. acknowledges support from the Junior  grant of the Czech Science Foundation No:16-03564Y.
Z. S. acknowledges  excellence grant of the Albert Einstein Centre for Gravitation and Astrophysics of Czech Science Foundation  No.
 14-37086G.
 The authors thank the anonymous referees for valuable comments on the manuscript.
 \end{acknowledgements}

\begin{appendix}
\section{Notes on  significant spin-mass ratios $a/M$}\label{Appendx:unique}
In this Appendix, we list further  relevant limiting values of the Kerr \textbf{BH} spin, discussing the main features of the \textbf{BHs}   in relation to the \textbf{RAD} toroidal  components.
We  refer to Table\il\ref{Table:nature-Atcol} and Table\il\ref{Table:nature-pics}.
Moreover, Table \ref{Table:coud-many} lists some general  properties of the  tori  couples,  some of these were discussed in  Sec.\il\ref{Sec:RADs-sis}.
\begin{table*}[h!]
\caption{\label{Table:nature-pics}Classes of attractors. Kerr \textbf{BH}  spin regulating  the  magnitude of ratio  $\ell_i/\ell_o$  for a seed of orbiting tori, in a defined regions of $(\ell-a)$ plane as in Fig.\il\ref{Figs:GROUND-StA0}.  Relations between the range of variation of the fluid specific angular momentum  having relevance in the collisional problems.  Arrows are in accordance to the decreasing spin (left column)  or  increasing spin (right column),
starting from the initial spin in one column  of the considered range  of spin and ended by a second arrow of the other column.}
\centering
\begin{tabular}{c|c|c|c|c}
\hline\hline
\textbf{\upshape Classes of attractors}
&\textmd{\scriptsize\textbf{down-to-top}}& \textbf{\upshape Decreasing BH spins}& \textmd{\scriptsize\textbf{top-to-down}}& \textbf{\upshape Increasing BH spins}\\
\hline
${{a_{I}}}\equiv 0.172564M:-\ell_{\mso}^+=\ell_{\mbo}^-$&$\uparrow$&$|\mathbf{L1}^+|\subset (\mathbf{L2}^-\cup\mathbf{L1}^-)$& $\downarrow$&$ |\mathbf{L1}^+|\subset\mathbf{L2}^-,\;|\mathbf{Li}^+|> \mathbf{L1}^- $
\\
    \hline
    ${a_{III}}\equiv0.390781M:\ell_{\gamma}^-=-\ell_{\mbo}^+$&$\uparrow$&$|\mathbf{L1}^+|\subset \mathbf{L2}^- $&$\downarrow$&$ |\mathbf{L1}^+|\cap \mathbf{L2}^-\neq\emptyset,\;|\mathbf{L1}^+|<\mathbf{L2}^-$
     \\
    \hline
    ${a_{V}}\approx0.5089M:-\ell_{\mso}^+=\ell_{\gamma}^-$&$\uparrow$&$ |\mathbf{L1}^+|\cap \mathbf{L2}^-\neq\emptyset, \;|\mathbf{L1}^+|<\mathbf{L2}^-$& $\downarrow$&$ |\mathbf{L1}^+|\cap \mathbf{L2}^-=\emptyset$, $|\mathbf{L1}^+|>!\mathbf{L2}^-$
\\
\hline
\end{tabular}
\end{table*}
\begin{table*}[h!]
\caption{\label{Table:nature-Atcol}Classes of Kerr attractors. For a spin value $a_{\bullet}$, the classes $\mathbf{A}_{\bullet}^{\lessgtr}$ stand for  the ranges $0\leq a <a_{\bullet}$ and $ a_{\bullet}<a\leq M$ respectively. Definition of angular momenta $({\ell}_*,{\ell}_{\varrho}^{\pm},{\ell}_{\beta}^-,\ell_{\Gamma}^{-},\ell_{\mu}^-,\ell_{q}^-)$ are in Table\il\ref{Table:def-flour}-- see also Table\il\ref{Table:nature-Att}.}
\centering
\begin{tabular}{c|c|c|c}
\hline\hline
\textbf{\upshape SMBHs Spins}
& &\textbf{\upshape SMBHs Spins}
&\\
\hline
$a_{\theta} = 0.201697M\in ]a_{I},a_{\iota}[:$&$\ell_{\mso}^-=\ell^-(r_{\mbo}^+)$ and $ r_{\mso}^+=r_{cent}^ {1^-}$, $r_{\mbo}^+=r_{\times}^{1^-}$
&
   $a_{II}=0.382542M\in]a_{K},a_{III}[:$&$\ell_{\gamma}^-=-\ell_{+}(r_{\mso}^-)$,
\\
\hline
    ${a_{\mu}} =0.618034 M:$&$\ell^{+}_{\varrho}=\ell_{\mu}^-$ $ r_{\mso}^-\non{\in} \cc_2^{+} (*) $ $ r_{\mso}^-\non{\in} \cc_2^{+} (**)$
 &
$ a_o = 0.728163M\in]a_1, a_{VI}[:$&$\ell_{\mbo}^-=\ell^-(r_{\mbo}^+)$
\\
\hline
$a_{\varsigma} = 0.867744M\in]a_b,a_{\mathcal{M}}^{(3)}[:$&$\ell_{\gamma}^-=\ell^-(r_{\mbo}^+)$.
\\
\hline
\end{tabular}
\end{table*}
\textbf{Black hole spins: $({{a_{I}}},{a_{III}},{a_{V}})$}
 Definitions of    Kerr \textbf{BH} spins ${{a_{I}}}, {a_{III}}$ and ${a_{V}}$,    introduced    in Table\il\ref{Table:nature-pics},   follow  the  analysis  of the  regions of the  $(\ell-a)$ plane   in  Fig.\il\ref{Figs:GROUND-StA0} and Fig.\il\ref{Figs:pranz}.  Relations between the angular momentum ranges   are relevant in the context of tori collision   \citep{long,dsystem}. If  the   tori of the   $\ell$corotating sequences   have close  values of the   specific angular momentum in one of the $\mathbf{Li}$  range (see  Sec.\il\ref{Sec:notation-sec}),   then tori  are very close. For the $\ell$counterrotating subsequences  the situation is more complex, and to fix constraints on the relative location of the tori it is necessary to consider  the ratios $\ell_i/\ell_o$ for the adjacent configurations-- \citep{ringed,open,long}.
\begin{table*}[h!]
\caption{General considerations on the tori couples  in the Kerr spacetimes.}\label{Table:coud-many}
\centering
\begin{tabular}{l}
\hline\hline
\textbf{\texttt{a:}} Couples of tori
\textbf{(-)}$\cc_{\times}^{\pm}<\cc^{\pm}$, \textbf{(-) }  $\cc_{\times}^-<\cc^{+}$ and\textbf{ (-)}  $\pp^+<\cc^-$  can be observed  in all Kerr spacetimes: $0\leq a\leq M$.
\\\\\hline
\textbf{\texttt{b:}} The   couple $\cc_{\times}^{-}< \cc_{\times}^{+}$  can be observed around any Kerr \textbf{BH} only ($a\neq0$).
\\ \hspace{0.5 cm}For these  tori,  the lower  is the  \textbf{BH}  dimensionless spin, i.e., $a\lessapprox a_u$,
\\
\hspace{0.5 cm}the lower must be the  specific angular momentum $\ell^-$.
\\\\\hline
\textbf{\texttt{c:}}  There is no correlation for the couples   $\cc^+< \oo_{\times}^- $.
\\\\\hline
\textbf{\texttt{d:}} Relations \textbf{(-)} $\cc_{\times}^+\prec \cc_{\times}^+$ and \textbf{(-)} $\oo_{\times}^-\prec \cc_{\times}^+$ holds in any spacetime.
\\\\\hline
\textbf{\texttt{e:}} The tori couple  $\pp_3^+<\pp^-$  can be observed only as    $\pp_3^+<\pp_3^-$.
\\\\
\hline
\end{tabular}
\end{table*}
\section{Comments on the  \textbf{SMBHs} classes $\mathbf{A}_{K}^>$ and $\mathbf{A}_{K}^<$}\label{Appendix:sub-reference}
\begin{figure}[h!]
  \includegraphics[width=1\columnwidth]{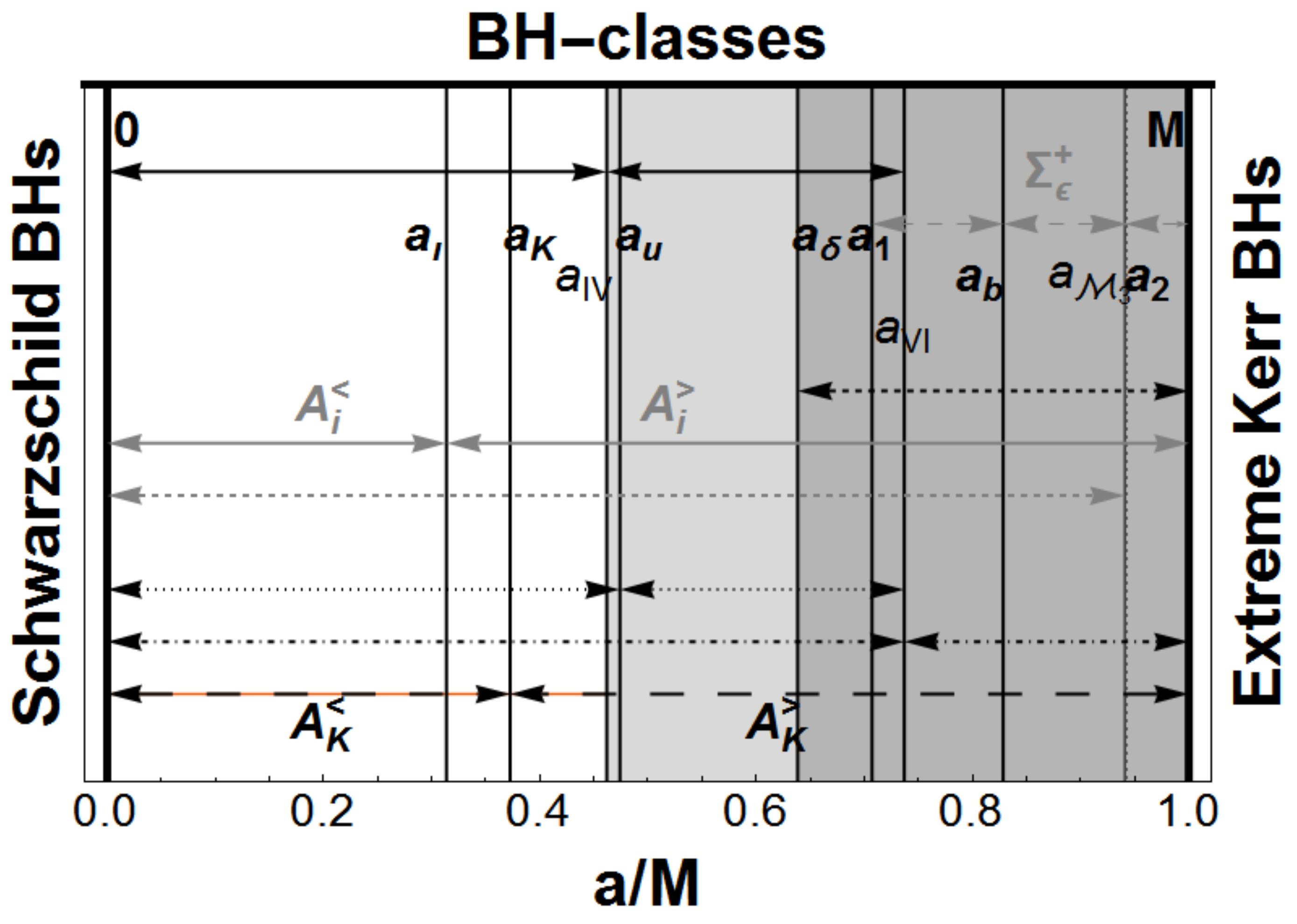}
  \caption{Representations of the main classes of attractors defined in accordance with the dimensionaless spins. It follows the analysis of the Sec.\il\ref{Sec:RADs-sis}. See also Figs\il\ref{Figs:GROUND-StA1}.}\label{Fig:Relevany}
\end{figure}

We examine some general properties of the  $\mathbf{A}_{K}^>$ and $\mathbf{A}_{K}^<$ classes as in Figs\il\ref{Figs:GROUND-StA1}, \ref{Fig:Relevany}  and Table\il\ref{Table:nature-Att}, depending on  the location of the relevant radii $R$.
Results discussed here refer also to the analysis presented in \cite{open} and detailed in \cite{long}.
We  focus  first on counterrotating fluids. According to  Table\il\ref{Table:nature-Att},  considering the launching point of a proto-jet, $r_{\jj}^+$, it can be  $r_{\jj}^+<r_{\mbo}^-$  for sufficiently low \textbf{BH} spin, i.e. \textbf{BH} in the  $\mathbf{A}_{\iota}^<$ class,  having  low magnitude of the specific angular  momentum.
Then the launching point    $r_{\jj}^+>r_{\mso}^-$ in the geometries  $a>a_{\delta}$, and
$r_{\jj}^+<r_{\mso}^-$
  for slower spinning  attractors (but still in  the geometries  $\mathbf{A}_{K}^>$), and   for sufficiently high angular momentum  magnitude.
  On the other hand,  for low \textbf{BH} spins,  $\mathbf{A}_{K}^<$, there is $r_{\jj}^+<r_{\mso}^-$
 for all values of angular momentum.
 The \emph{jet-accretion} correlation  is always possible, except in the $\ell$counterrotating  couples  in the geometries of the fastest spinning \textbf{BHs}, $\mathbf{A}^>_{K}$, made up by a corotating fluid in accretion and a counterrotating jet,  where there is $r_\times^-<r^+_J$.
The  \emph{accretion-equilibrium}  correlation is impossible or
	 subjected to particularly restrictive constraints   for   the counterrotating tori  orbiting in the geometries of   attractors with large spins ($\mathbf{A}_{K}^>$); these tori  must be the outer of a  couple with an inner corotating accreting torus. In the case of corotating equilibrium torus,  a correlation is always possible.
The most significant aspect of this case   is that,   in the $\mathbf{A}_{K}^>$ geometries, the point of accretion  satisfies   $r_{\mso}^+<r_{\times}^-<r_{\mso}^-$.

Focusing on the inner and outer $(r_{in}^+, r_{out}^+)$  edges of the  quiescent $\cc^+_1$ torus,  there is  $r_{in}^+<r_{\mso}^+<r_{out}^+$ without being unstable,
but it cannot ``contain'' the radius  $r_{\mbo}^-$. The torus can contain $r_{\mso}^-$, i.e. $r_{in}^+<r_{\mso}^-<r_{out}^+$, in the spacetimes  of $\mathbf{A}_{K}^>$ class  or  in the case
or   $\mathbf{A}_{K}^<$ with some constraints of the  angular momentum magnitude.
Concerning the inner and outer $(r_{in}^+, r_{out}^+)$  edges of the  quiescent   $\cc_2^+$ (whose unstable topology is that of a proto-jet), there is $r_{in}^+<r_{\mso}^-$
only for  slower  spinning attractors, and low magnitude of the $\ell$ parameters.
Again, we emphasize  the role of the  $\ell/a$ ratio especially in relation to the instability.

We examine now the class of attractors with $a<a_{\Gamma}$, when $r_{\gamma}^+$ can always be  contained (in the sense $r_{in}^-<r_{\gamma}^+<r_{our}^-$) in  $\pp^-_1$, but never in $\cc^-_3$,
while it can be contained in a  $\pp^-_2$ configuration for  a sufficiently low angular momentum.
 In the geometries of the  faster spinning attractors, radius $r_{\gamma}^+$ can be always contained in  $\pp^-_1$ and  $\pp^-_2$,
  but  for the  $ \cc^-_3$ torus  this can occur only for  low angular momentum.
  For $a\in\mathbf{A}^<_{K}$, the orbit $r_{\mbo}^+$  is never contained in the  $\cc^-_3$ configuration;
   in this class of \textbf{BH} attractors, $r_{\mbo}^+$  is contained in the tori $\cc^-_2$ only for  low specific angular  momentum.
    The situation is  in general more articulated for the tori $ \cc^-_1$ and  $\cc^-_2$, depending on the \textbf{BHs} spin, related to the limiting spin value   $a_{\beta}$.  In  the geometries $\mathbf{{A}}_*^<$, the marginally stable orbit $r_{\mso}^+>r_{\mso}^-$  can always be  contained in a $\pp_1^-$ configuration,   while  in $\pp_2^-$ only for  low values of specific angular momentum.  For large  spin attractors  instead, it can be included in $\pp_1^-$  and  $\pp_2^-$, and  in $\pp_3^-$  only for low spins.
\medskip

{ We conclude this section by noting that, as demonstrated  in  \citet{ringed}, we can generalize the definition of an effective potential function  $V_{eff}$  in \il(\ref{Eq:scond-d})  for each   orbiting torus,  to the system  of multiple tori\footnote{The \textbf{RADs} effective potential may be derived from  composite energy-momentum tensor made by  collections of  each fluid tensors  decomposed in each fluid adapted frame.  They  will be naturally coupled  through the  unique  background metric tensor $g_{\mu \nu}$ and certain boundary conditions imposed on the fluid density and pressure.{Clearly  the projection after $3 + 1$ decomposition defining the  3D hyperplane $h^{(n)}_{ij}$ as in (\ref{E:1a0}) has to be done according to the orthogonality condition defining fluids  field velocity vectors  $\mathbf{u}^{(n)}$ respectively, where $(n)$ is the configuration index, different for each torus of the \textbf{RAD}. Boundary condition defining the \textbf{RAD} in the two forms of the \textbf{RAD} potential in (\ref{Eq:def-partialeK}),   by the  step-functions cuts $H(\theta)$ will be included in the energy momentum tensor. There are however some special cases when $\mathbf{u}^{(n_i)}=\mathbf{u}^{(n_j)}$ (note that the fluids  four-velocity here  has for any torus an azimuthal and temporal $t$ component only), occurring for example in excretion tori which are not possible here.}}, eventually producing  a  \textbf{RADs} effective potential. More specifically, we  introduce the  effective potential $\left.V_{eff}^{\mathbf{C}^n}\right|_{K_i}$ of the  \emph{decomposed} $\mathbf{C}^n$ macro-structure and {the effective potential  $V_{eff}^{\mathbf{C}^n}$ of the configuration}
defined respectively   as
\bea&&\label{Eq:def-partialeK}
\left.V_{eff}^{\mathbf{C}^n}\right|_{K_i}\equiv\bigcup_{i=1}^n V_{eff}^{i}\Theta(-K_i), \
\\&&\nonumber
V_{eff}^{\mathbf{C}^n}\equiv\bigcup_{i=1}^n V_{eff}^{i}(\ell_i)\Theta(r_{min}^{i+1}-r)\Theta(r-r_{min}^{i-1}),\
\\&&\nonumber  r_{min}^{0}\equiv r_{+},\quad r_{min}^{n+1}\equiv+\infty,
\eea
where $\Theta(-K_i)$ is the Heaviside (step) function  such that $\Theta(-K_i)=1$  for $V_{eff}^{i}<K_i$ and $\Theta(-K_i)=0$ for $V_{eff}^{i}>K_i$, so that the curve $V_{eff}^{\mathbf{C}}(r)$ is the union of each curve $V_{eff}^{i}(r)<K_i$ of its decomposition.
 Potential  $\left.V_{eff}^{\mathbf{C}^n}\right|_{K_i}$ regulates  behavior of each ring, taking into  account the  gravitational effects induced by the background, and  the centrifugal effect induced by the motion of the fluid, while the potential  $V_{eff}^{\mathbf{C}^n}$  governs the individual configurations considered as part of the macro-configuration.}

\begin{table}[h!]
\centering
\caption{\label{Table:def-flour}Definition of angular momenta $(\ell^{\pm}_{\varrho},\ell_{\mu}^-)$. {See also Table\il\ref{Table:nature-Att}, Figs\il\ref{Figs:GROUND-StA1} and Table\il\ref{Table:nature-Atcol}.}}
\begin{tabular}{c|c}
\hline
${\ell}_*: V_{eff}({\ell}_*,r_{\mso}^+)=1$
&
$\ell^{\pm}_{\varrho}(a/M):\; V_{eff}(\ell^{\pm}_{\varrho},r_{\mso}^{\pm})=1$
\\\\
\hline
 ${\ell}_{\beta}^-:\; V_{eff}(\ell_{\beta}^-,r_{\mbo}^+)<1$
&
$\ell_{\Gamma}^{-}:\; V_{eff}(\ell_{\Gamma}^{-},r_{\gamma}^+)=1$
\\\\
\hline
$\ell_{\mu}^-:\;V_{eff}(\ell_2^+, r_{\mso}^-)=1$
&$\ell_{q}^-\equiv{\ell_{\varrho}}^-\in \mathbf{L2}$
\\
\hline
\end{tabular}
\end{table}
\end{appendix}

\end{document}